\documentclass[twocolumn]{aastex631}

\usepackage{bm}
\usepackage{amsmath}
\usepackage{graphicx}       
\usepackage{multirow}
\usepackage{soul}

\begin{document}

\title{Measurements of the Low-Acceleration Gravitational Anomaly from the Normalized Velocity Profile of Gaia Wide Binary Stars and Statistical Testing of Newtonian and Milgromian Theories}

\correspondingauthor{Kyu-Hyun Chae}
\email{chae@sejong.ac.kr, kyuhyunchae@gmail.com}

\author[0000-0002-6016-2736]{Kyu-Hyun Chae}
\affiliation{Department of Physics and Astronomy, Sejong University, 209 Neungdong-ro Gwangjin-gu, Seoul 05006, Republic of Korea}



\begin{abstract}
Low-acceleration gravitational anomaly is investigated with a new  method of exploiting the normalized velocity profile $\tilde{v}\equiv v_p/v_c$ of wide binary stars as a function of the normalized sky-projected radius $s/r_{\rm{M}}$ where $v_p$ is the sky-projected relative velocity between the pair, $v_c$ is the Newtonian circular velocity at the sky-projected separation $s$, and $r_{\rm{M}}$ is the MOND radius. With a Monte Carlo method Gaia observed binaries and their virtual Newtonian counterparts are probabilistically distributed on the $s/r_{\rm{M}}$ versus $\tilde{v}$ plane and a logarithmic velocity ratio parameter $\Gamma$ is measured in the bins of $s/r_{\rm{M}}$. With three samples of binaries covering a broad range in size, data quality, and implied fraction of hierarchical systems including a new sample of 6389 binaries selected with accurate distances and radial velocities, I find a unanimous systematic variation from the Newtonian flat line. With $\Gamma=0$ at $s/r_{\rm{M}}\lesssim 0.15$ or $s\lesssim 1$~kilo astronomical units (kau), I get $\Gamma=0.068\pm 0.015$ (stat) $_{-0.015}^{+0.024}$ (syst) for $s/r_{\rm{M}} \gtrsim 0.7$ or $s\gtrsim 5$~kau. The gravitational anomaly (i.e.\ acceleration boost) factor given by $\gamma_g = 10^{2\Gamma}$ is measured to be $\gamma_g = 1.37_{-0.09}^{+0.10}$ (stat) $_{-0.09}^{+0.16}$ (syst). With a reduced $\chi^2$ test of Newtonian and Milgromian nonrelativistic theories, I find that Newtonian gravity is ruled out at $5.8\sigma$ ($\chi^2_\nu=9.4$) by the new sample (and $9.2\sigma$ by the largest sample used). The Milgromian AQUAL theory is acceptable with $0.7\lesssim \chi^2_\nu\lesssim 3.1$. These results agree well with earlier results with the ``acceleration-plane analysis'' for a variety of samples and the ``stacked velocity profile analysis'' for a pure binary sample.
\end{abstract}

\keywords{:Binary stars (154); Gravitation (661); Modified Newtonian dynamics (1069); Non-standard theories of gravity (1118); Wide binary stars (1801)}


\section{Introduction} \label{sec:intro}

The nature of gravity in the low acceleration limit can be directly probed by wide binaries (widely separated, long-period, gravitationally bound binary stars) since any conceivable dark matter density (even if it were detected) in the Milky Way cannot affect their internal dynamics (e.g., \citealt{hernandez2012,banik2018,pittordis2018,banik2019,pittordis2019,hernandez2019,elbadry2019,clarke2020,hernandez2022,pittordis2023,hernandez2023}). For this reason and thanks to the state-of-the-art data provided by the Gaia satellite several statistical analyses of wide binaries have been recently carried out aiming at testing gravity quantitatively and even definitely (\citealt{chae2023a,banik2024,chae2024,hernandez2024}) based on the Gaia data release 3 \citep[DR3;][]{dr3}.

\cite{chae2023a} put forward an algorithm \citep{chae2023b} that calculates the probability distribution of a kinematic acceleration $g=v^2/r$ with respect to the Newtonian gravitational acceleration $g_{\rm{N}}$ between the two stars where $v$ is the relative velocity and $r$ is the separation in the three-dimensional real space, and compares it with the corresponding Newtonian prediction. One key aspect of this algorithm is that the occurrence rate ($f_{\rm{multi}}$) of multiplicity higher than two (i.e.\ harboring hidden additional components) can be self-calibrated at a high enough acceleration and checked at another high acceleration. Through this algorithm, to be referred to as the ``acceleration-plane analysis'', \cite{chae2023a} found that the observed acceleration started to get boosted from the Newtonian prediction for $g_{\rm{N}}\la 10^{-9}$~m~s$^{-2}$ with a boost factor of $\approx 1.4$ for $g_{\rm{N}}\la 10^{-10}$~m~s$^{-2}$, at an extremely high ($>5\sigma$) statistical significance. \cite{chae2023a} considered various samples by varying selection criteria and noted that the low-acceleration gravitational anomaly persisted.

\cite{chae2024} further considered a sample of statistically pure binaries (i.e.\ the limiting case of $f_{\rm{multi}}=0$) for an independent test. For this sample \cite{chae2024} employed another algorithm to be referred to as ``stacked velocity profile analysis'' as well as the acceleration-plane analysis. The stacked velocity profile analysis compares the observed distribution of the sky-projected relative velocities against the sky-projected separations with the corresponding Newtonian prediction. \cite{chae2024} found that the results for the pure binary sample through the two independent algorithms agreed well with each other and the \cite{chae2023a} results for general or ``impure'' samples. This means that the gravitational anomaly at low acceleration is robust with respect to a large variation of the sample selection between $f_{\rm{multi}}\sim 0.5$ \citep{chae2023a} and $f_{\rm{multi}}= 0$ \citep{chae2024}. 

\cite{hernandez2024} carried out a statistical analysis of the distribution of normalized velocities in a pure binary sample that was selected by \cite{hernandez2023}. Here the normalized velocity on the sky plane is defined  \citep{pittordis2018,banik2018} by
\begin{equation}
  \tilde{v} \equiv \frac{v_p}{v_c(s)},
  \label{eq:vtilde}
\end{equation}
where $v_p$ is the observed sky-projected relative velocity between the pair and $v_c(s)$ is the theoretical Newtonian circular velocity at the sky-projected separation $s$. The \cite{hernandez2024} algorithm and sample are completely independent of \cite{chae2023a} and \cite{chae2024} though the sample was drawn from the same Gaia DR3 database. \cite{hernandez2024} obtained a gravitational anomaly that was well consistent with those obtained by \cite{chae2023a} and \cite{chae2024}.

These recent quantitative analyses \citep{chae2023a,chae2024,hernandez2024} consistently show that the gravitational acceleration is boosted by a factor from $\approx 1.3 - 1.5$ and the relative velocity is boosted by a factor of $\approx 1.2$ for $g_{\rm{N}}\la 10^{-10}$~m~s$^{-2}$ or $s\ga 5$~kau (kilo astronomical units). Despite these consistent results from various samples of wide binaries with various independent methods, \cite{banik2024} claimed an opposite conclusion based on their own statistical method and raised a concern for data quality control in \cite{chae2023a} samples.

However, \cite{hernandez2024a} soon pointed out the problems of their methodology. Above all, they knowingly excluded the Newtonian-regime ($g_{\rm{N}}> 10^{-9}$~m~s$^{-2}$ or $s< 2$~kau) binaries that are essential for an accurate determination of $f_{\rm{multi}}$ in any sample.\footnote{The value of $f_{\rm{multi}}$ depends on the selection criteria used to select binaries, and there is an intrinsic degeneracy between $f_{\rm{multi}}$ and the inferred gravity. Thus, the Newtonian-regime binaries are needed to fix $f_{\rm{multi}}$ with the known gravity for the sample under consideration.} Then, they employed a statistical method of fitting the distribution (or number count) of $\tilde{v}$ (Equation~(\ref{eq:vtilde})) in cells in an attempt to simultaneously constrain gravity, $f_{\rm{multi}}$, and other parameters without the Newtonian-regime data. However, in such a approach they improperly used cells smaller than the errors of $\tilde{v}$ although their approach relies on number counts in cells. The reader is referred to \cite{hernandez2024a} for further details.

The concern for data quality control raised by \cite{banik2024} based on the uncertainty of $\tilde{v}$ is not relevant for the acceleration-plane analysis of \cite{chae2023a} because $\tilde{v}$ is not used, but probability distributions of $g$ and $g_{\rm{N}}$ are directly derived on the acceleration plane with sufficiently precise projected relative velocities, and taking into account possible ranges of parameters such as eccentricity, inclination, and orbital phase in the deprojection to the three-dimensional space. At any rate, the pure binary sample of \cite{chae2024,chae2024b} already meets high signal-to-noise ratios ($S/N\ga 10$) for $\tilde{v}$. Moreover, Appendix~B of \cite{chae2024} shows that even with an artificial error cut of $\tilde{v}$ imposed on the \cite{chae2023a} main samples the acceleration-plane analysis returns similar results for the gravitational anomaly for $g_{\rm{N}}\la 10^{-9}$~m~s$^{-2}$. 

Although the \cite{banik2024} criticism of the \cite{chae2023a} acceleration-plane results based on their error cut of $\tilde{v}$ is baseless, it is interesting to investigate the profile of $\tilde{v}$ with respect to the projected separation $s$ normalized by the MOND radius
\begin{equation}
  r_{\rm{M}}\equiv \sqrt{\frac{GM_{\rm{tot}}}{a_0}}
  \label{eq:rmond}
\end{equation}
where $G$ is Newton's constant, $M_{\rm{tot}}$ is the total mass of the system, and $a_0$ is the critical acceleration whose value is taken to be $1.2\times 10^{-10}$~m~s$^{-2}$. \cite{banik2024} showed that the observed $\tilde{v}$ profiles for the \cite{chae2023a} main samples were significantly affected by a cut on the uncertainty of $\tilde{v}$, and used this fact to question the gravitational anomaly reported by \cite{chae2023a}. However, they did not calculate how the Newtonian prediction of the $\tilde{v}$ profile in a sample was affected by modifying the sample with such a cut and thus did not make a valid comparison.

Here I propose an analysis of the $\tilde{v}(s/r_{\rm{M}})$ profile, to be referred to as ``normalized velocity profile analysis'', as a method different from the other methods employed in \cite{chae2023a,chae2024}. This analysis is different from the stacked velocity profile analysis in that both velocity and separation are normalized. It is different from the acceleration-plane analysis in that no deprojection to the 3D space is considered. This method is interesting in its own right as an intermediate method between the acceleration-plane and stacked velocity profile analyses. It can also be used to directly and precisely address the effect of a cut on the uncertainty of $\tilde{v}$. 

In Section~\ref{sec:method}, I describe the wide binary samples and the methodology. In Section~\ref{sec:result}, I present the results. In Section~\ref{sec:disc}, I compare the results with relevant previous results and discuss possible systematic errors. I conclude in Section~\ref{sec:conc}. In Appendix~\ref{sec:accel}, I present the acceleration-plane analyses results from which relevant $f_{\rm{multi}}$ values are derived and used in the main analyses.  {Appendix~\ref{sec:pmscat} addresses effects of the proper motion (PM) measurement uncertainties in the gravity test methods used in this work.} Revised and new python scripts used for this work and the sample of binaries can be accessed at \cite{chae2023b}.  {Most analyses of this work are based on the medians of $\tilde{v}$ (Equation~(\ref{eq:vtilde})) in various bins. Throughout this paper they are denoted by $\langle\tilde{v}\rangle$. This work uses several samples of wide binaries with various implied fractions of hierarchical systems including the limiting sample with $f_{\rm{multi}}\rightarrow 0$. The latter sample was obtained in a systematic way by \cite{chae2024} and revised by \cite{chae2024b}. It will be referred to as the `pure binary' sample as in \cite{chae2024}. However, it should be understood that the sample may include a tiny fraction of individual exceptions that evaded the stringent restrictions that \cite{chae2024,chae2024b} applied statistically.}

\section{Data and Methodology} \label{sec:method}

\subsection{Binary Samples} \label{sec:sample}

As in \cite{chae2023a,chae2024}, this work is based on a sample of binaries within 200~pc from the Sun that are free of clusters, background pairs, and resolved ($1''$) triples or higher-order multiples.  {The sample was drawn from the \cite{elbadry2021} catalog of 1.8 million binary candidates within 1 kpc from the Sun.} The sample\footnote{The sample can be accessed at \cite{chae2023b}.} of 81,880 binary candidates defined by \cite{chae2023a} has the range $0.05 < s < 50$~kau. It is not free of chance-alignment (fly-by) cases or unresolved close companions. Thus, they need to be dealt with carefully in using the \cite{chae2023a} catalog of 81,880 binary candidates.

 {Chance-alignment cases (i.e.\ false binaries) and binaries including undetected close companions (i.e.\ hierarchical systems) are two main issues in assembling a sample and carrying out gravity tests with it. There can be an infinite number of ways of defining a sample with a varying degree of including chance-alignment cases and hierarchical systems, and in principle any sample is permissible as long as chance-alignment cases and hierarchical systems are consistently modeled for the sample.}

 {The \cite{elbadry2021} catalog has unique features. They introduced and estimated a chance-alignment probability $\mathcal{R}$ for each system. For this they used abundant observational information including angular separation, distance, parallax difference uncertainty, local sky density, relative tangential velocity, parallax difference over error, and PM difference over error. In particular, for the relative tangential velocity, i.e.\ the sky-projected relative velocity between the pair $v_p$, to be consistent with a gravitationally bound system, they required that $v_p$ be smaller than the Newtonian expectation for a $5M_\odot$ system (considering that most systems have masses well below this limit) with a tolerance of $3\sigma$ or $6\sigma$ combined uncertainties depending on the separation (their Equations~(3) through (7)).}

\begin{figure*}
  \centering
  \includegraphics[width=1.0\linewidth]{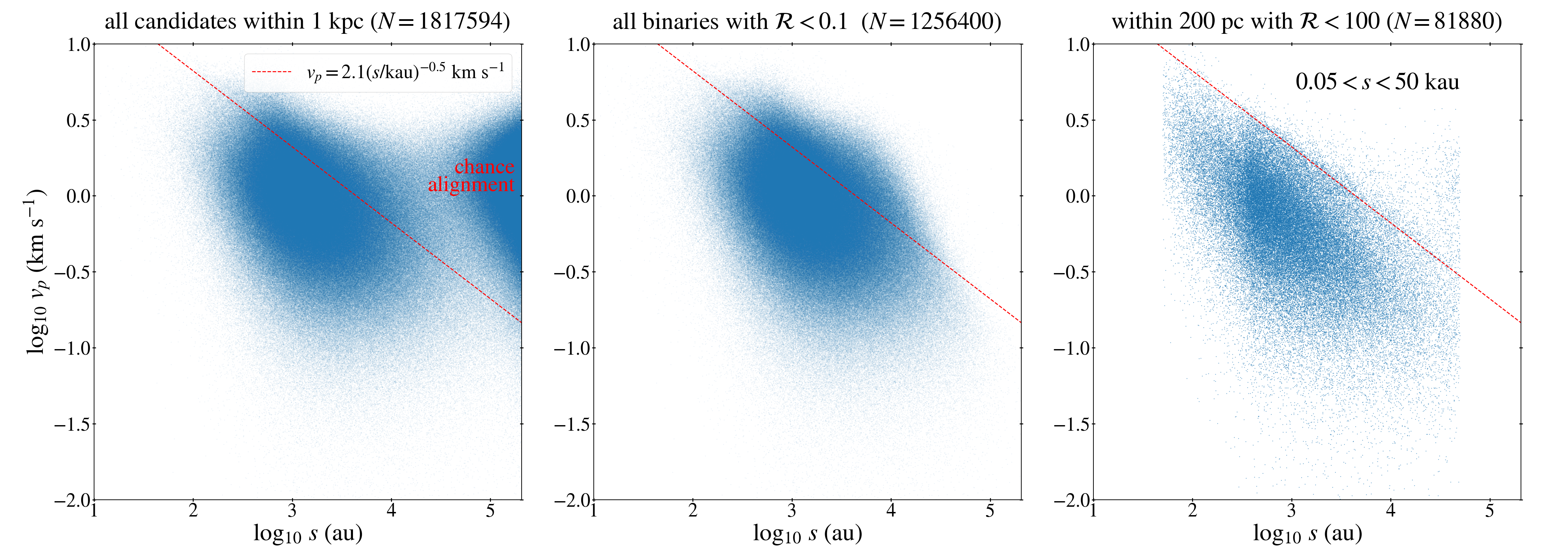}
    \vspace{-0.3truecm}
    \caption{\small 
     {The left panel exhibits all 1.8 million binary candidates assembled by \cite{elbadry2021}. The diagonal dashed red line indicates a kinematic criterion used (with tolerance of $3-6\sigma$ combined measurement uncertainties) by \cite{elbadry2021} to identify chance-alignment cases. Indeed, most chance-alignment cases are above this line. The middle panel shows cases with low chance-alignment probabilities satisfying $\mathcal{R}<0.1$ where $\mathcal{R}$ is the parameter introduced by \cite{elbadry2021}. Note that a significant number of systems are still above the red dashed line. Those systems are more likely to be hierarchical systems hosting hidden close companion stars although hierarchical systems do exist even below the line. The right panel shows the sample within 200~pc from the Sun defined by \cite{chae2023a}.}
    } 
   \label{s_vp}
\end{figure*} 

 {The left panel of Figure~\ref{s_vp} displays all 1.8 million candidates in the plane spanned by the sky-projected separation $s$ and $v_p$. Clearly, this base catalog includes chance-alignment cases above the red dashed line given by Equation~(7) of \cite{elbadry2021}. The middle panel with $\mathcal{R}<0.1$ excludes chance-alignment cases but still includes many systems above the red dashed line. The right panel exhibits the base sample defined by \cite{chae2023a}, which includes some chance-alignment cases as well as hierarchical systems. Here I note that the \cite{elbadry2021} Equation~(7) primarily plays the role of removing chance-alignment cases but may also remove some pronounced hierarchical systems. }

 {Considering the uncertainties of chance-alignment and hierarchical cases in any sample, it is necessary to consider a range of samples rather than relying on one specific sample and check whether gravity test results are robust against the sample variation.\footnote{This is indeed the approach already taken from \cite{chae2023a,chae2024}.} Here taking advantage of the unique features of the \cite{elbadry2021} sample, I remove chance-alignment cases but allow a varying fraction of hierarchical systems. }

\cite{chae2023a} considered a `clean' range of $4<M_G<14$\footnote{ {Because of this limited magnitude range, the total masses are mostly in the range $0.5\la M_{\rm{tot}}/M_\odot\la 2.5$ well below $5M_\odot$ that was used to define Equation~(7) of \cite{elbadry2021}.}} where $M_G$ is the Gaia $G$-band absolute magnitude so that a relatively more reliable part of an empirical magnitude-mass ($M_G-M$) relation could be used (see Figure~7 of \cite{chae2023a}). \cite{chae2023a} presented a couple of $M_G-M$ relations based on the \cite{pecaut2013} results. Since the two $M_G-M$ relations give similar results as shown in \cite{chae2023a,chae2024}, here I consider the standard choice only (the first choice given in Table 1 of \cite{chae2023a}).

In this work I consider two approaches to remove chance-alignment cases. The first approach is to use the cut $\mathcal{R}<0.01$ following \citep{chae2023a}. Specifically, I use a main sample of 19,716 binaries defined in \cite{chae2023a} by requiring that the PM relative (i.e.\ fractional) errors are $< 0.005$ for all components and $0.2<s<30$~kau.\footnote{For the other criteria the reader is referred to \cite{chae2023a}. I note that the larger sample with  {the PM fractional errors} $< 0.01$ is not considered here because its sample is larger by just 35\% though its allowed maximum PM uncertainty is twice.} This sample will be referred to as the `Chae (2023a) sample'. Table~\ref{tab:sample} gives a summary of the samples to be used in this work.

\begin{table*}
  \caption{Samples of binaries used in this study}\label{tab:sample}
\begin{center}
  \begin{tabular}{cccc}
  \hline
 sample   & $N_{\rm{binary}}$ &  key selection criteria & reference/comments  \\
 \hline
 Chae (2023a) & 19716   &  $\mathcal{R}<0.01$, PM relative errors $< 0.005$  &  \cite{chae2023a}  \\
 new & 6389  & relative errors: PM  $< 0.005$, dist  $< 0.01$, RV $<0.5$  &   $\mathcal{R}$ not used (this work)  \\
 pure binary & 3557   &  $\mathcal{R}<0.01$, relative errors: PM  $< 0.005$, dist  $< 0.005$, RV $<0.2$  &  \cite{chae2024,chae2024b}  \\
 Chae (2023a) limited & 5635  &  $\mathcal{R}<0.01$, PM relative errors $< 0.005$, $2<s<30$~kau &  limited range of $s$  \\
\hline
\end{tabular}
\end{center}
Note. (1) The Chae (2023a) limited sample is considered for the purpose of investigating/illustrating the effects of a limited dynamic range.  {(2) The pure binary sample needs to be understood as a statistical sample satisfying $f_{\rm{multi}}=0$ in the Newtonian regime (it may include rare individual exceptions of hierarchical systems).}
\end{table*}

The second approach is to use RVs as well as distances to remove chance-alignment cases for those with measured RVs. Because two stars orbiting each other must be at the same distance from the Sun up to the minor difference of the orbit size and have the same system RV up to the minor difference from the orbital relative motion\footnote{Note that the median $s/d$ is $\approx 4\times 10^{-5}$ and relative radial velocities ($\la 1$~km~s$^{-1}$) are much less than the typical system radial velocity dispersion of $\approx 33$~km~s$^{-1}$ (see Figure~5 of \cite{chae2023a}).}, a requirement of consistent distances and RVs can effectively remove chance-alignment cases detected on the sky. For this, I first select binaries with well-measured PMs, distances, and RVs. The required relative errors are respectively $<0.005$, $<0.01$, and $<0.5$ for PMs, distances, and RVs. Then, I require that distances satisfy
  \begin{equation}
    \left|d_{A} - d_{B}\right| < \sqrt{9(\sigma_{d_{A}}^2+\sigma_{d_{B}}^2)+(6s)^2},
    \label{eq:deld}
    \end{equation}
  and RVs satisfy
  \begin{equation}
    \left|v_{r,A}-v_{r,B}\right| < \sqrt{9(\sigma_{v_{r,A}}^2+\sigma_{v_{r,B}}^2)+(\Delta v_{r,\rm{orbit}}^{\rm{max}})^2}
    \label{eq:delvr}
  \end{equation}
  with
  \begin{equation}
   \Delta v_{r,\rm{orbit}}^{\rm{max}} = 0.9419{\text{ km s}^{-1}}\sqrt{\frac{M_{\rm{tot}}}{s}}\times 1.3\times 1.2,
    \label{eq:vmax}
  \end{equation}
  where $M_{\rm{tot}}$ is the total mass of the system in units of solar mass. Equations~(\ref{eq:deld}) and (\ref{eq:delvr}) are similar to Equations~(1) and (2) of \cite{chae2024} differing only in that $3\sigma$ (instead of $2\sigma$) differences are allowed. A sample based on this selection will be referred to as the `new sample' (Table~\ref{tab:sample}).

  The new sample is interesting for a few reasons. First, the selection is not based on $\mathcal{R}$ that has been used from \cite{chae2023a}. I note that this selection is considered not because the \cite{elbadry2021} $\mathcal{R}$ has any problems in a statistical sense but to provide an independent sampling. It turns out that 99.5\% of the binaries in the new sample satisfy $\mathcal{R}<0.01$ automatically, proving that two independent methods are reliable and consistent. Second, the new sample includes only binaries with measured RVs with $S/N>2$ and thus system velocities are known. Lastly, the new sample has the additional requirement on the precision of distances (i.e.\ parallaxes). Overall, this sample is intermediate between the \cite{chae2023a} main sample and the pure binary sample defined in \cite{chae2024} that has more stringent requirements.

  In this work, I also use the pure binary sample of \cite{chae2024,chae2024b} for the normalized velocity profile analysis. Finally, I consider a subsample from the `Chae (2023a) sample' that has a limited separation range of $2<s<30$~kau. This sample will be referred to as the `Chae (2023a) limited sample'. This sample is considered only for the purpose of illustrating or investigating the effects of limiting the dynamic range because \cite{banik2024} (see also \citealt{pittordis2023}) recently considered such a narrow dynamic range.

\subsection{Empirical Relations to Test Gravity: from Acceleration Relation to Normalized Velocity} \label{sec:relations}

  For the observed R.\ A.\ ($\alpha$) and decl.\ ($\delta$) components of the PMs in a binary, $(\mu_{\alpha,A}^\ast, \mu_{\delta,A})$ and $(\mu_{\alpha,B}^\ast, \mu_{\delta,B}),$\footnote{Here $A$ and $B$ denote the brighter and the fainter components, and $\mu_\alpha^\ast\equiv \mu_\alpha \cos\delta$ for the PM component $\mu_\alpha$.} the plane-of-sky scalar relative velocity $v_p$ is given by
\begin{equation}
  v_p =  4.7404\times 10^{-3}\text{ km s}^{-1}\times \Delta\mu \times  d
  \label{eq:vp_ob}
\end{equation}
where $d$ is the distance in pc to the binary system, and
\begin{equation}
 \Delta\mu = \left[(\mu_{\alpha,A}^\ast - \mu_{\alpha,B}^\ast )^2 + (\mu_{\delta,A} - \mu_{\delta,B} )^2\right]^{1/2}
  \label{eq:PM}
\end{equation}
with all PM components given in units of mas~yr$^{-1}$. Once the system is known/determined to be gravitationally bound, $d$ is given by an error-weighted mean of the measured distances of the two components.

The uncertainty of $\Delta\mu$ (Equations~(\ref{eq:PM})) is estimated as
\begin{equation}
  \begin{array}{ccl}
    \sigma_{\Delta\mu} & = & \left[(\sigma_{\mu_{\alpha,A}^\ast}^2 + \sigma_{\mu_{\alpha,B}^\ast}^2 )(\Delta\mu_\alpha)^2\right. \\
      &  & \left. + (\sigma_{\mu_{\delta,A}}^2 + \sigma_{\mu_{\delta,B}}^2 )(\Delta\mu_\delta)^2 \right]^{1/2}/\Delta\mu, \\
    \end{array}
  \label{eq:PMerr}
\end{equation}
where
\begin{equation}
  \begin{array}{ccl}
  (\Delta\mu_\alpha)^2 & = & (\mu_{\alpha,A}^\ast - \mu_{\alpha,B}^\ast )^2, \\
  (\Delta\mu_\delta)^2 & = & (\mu_{\delta,A} - \mu_{\delta,B})^2. \\
    \end{array}
  \label{eq:PMcomp}
\end{equation}
Then, the uncertainty of $v_p$ (Equation~(\ref{eq:vp_ob})) follows from the uncertainty of $\Delta\mu$.

The normalized velocity $\tilde{v}$ (Equation~(\ref{eq:vtilde})) follows from the measured value of $v_p$ (Equations~(\ref{eq:vp_ob})) and the theoretical Newtonian circular velocity at $s$ given by
\begin{equation}
  v_c(s) \equiv \sqrt{\frac{GM_{\rm{tot}}}{s}}, 
  \label{eq:vc}
 \end{equation}
where $M_{\rm{tot}}$ is the total mass of the system including any hidden additional component(s).

The observed distribution of $\tilde{v}$ with respect to $s/r_{\rm{M}}$ can test the low-acceleration behavior of gravity. This can be understood as follows. Consider a binary with a scalar relative velocity $v$ and a relative separation $r$ in the real 3D space so that $g=v^2/r$ and $g_{\rm{N}}=GM_{\rm{tot}}/r^2=v_c^2(r)/r$. We then have
\begin{equation}
  \frac{g}{g_{\rm{N}}} = \left( \frac{v}{v_c(r)}  \right)^2
  \label{eq:gratio}
\end{equation}
and
\begin{equation}
  \frac{g_{\rm{N}}}{a_0} = \left(\frac{r}{r_{\rm{M}}}\right)^{-2}.
  \label{eq:rratio}
 \end{equation}

As done in \cite{chae2023a}, a direct test of gravity can be done by the distribution of $g$ with respect to $g_{\rm{N}}$ in the acceleration plane. This requires a deprojection from the observed quantities $v_p$ and $s$ to $v$ and $r$ with a Monte Carlo (MC) method. This test is reminiscent of the radial acceleration relation \citep{mcgaugh2016} test in galaxies. Now one can alternatively consider the ratio $g/g_{\rm{N}}$ with respect to $g_{\rm{N}}$, or equivalently $v/v_c(r)$ with respect to $r/r_{\rm{M}}$ using Equations~(\ref{eq:gratio}) and (\ref{eq:rratio}). Finally, without a deprojection one can use $\tilde{v}$ as a proxy for $v/v_c(r)$ and $s/r_{\rm{M}}$ for $r/r_{\rm{M}}$. 

\begin{figure*}
  \centering
  \includegraphics[width=0.8\linewidth]{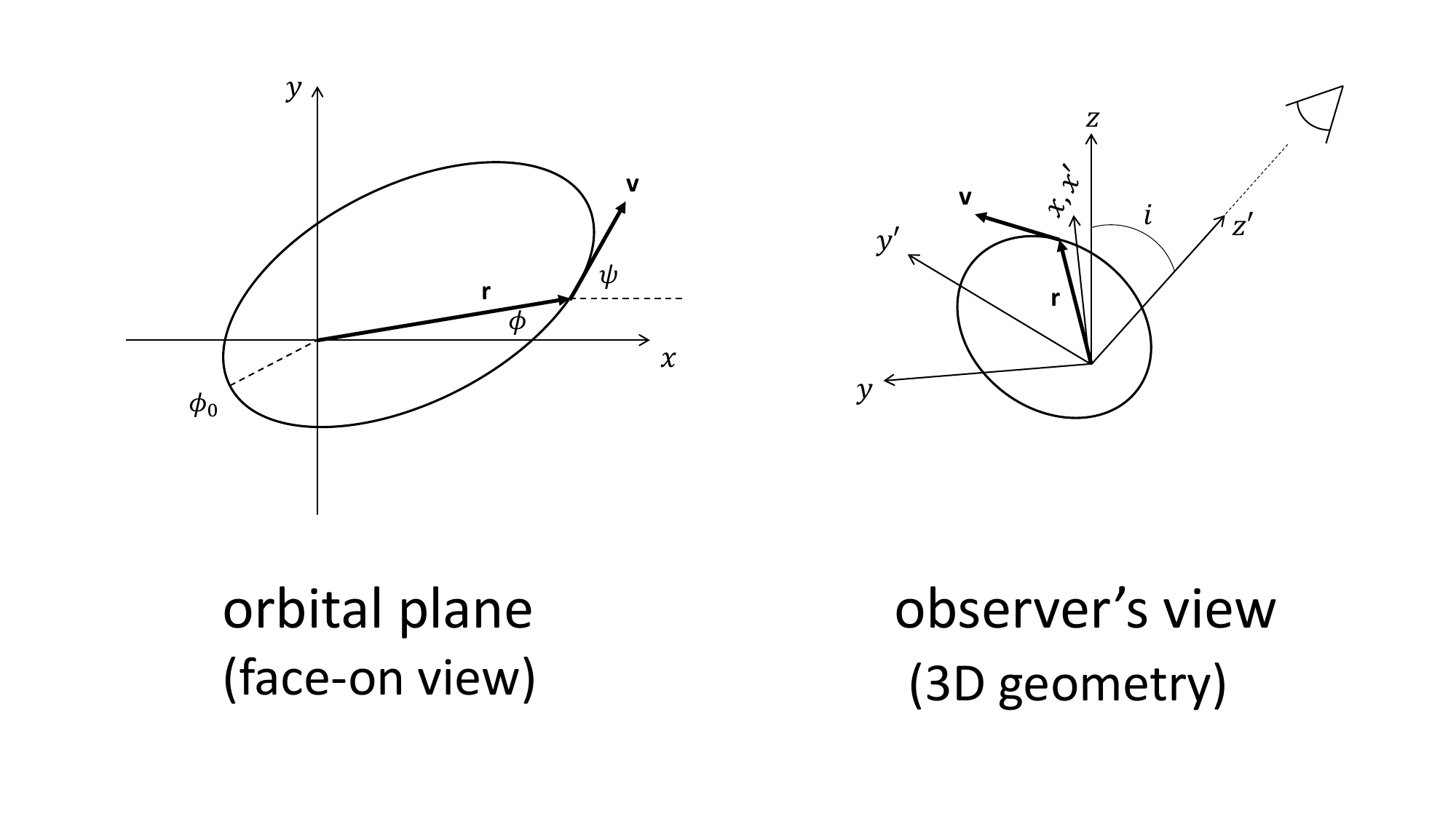}
    \vspace{-0.2truecm}
    \caption{\small 
    Adapted from \cite{chae2023a}. The left panel shows a one-particle equivalent description of the relative motion between the two stars in a binary system. The right panel defines the observer's viewpoint at an inclination $i$. 
    } 
   \label{orbit}
\end{figure*} 

\subsection{ {Normalized Projected Velocities from Proper Motions: Measured versus Newtonian Simulation}}  \label{sec:calcvt}

The observed normalized velocities $\tilde{v}(=v_p/v_c(s))$ will be statistically compared with their Newtonian counterparts that are generated by a Monte Carlo method. Here I describe how normalized velocities are calculated for the Gaia observed PMs and mock Newtonian PMs. The methodology follows from \cite{chae2023a} with a minor correction described in Appendix~A of \cite{chae2024}. I summarize the essential concepts and elements while referring the reader to the relevant further details in \cite{chae2023a,chae2024}.

\subsubsection{Measured Proper Motions} \label{sec:PMobs}

The projected relative velocity $v_p$ (Equation~(\ref{eq:vp_ob})) follows from the observed PMs through Equation~(\ref{eq:PM}). The Newtonian circular velocity (Equation~(\ref{eq:vc})) depends on the total mass that may include the mass(es) of any unresolved hidden component(s). In each MC, a randomly selected fraction of binaries with probability $f_{\rm{multi}}$ are given additional components: 70\% of the hierarchical occurrences are assumed to be triples while 30\% are quadruples. The mass of a close companion is probabilistically assigned with the empirical distribution of magnitude difference between the host and the companion as described in Section~3.2 of \cite{chae2023a}.

\subsubsection{Newtonian Proper Motions: without hidden companions} \label{sec:PMpure}

Each wide binary was observed by Gaia only at one phase of the orbital motion (in its long period) of an unknown eccentricity at an unknown inclination. Lacking the detailed information on the orbit, it is not possible to obtain a Newtonian mock PM as a precise counterpart to the observed PM. One can obtain only a random realization out of broad possible ranges. Thus, an individual mock Newtonian PM cannot be individually compared with the observed counterpart PM. Only a sufficiently large collection of mock Newtonian PMs can be statistically compared with the counterpart collection of observed PMs.

Keeping the above statistical nature in mind, now I describe how a random Newtonian PM is obtained for the observed binary system with masses $M_A$ and $M_B$ and the sky-projected separation $s$. Consider the one-particle equivalent motion described in Figure~\ref{orbit} taken from \cite{chae2023a}. The one-particle equivalent motion can be conveniently used for the Newtonian simulation of the relative motion between the two stars.

The Newtonian mock PMs are obtained as follows.
\begin{enumerate}
\item For each system, eccentricity ($e$) is assigned using the observational input as described in Section~\ref{sec:input}. The longitude of the periastron ($\phi_0$) is assigned from the range $(0,2\pi)$. The phase angle $\phi$ is obtained by solving Equation~(10) of \cite{chae2023a} for a time $t$ randomly drawn from $(0,T)$ where $T$ is the period also determined from the same equation. The inclination angle $i$ is drawn from $(0,\pi/2)$ with a probability density function $p(i)=\sin i$.
\item The 3D separation $r$ is obtained by $s/\sqrt{\cos^2\phi+\cos^2i\sin^2\phi}$. The semi-major axis follows from $a=r(1+e\cos(\phi-\phi_0))/(1-e^2)$.
\item The magnitude of the relative 3D velocity is given by
  \begin{equation}
    v(r) = \sqrt{\frac{GM_{\rm{tot}}}{r} \left(2- \frac{r}{a} \right)}.
    \label{eq:vN}
  \end{equation}
  The sky-projected relative velocity components are then given by
  \begin{equation}
    \begin{array}{lll}
    v_{p,x^\prime} & = & v(r) \cos \psi, \\
    v_{p,y^\prime} & = & v(r) \cos i \sin\psi, \\
    \end{array}
    \label{eq:vpcomp}
  \end{equation}
  with
  \begin{equation}
  \psi = \tan^{-1} \left( - \frac{\cos\phi+e\cos\phi_0}{\sin\phi+e\sin\phi_0} \right)+\pi.
  \label{eq:psi}
\end{equation}
\item The mock PM components are given by
  \begin{equation}
    \begin{array}{lll}
      \mu^\ast_{\alpha,A} & = & \mu^\ast_{\alpha,M}\pm(M_B/M_{\rm{tot}})v_{p,x^\prime}/d_A, \\
      \mu^\ast_{\alpha,B} & = & \mu^\ast_{\alpha,M}\mp(M_A/M_{\rm{tot}})v_{p,x^\prime}/d_B, \\
      \mu_{\delta,A} & = & \mu_{\delta,M}\pm(M_B/M_{\rm{tot}})v_{p,y^\prime}/d_A, \\
      \mu_{\delta,B} & = & \mu_{\delta,M}\mp(M_A/M_{\rm{tot}})v_{p,y^\prime}/d_B, \\
    \end{array}
    \label{eq:mockPM}
  \end{equation}  
  where $d_A$ and $d_B$ are distances to the components, which can be taken to be the same as the error-weighted mean of the observed distances,\footnote{Because $s/d\ll 1$, it makes practically no difference whether one uses a common $d$ or individual distances as detailed in \cite{chae2023a}.} and $\mu^\ast_{\alpha,M}$ and $\mu_{\delta,M}$ are physically irrelevant constants chosen to be the error-weighted means of the observed PM components.  
\end{enumerate}

Once the mock PMs (Equation~(\ref{eq:mockPM})) are obtained, $v_p$ and then $\tilde{v}$ can be calculated in the same manner as used for the observed PMs.

\subsubsection{Newtonian Proper Motions: with hidden companions} \label{sec:PMhir}

If the binary system has a hidden inner binary for one component (or two inner binaries for both components), the apparent motion of a photocenter with respect to the barycenter in each inner binary is calculated and added to the outer velocity components. The details can be found in step \#{7} of Section~3.4 of \cite{chae2023a}. Here I give a brief description.

The same companion masses randomly assigned for the observed binaries (as described in Section~\ref{sec:PMobs}) are used for the simulation. Thus, mock and real systems share the same masses and the same projected separation. The semi-major axis of the inner orbit $a_{\rm{in}}$ is sampled from 0.01~au to ($d$/pc)~au with a steep power-law distribution of probability density $p \propto (a_{\rm{in}}/a_{\rm{out}})^{-0.8}$ (Figure~14 of \citealt{chae2023a}; a nearly uniform probability density in log space) that was derived from near-by surveys (e.g.\ \citealt{tokovinin2008}). The upper limit represents $1''$ \citep{lindegren2021} within which unresolved component may be hidden. Other parameters such as $i$, $e$, $\phi_0$ and $\phi$ are assigned in a similar way as for, but independently of, the main outer orbit (see \citealt{chae2023a} for the details).

Once the sky-projected relative velocities for the inner orbit are obtained, they are multiplied by a dimensionless factor $\eta_{\rm{phot}}$ that mainly accounts for the distance of the photocenter from the barycenter normalized by the 3D separation between the host and the companion. The value of $\eta_{\rm{phot}}$ is given by
  \begin{equation}
    \eta_{\rm{phot}} = \left\{
    \begin{array}{l}
      0    \text{\hspace{18ex} ($P_{\rm{in}}<3$ yr)}, \\
      \frac{M_h M_c (M_h^{\alpha-1} - M_c^{\alpha-1})}{(M_h+M_c)(M_h^\alpha + M_c^\alpha)} \text{\hspace{1ex}($\theta_{\rm{in}}<1''$)}, 
    \end{array} \right.
    \label{eq:rphot}
  \end{equation}  
where $P_{\rm{in}}$ is the period, $\theta_{\rm{in}}$ is the projected angular separation, and $M_h$ and $M_c$ are the masses of the host and the companion, and luminosity $\propto$ (mass)$^\alpha$ with $\alpha=3.5$. As indicated in Equation~(\ref{eq:rphot}), if the period of the inner orbit is shorter than the time span of 3 years of Gaia DR3 observations, the motion does not make a fixed contribution to PM components but may have contributed to the reported uncertainties of PM components and parallax.\footnote{ {Incidentally, some of hierarchical systems hosting sufficiently massive very close hidden companions may have already been excluded from the binary sample either because of large uncertainties in PMs (and/or RVs) or unusually large relative velocities $v_p$ or $v_r$.}} Finally, the PM components from the inner orbit follow from the adjusted velocities, and are component-by-component added to the PM components from the outer orbit.

\subsection{Observational Inputs}  \label{sec:input}

The MC procedure requires various observational inputs including the magnitude-mass relation, orbital eccentricities, and the statistical properties of hidden close companions (see \cite{chae2023a} for the details). As noted in Section~\ref{sec:sample}, it is sufficient to consider the standard magnitude-mass relation for the considered magnitude range. \cite{chae2023a} considered several variations in the statistical properties of hidden close companions and found no significant systematic concern as long as $f_{\rm{multi}}$ is self-calibrated for each sample.

Eccentricities can have relatively large impact on wide binary tests of gravity because the Newtonian prediction on the relative velocity and the ``kinematic acceleration'' depend significantly on the orbital eccentricity for the observed mass and sky-projected separation of the binary system. Eccentricities for all binaries of the \cite{elbadry2021} sample were investigated by \cite{hwang2022} by analyzing the angle between the relative displacement vector and the relative velocity vector on the sky. They provide binary-specific eccentricity ranges for all binaries.

As in \cite{chae2023a,chae2024} the \cite{hwang2022} eccentricity ranges will be used as the standard input. From the ranges, eccentricities are drawn using a truncated asymmetric Gaussian function as described in \cite{chae2023a,chae2024}. However, I update the code \citep{chae2023b} so that the truncated part is handled more properly. In the previous version, when a sampled $e$ falls outside the range $(0.001,0.999)$, its value is replaced by the boundary value. This procedure has the caveat that some occurrences can accumulate at the boundaries. To remedy this caveat, in the updated version $e$ is resampled from an initial pool of $e$ values that are within the range $(0.001,0.999)$. I note that this modification has a small effect. Unless specified otherwise, eccentricities from the updated code are used in this work as the standard/default choice.

While the \cite{hwang2022} individual ranges are most informative at present, the ranges are broad and may not be so accurate in some cases. Thus, I also consider sampling eccentricities from a power-law probability distribution of the form $p(e;\alpha)=(1+\alpha)e^\alpha$. I consider two cases: (1) an empirically determined $\alpha$ varying as a function of $s$ as given by Equation~(18) of \cite{chae2024}, (2) the ``thermal'' eccentricity distribution with $\alpha=1$. In the former, $\alpha$ is close to thermal at relatively smaller separation ($0.2<s\la 1$~kau) but becomes superthermal at larger separations. Eccentricities sampled from this systematically varying power-law distribution will be referred to as ``statistical eccentricities'' while those sampled from the $\alpha=1$ distribution ``thermal eccentricities''. Note that based on the \cite{hwang2022} measurements, thermal eccentricities are systematically biased.  

\subsection{Profiles of Normalized Projected Velocity and Mass with a Normalized Projected Radius} \label{sec:profile}

\begin{figure*}
  \centering
  \includegraphics[width=0.9\linewidth]{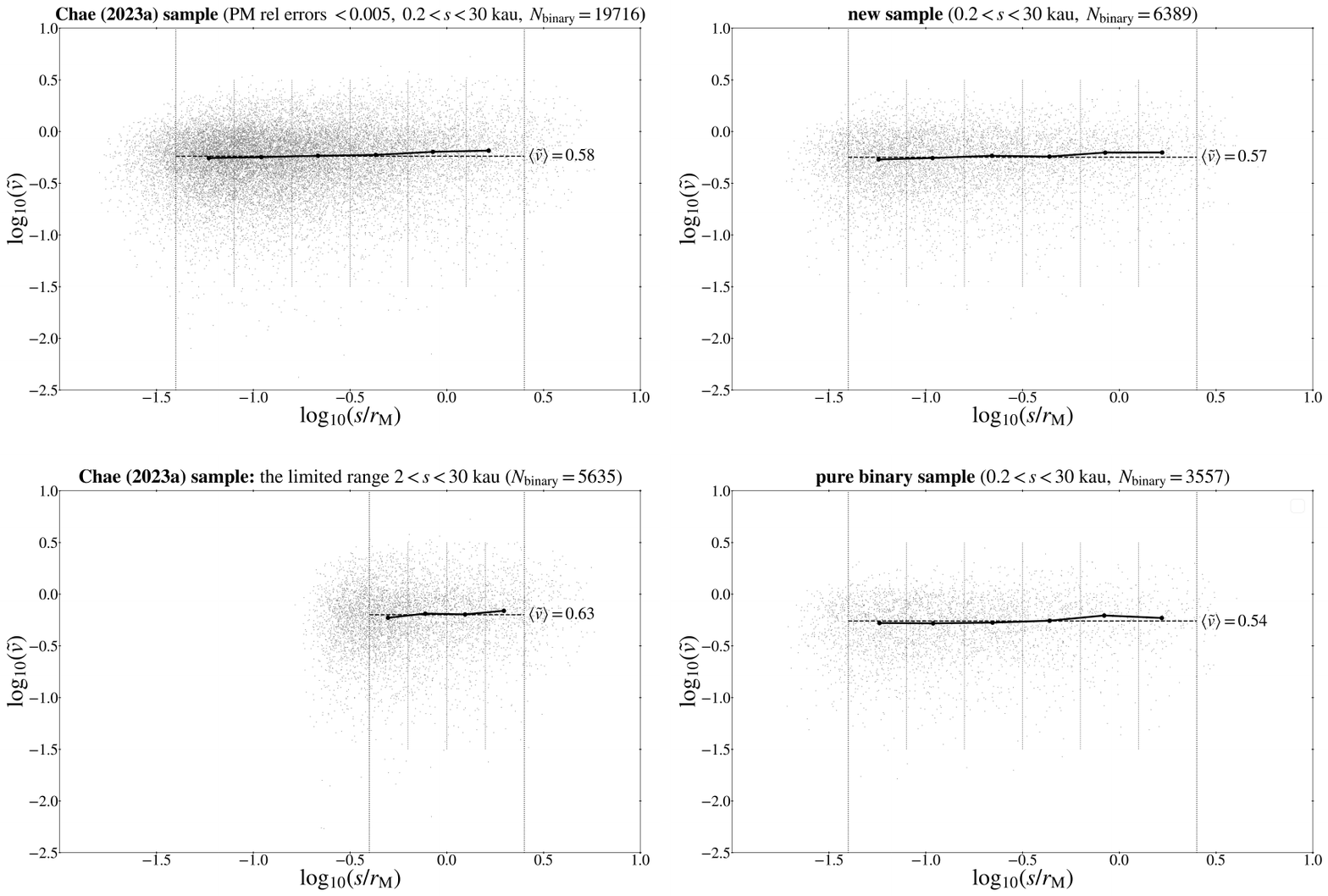}
    \vspace{-0.5truecm}
    \caption{\small 
    {Each panel shows a distribution of binaries in the plane spanned by logarithmic values of two dimensionless quantities $s/r_{\rm{M}}$ (sky-projected separation over the MOND radius) and $\tilde{v}$ (Equation~(\ref{eq:vtilde})) from one MC realization of binary total masses including probabilistically assigned masses of hidden close binaries. Four samples summarized in Table~\ref{tab:sample} are shown. The bins and the range used for analyses in $\log_{10}(s/r_{\rm{M}})$ are indicated by vertical lines. For each sample the overall median value of $\tilde{v}$ is indicated by the horizontal dashed line and the binned medians are indicated by big dots and the thick solid line. I emphasize that the scattered points should be regarded as a probability distribution from all binaries with all uncertainties and possible ranges of parameters taken into account in a natural way through the MC. See the text for further. }
    } 
   \label{vtilde_merged}
\end{figure*} 

\begin{figure}
  \centering
  \includegraphics[width=0.9\linewidth]{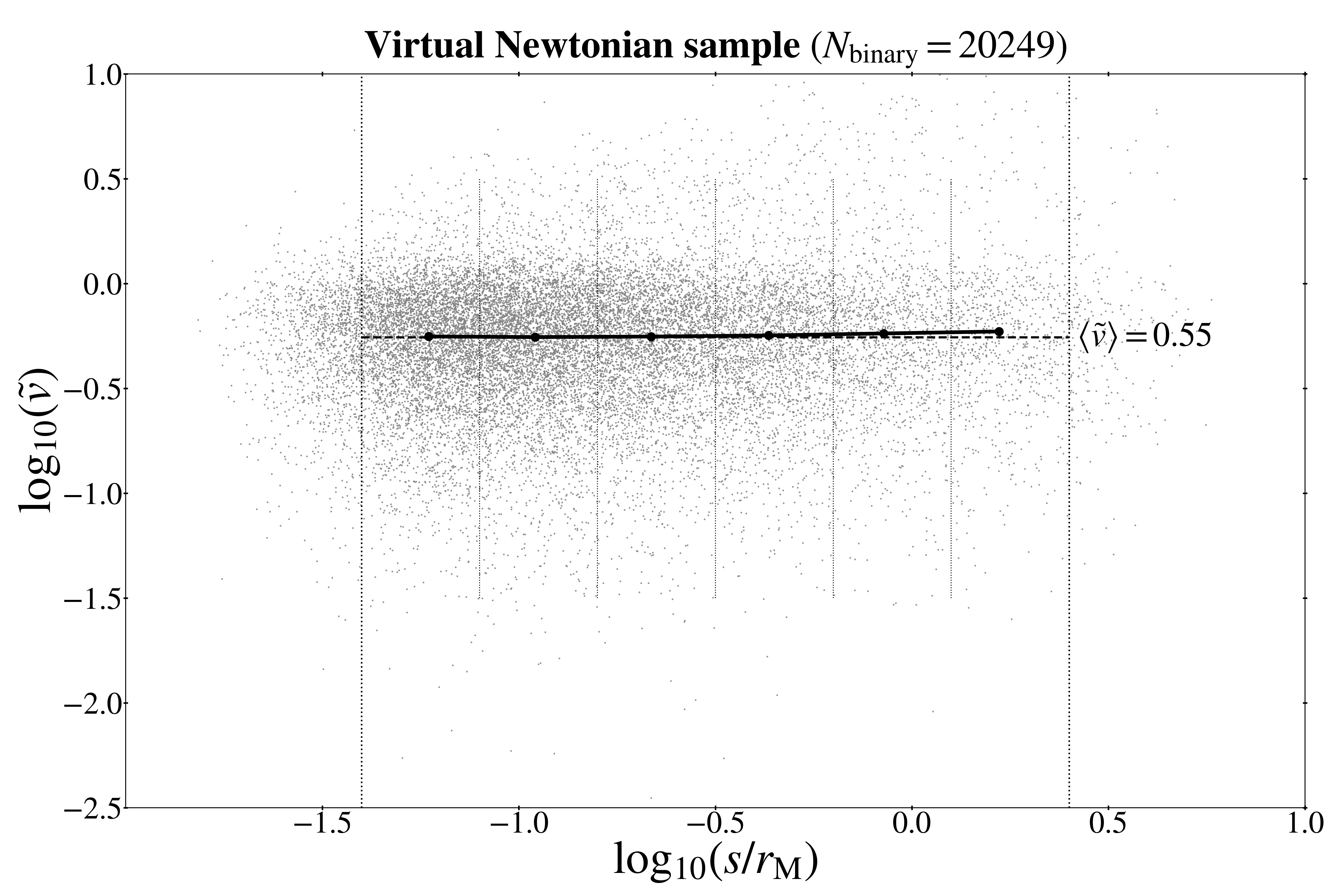}
    \caption{\small 
    {This figure is the same as the upper left panel of Figure~\ref{vtilde_merged} but for a virtual Newtonian sample as a counterpart of the Chae (2023a) Gaia sample. Note that the binned medians of $\tilde{v}$ are  {nearly} flat in the virtual sample.}
    } 
   \label{vtilde_sim}
\end{figure} 

As described in Section~\ref{sec:relations}, the relation between $\tilde{v}$ and $s/r_{\rm{M}}$ can be used as a proxy for the relation between $g/g_{\rm{N}}$ and $g_{\rm{N}}$ without a deprojection of the observed 2D quantities to the 3D space. Before considering a statistical method of testing gravity with the relation between $\tilde{v}$ and $s/r_{\rm{M}}$, in this subsection I investigate the observed properties of $\tilde{v}$ and associated quantities  with respect to $s/r_{\rm{M}}$. Here four quantities are involved: $v_p$ (Equation~(\ref{eq:vp_ob})), $v_c(s)$ (Equation~(\ref{eq:vc})), $s$, and $r_{\rm{M}}$ (Equation~(\ref{eq:rmond})). But, both $v_c(s)$ and $r_{\rm{M}}$ depend on $M_{\rm{tot}}$, and thus the key quantities are $v_p$, $s$, and $M_{\rm{tot}}$.

The projected separation $s$ has negligible uncertainty. Also, because the required fractional uncertainty for  {the scalar relative PM} is $<0.01$  {(since the four PM components have fractional errors $<0.005$)} from the sample selection,  {the measurement uncertainty of $v_p$}\footnote{ {Even for the largest Chae (2023a) sample with least stringent requirements, the median measurement uncertainty of $v_p$ is 0.03~km~s$^{-1}$ and 94.5\% of the binaries have uncertainties less than 0.1~km~s$^{-1}$ well below typical $v_p$ values.}} is practically negligible compared with  {the uncertainty arising from the uncertainties of hidden companions}.  {In other words,} the mass uncertainty  {associated with hidden components} dominates the uncertainties of both $\tilde{v}$ and $s/r_{\rm{M}}$. In obtaining and using the relation between $\tilde{v}$ and $s/r_{\rm{M}}$, i.e.\ the profile $\tilde{v}(s/r_{\rm{M}})$, the uncertainties need to be taken into account.  {When hierarchical systems are present as in the Chae (2023a) sample and the new sample, the nominal analyses will be carried out including the uncertainties associated with hierarchical systems but not the PM measurement uncertainties. However, for the pure binary sample the PM measurement uncertainties are always included. In Appendix~\ref{sec:pmscat}, I explicitly show that the PM measurement uncertainties can only have a minor effect when hierarchical systems are present.} 

I use  {the MC method described in Section~\ref{sec:calcvt} that is} similar to those used in \cite{chae2023a,chae2024} to take into account the uncertainties. The mass of additional component(s) is assigned for a given value of $f_{\rm{multi}}$ with the method presented in Section~3.2 of \cite{chae2023a}. One MC draws one set of values from the respective probability distributions of $\tilde{v}$ and $s/r_{\rm{M}}$. By drawing many MC sets, the uncertainty of the distribution of ($\tilde{v},s/r_{\rm{M}}$) will be derived and used in the statistical analysis to be described in the next subsection.

Figure~\ref{vtilde_merged} shows distributions of ($\tilde{v}$, $s/r_{\rm{M}}$) for the four samples summarized in Table~\ref{tab:sample} from one MC set for each sample.  Here I use $f_{\rm{multi}}=0.43$ for the Chae (2023a) sample, and $f_{\rm{multi}}=0.18$ for the new sample as determined from this work (see Appendix~\ref{sec:accel}). I consider six bins uniformly spaced in $\log_{10}(s/r_{\rm{M}})$ within the specified range bounded by the vertical dotted lines and calculate the medians of $\tilde{v}$ in the bins. As will be shown below, the lower and upper limits are set to exclude data that give biased distributions of $M_{\rm{tot}}$ at $s/r_{\rm{M}}$.

Figure~\ref{vtilde_sim} shows one MC distribution for a virtual Newtonian sample as a Newtonian counterpart of the Chae (2023a) sample. The virtual Newtonian sample is obtained by replacing the observed PMs with simulated PMs as described in {Section~\ref{sec:calcvt}}.
  
The derived binned medians $\langle\tilde{v}\rangle$ for the Gaia samples increase slowly with $s/r_{\rm{M}}$ and an elevation is apparent for the bins of $s/r_{\rm{M}}\ga 0.5$ compared with the lower $s/r_{\rm{M}}$ bins. On the contrary, the virtual Newtonian sample  {is nearly flat} from the lowest to the highest bin of $s/r_{\rm{M}}$. Consequently, the median $\langle\tilde{v}\rangle=0.55$ of the virtual Newtonian sample for the entire range is lower than the value $0.58$ of the counterpart Gaia sample.

The overall medians of $\tilde{v}$ for the new and the pure binary samples differ somewhat from that for the Chae (2023a) sample. This can be attributed to the fact that the three samples have different $f_{\rm{multi}}$ and $\langle M_{\rm{tot}}\rangle$ (as shown below). The sample with the limited range of $2<s<30$~kau exhibits a minor variation with $s/r_{\rm{M}}$. The overall median $\langle\tilde{v}\rangle$ for this sample is clearly higher than those for the other samples. This is consistent with the fact that  the binned median increases with $s/r_{\rm{M}}$.

Because each MC draw generates values of $\tilde{v}$ and $s/r_{\rm{M}}$ from their respective probability distributions taking fully into account the uncertainties and the values from all binaries are stacked on the plane, there is no need in the present approach, as in the acceleration-plane approach of \cite{chae2023a}, to remove binaries based on a fractional uncertainty of $\tilde{v}$ as suggested by \cite{banik2024}. In other words, if a binary system has a relatively larger uncertainty (i.e.\ a broader probability distribution) of $\tilde{v}$, its MC-drawn values are scattered more broadly than others and thus naturally given a lower weight at fixed $s/r_{\rm{M}}$. 

In the approach taken by \cite{banik2024}, they applied a hard cut on the fractional uncertainty of $\tilde{v}$ and then treated all values equally ignoring their individual uncertainties. As \cite{chae2024} has already demonstrated, applying such a cut on a binary sample does not change the gravitational anomaly. In other words, the MC-based acceleration-plane analysis is largely immune to any sampling (excluding of course erroneous data) because all uncertainties are naturally taken into account. The same is true for the MC-based analysis of the profile $\tilde{v}(s/r_{\rm{M}})$ as will be demonstrated in this work.

However, I consider a kinematic cut based on an uncertainty of $\tilde{v}$ for the purpose of illustrating its effects in an analysis directly using $\tilde{v}$. For this purpose, a formal uncertainty is estimated as
\begin{equation}
  \sigma_{\tilde{v}} = \tilde{v}\sqrt{\left(\frac{\sigma_{v_p}}{v_p}\right)^2+\left(\frac{\sigma_{v_c}}{v_c}\right)^2},
  \label{eq:sigvtilde}
\end{equation}
where $\sigma_{v_c}$ follows from the uncertainty of $M_{\rm{tot}}$. I consider the kinematic cut suggested by \cite{banik2024}
\begin{equation}
  \sigma_{\tilde{v}} < 0.1 \max{(1,\tilde{v}/2)}.
  \label{eq:vtildecut}
\end{equation}
I take $\sigma_{v_c}/v_c=0.05$ assuming a uniform uncertainty of 10\% in $M_{\rm{tot}}$. I note that this formal uncertainty is of limited value because it is not known which systems are hierarchical and thus system-specific realistic uncertainties cannot be determined. \cite{banik2024} considered an MC error propagation to estimate the uncertainties of $\tilde{v}$ in their sample. However, they did not take into account the uncertainties of hierarchical systems either. Thus, the uncertainty given by Equation~(\ref{eq:sigvtilde}) or that estimated by \cite{banik2024} cannot be considered accurate, and it is necessary to consider an MC method to take into account hierarchical systems as done in this work.

\begin{figure*}
  \centering
  \includegraphics[width=0.9\linewidth]{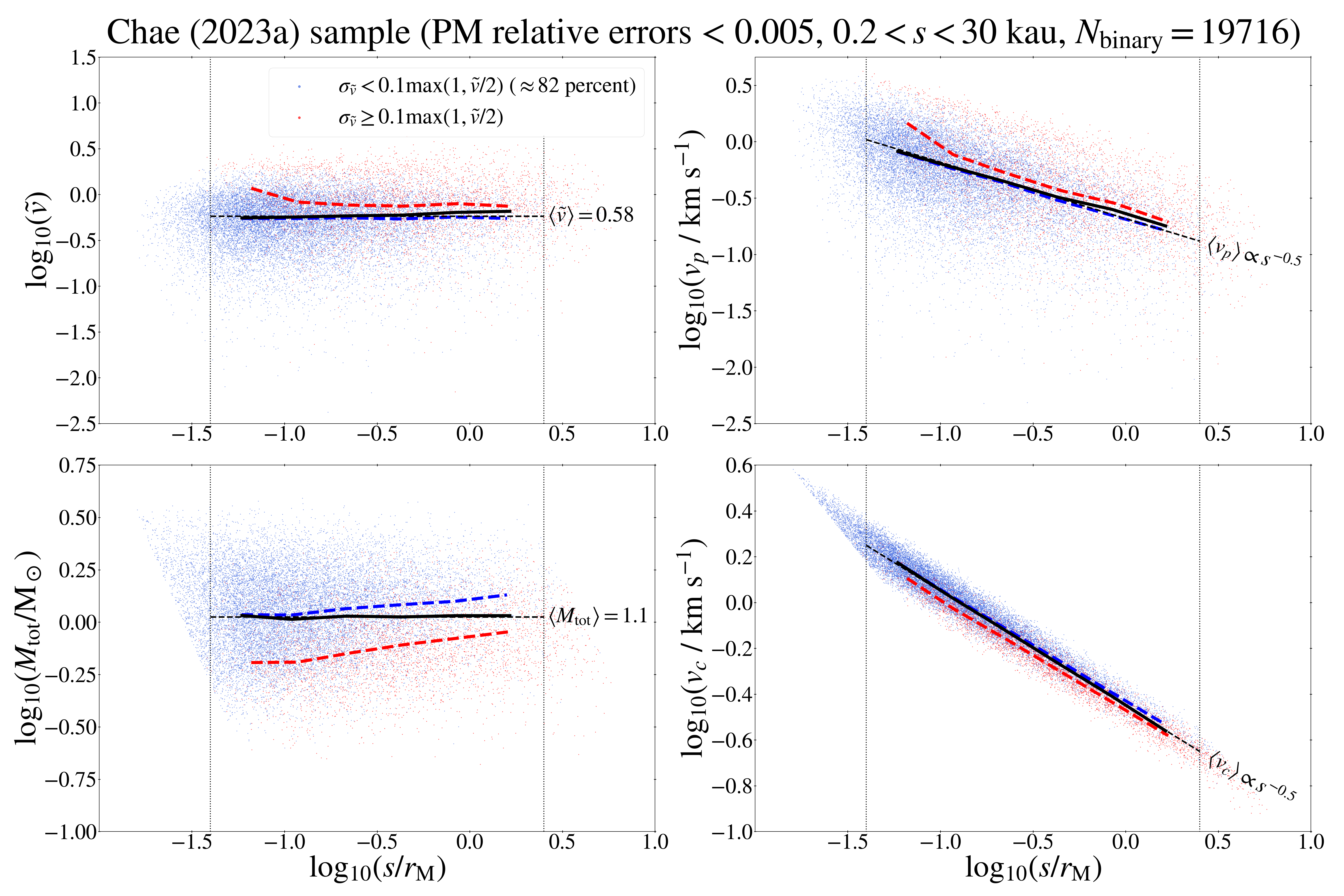}
    \vspace{-0.3truecm}
    \caption{\small 
      {This figure shows the scaling of four quantities with $s/r_{\rm{M}}$ from one MC result for the Chae (2023a) sample. They are $\tilde{v}$ (that was already shown in Figure~\ref{vtilde_merged}), $v_p$ (sky-projected relative velocity between the pair of stars), $M_{\rm{tot}}$ (total mass of the binary system including probabilistically assigned mass of any hidden close binary), and $v_c$ (the Newtonian circular velocity due to $M_{\rm{tot}}$ at the sky-projected separation $s$). For the bins defined in the upper left panel of Figure~\ref{vtilde_merged}, binned medians are shown for all (black solid line) and those (red and blue dashed lines) split by a kinematic constraint suggested by \cite{banik2024}. Once about 18\% of binaries (small red dots) are removed by the kinematic cut, the binned medians of $\tilde{v}$ and $v_p$ (blue lines) for the remaining binaries follow the Newtonian predictions. However, as the bottom panels show, binned masses for these binaries rise systematically with $s/r_{\rm{M}}$ and thus $v_c$ systematically deviates from the Newtonian prediction showing that the subsamples are biased in the distribution of masses. See the text for further. }
    } 
   \label{scaling_main}
\end{figure*}

In Figure~\ref{scaling_main}, binaries from the Chae~(2023a) sample are shown with different colors split by the kinematic cut of Equation~(\ref{eq:vtildecut}). Figure~\ref{scaling_main} also shows the distributions of $v_p$, $v_c$, and $M_{\rm{tot}}$ with respect to $s/r_{\rm{M}}$. Figure~\ref{scaling_main} shows that when all binaries are considered (black solid lines), $\langle\tilde{v}\rangle$ (i.e.\ the binned median of $\tilde{v}$) increases with $s/r_{\rm{M}}$ as a consequence of $\langle v_p\rangle$ deviating from the Keplerian scaling $\propto s^{-1/2}$ while $\langle v_c\rangle$ follows the Keplerian scaling as $\langle M_{\rm{tot}}\rangle$ is flat. However, for the binaries satisfying the kinematic cut, $\langle\tilde{v}\rangle$ (blue dashed line) does not vary with $s/r_{\rm{M}}$ as was first noted by \cite{banik2024}. For those binaries $\langle v_p\rangle$ follows the Keplerian scaling. Thus, at first glance they appear to obey Newtonian dynamics. This was indeed the claim by \cite{banik2024}. This would be the case if $\langle M_{\rm{tot}}\rangle$ did not vary and thus $\langle v_c\rangle$ followed the Keplerian scaling.

However, as the bottom panels of Figure~\ref{scaling_main} show, the binary sample with the kinematic cut is biased so that $\langle M_{\rm{tot}}\rangle$ increases with $s/r_{\rm{M}}$ and thus $\langle v_c\rangle$ deviates from the Keplerian scaling. This means that the Newtonian prediction of $\langle\tilde{v}\rangle$ for this sample is different from the general sample for which $\langle M_{\rm{tot}}\rangle$ does not vary. This indicates that the apparently Newtonian flat behavior of $\langle\tilde{v}\rangle$ with the kinematic cut is driven by the variation of $\langle M_{\rm{tot}}\rangle$ with $s/r_{\rm{M}}$. This will be addressed quantitatively in this work. 

\begin{figure*}
  \centering
  \includegraphics[width=0.9\linewidth]{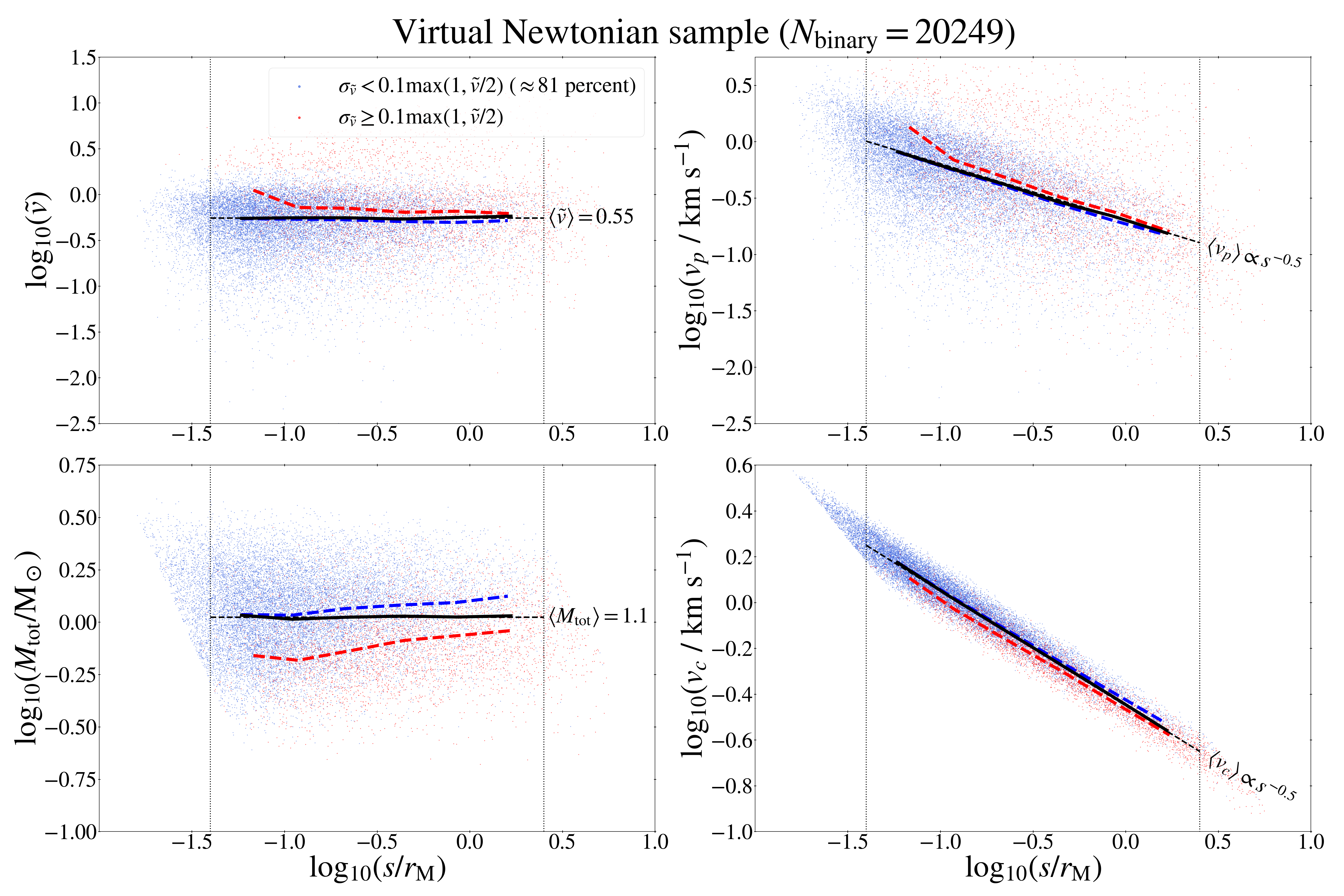}
    \vspace{-0.3truecm}
    \caption{\small 
    {Same as Figure~\ref{scaling_main} but for the virtual Newtonian sample as a counterpart of the Chae (2023a) Gaia sample. Note that unlike Figure~\ref{scaling_main}, for the virtual binaries satisfying Newtonian gravity the binned medians of $\tilde{v}$ (black solid line) for all binaries follow the flat Newtonian line while the blue dashed line deviates now downward from the Newtonian line. The behaviors of $M_{\rm{tot}}$ and $v_c$ are the same as in Figure~\ref{scaling_main} because virtual binaries are statistically equivalent to true binaries except that Newtonian gravity is assumed for the virtual sample.}
    } 
   \label{scaling_sim}
\end{figure*}

Further insights may be gained from considering the virtual Newtonian sample. Figure~\ref{scaling_sim} shows the distributions in the virtual Newtonian sample. For this sample, when all binaries are considered (black solid lines), $\langle\tilde{v}\rangle$ does not vary with $s/r_{\rm{M}}$ as a consequence of both $\langle v_p\rangle$ and $\langle v_c\rangle$ following the Keplerian scaling. However, for the binaries satisfying the kinematic cut, $\langle\tilde{v}\rangle$ now decreases as a consequence of $\langle v_c\rangle$ deviating from the Keplerian scaling and $\langle v_p\rangle$ (nearly) following the Keplerian scaling. What is striking is that the effects of the kinematic cut on the scalings of the four quantities are the same for the Gaia and the virtual Newtonian samples. In particular, due to the variation of the median mass with $s/r_{\rm{M}}$ when the kinematic cut is imposed, $\langle\tilde{v}\rangle$ is affected in the same manner by the cut however $\langle\tilde{v}\rangle$ scales with $s/r_{\rm{M}}$ without the cut. 

\begin{figure*}
  \centering
  \includegraphics[width=0.9\linewidth]{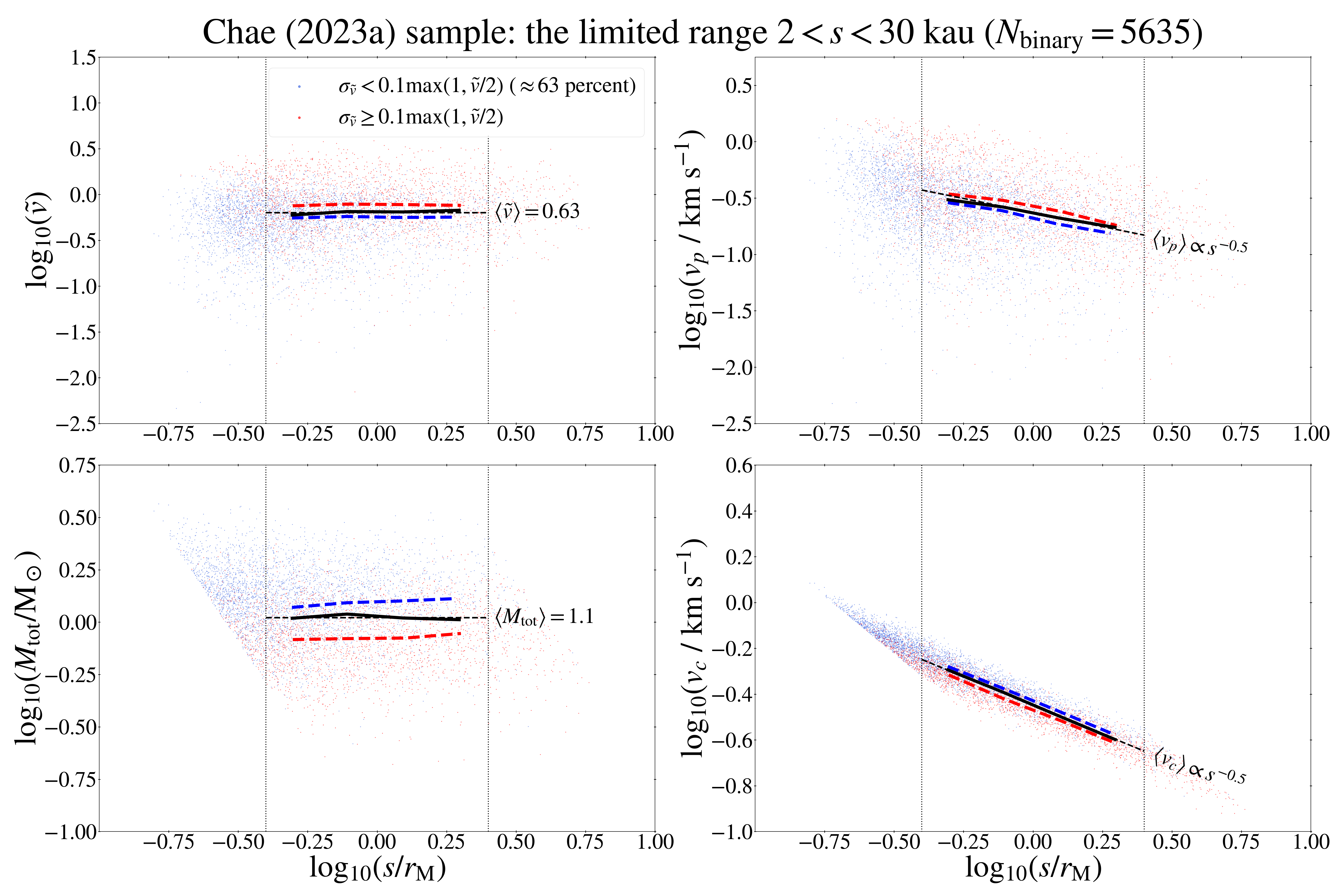}
    \vspace{-0.3truecm}
    \caption{\small 
    {Similar to Figure~\ref{scaling_main} but for the Chae (2023a) limited sample satisfying $s>2$~kau. Because of the narrow range of $s/r_{\rm{M}}$ after excluding binaries suffering from the edge effects, there is no obvious scaling of $\tilde{v}$, and its overall median is higher than that shown in Figure~\ref{scaling_main} for the sample including the $s<2$~kau binaries.}
    } 
   \label{scaling_sub}
\end{figure*} 

In the Chae~(2023a) sample, the fraction of binaries not meeting the kinematic cut is $\approx 18$\%. In the new sample this fraction decreases dramatically to $\approx 3$\%. Thus, the effects of the kinematic cut are minimal. In the pure binary sample, the fraction is just $\approx 1$\% and so its effects are negligible.

Figure~\ref{scaling_sub} shows the distributions for the Chae (2023a) limited sample of 5635 binaries with the narrow range $2 < s <30$~kau. For this sample, variations of the quantities with $s/r_{\rm{M}}$ are weak or absent. Since just the twice large dynamics range (in log scale) reveals a clear variation of $\langle\tilde{v}\rangle$ with $s/r_{\rm{M}}$ (Figure~\ref{scaling_main}), the weak/null variation in this range (that was used by \cite{banik2024}) is clearly misleading. It could easily be interpreted as an indication for Newtonian gravity because the flatness of the $\langle\tilde{v}\rangle$ profile is its generic prediction (Figure~\ref{scaling_sim}).

In principle, it would still be possible to test gravity with the flat $\langle\tilde{v}\rangle$ value from a limited dynamic range sample. However, such a test requires a precise value of $\langle M_{\rm{tot}}\rangle$ that in turn depends on $f_{\rm{multi}}$. This also means that it may well be possible to obtain a wrong gravity with a wrong value of $f_{\rm{multi}}$ if it is not properly determined for the sample under consideration.

\subsection{Statistical Analysis} \label{sec:stat}

The distribution of binaries in the plane spanned by two dimensionless quantities $\tilde{v}$ and $s/r_{\rm{M}}$ for a sample is used to test gravity. For the range of $s/r_{\rm{M}}$ within which $M_{\rm{tot}}$ has an unbiased distribution (Figure~\ref{vtilde_merged}), the bins of $s/r_{\rm{M}}$ are defined as shown in Section~\ref{sec:profile}. MC sets of ($\tilde{v}$, $s/r_{\rm{M}}$) are generated taking into account the uncertainties of $M_{\rm{tot}}$ with a probability of $f_{\rm{multi}}$ hosting undetected close companions (as already mentioned in Section~\ref{sec:profile} and described in Section~3.2 of \cite{chae2023a}). For each MC set, binned medians of $\tilde{v}$ are obtained.

Along with each MC set for the given sample of binaries, a corresponding MC set is derived from a mock Newtonian sample with simulated PMs replacing the given/measured PMs. For each pair of MC sets, medians $\langle\tilde{v}\rangle_{{\rm{obs}},i}$ and $\langle\tilde{v}\rangle_{{\rm{newt}},i}$ are obtained for the $i$-th bin, where $i=1,\cdots,N_{\rm{bin}}$. From the medians, values of the velocity ratio $\langle\tilde{v}\rangle_{{\rm{obs}},i}/\langle\tilde{v}\rangle_{{\rm{newt}},i}$ are obtained. Since the kinematic acceleration $g\propto v^2$ and $v_p \propto v$ statistically (averaged over all possible orientations), this velocity ratio probes the ratio $(g_{\rm{obs}}/g_{\rm{pred}})^{1/2}$ where $g_{\rm{obs}}$ and $g_{\rm{pred}}$ are the observed and Newtonian-predicted values of the kinematic acceleration. 

$N_{\rm{draw}}$ MC sets are generated and from them I obtain a distribution of a parameter defined by
\begin{equation}
  \Gamma \equiv \log_{10}\gamma_{\tilde{v}} \equiv \log_{10}\left(\frac{\langle\tilde{v}\rangle_{\rm{obs}}}{\langle\tilde{v}\rangle_{\rm{newt}}}\right),
  \label{eq:gamma}
\end{equation}
where $\gamma_{\tilde{v}}$ represents the velocity boost (or anomaly) factor and $\Gamma$ is its logarithmic value. I have $(\Gamma_i)_j$ for the $i$-th bin ($i=1,\cdots,N_{\rm{bin}}$) and the $j$-th draw ($j=1,\cdots,N_{\rm{draw}}$). Since the MC distribution of $(\Gamma_i)_j$ for each $i$ follows approximately a Gaussian distribution, I use the conventional $\chi^2$ statistics (e.g.\ \citealt{wall2012}) for gravity model testing. For this I define a reduced $\chi^2$
\begin{equation}
  \chi^2_\nu \equiv \frac{1}{\nu}\sum_{i=1}^{N_{\rm{bin}}}  \frac{\left(\mu_{\Gamma_i} - \log_{10}\gamma_{\tilde{v}_i}^{\text{model}}\right)^2}{(\sigma_{\Gamma_i})^2+(\sigma_i^{\rm{model}})^2},
   \label{eq:chisq}
\end{equation}
where $\nu$ is the degree of freedom, $\mu_{\Gamma_i}$ and $\sigma_{\Gamma_i}$ are the mean and standard deviation of $(\Gamma_i)_j$ for $j=1,\cdots,N_{\rm{draw}}$. Because each $(\Gamma_i)_j$ is derived in a fully Monte Carlo way from the Gaia observations along with possible ranges of all involved parameters, $\mu_{\Gamma_i}$ and $\sigma_{\Gamma_i}$ are the measurements for the bins. In Equation~(\ref{eq:chisq}), $\gamma_{\tilde{v}_i}^{\text{model}}$ is the theoretical velocity boost factor for the $i$-th bin predicted by a given model, and $\sigma_i^{\rm{model}}$ is the uncertainty of the model prediction due to the uncertainties of the external field effect (EFE; e.g., \citealt{chae2020b,chae2021}) in a model having it.

In Equation~(\ref{eq:chisq}), the degree of freedom $\nu$ is equal to the number of bins $N_{\rm{bins}}$ if $f_{\rm{multi}}$ is fixed as in the pure binary sample. Otherwise, I have $\nu=N_{\rm{bins}}-1$ because $f_{\rm{multi}}$ is fitted by the high acceleration (or low $s/r_{\rm{M}}$) data where both standard and nonstandard gravity theories converge.

By its definition (Equation~(\ref{eq:gamma})), $\gamma_{\tilde{v}_i}^{\text{model}}=1$ and $\sigma_i^{\rm{model}}=0$ for the standard gravity. For non-standard gravity I consider two nonrelativistic MOND \citep{milgrom1983} models, AQUAL \citep{bekenstein1984} and QUMOND \citep{milgrom2010} that permit numerical solutions of the gravity boost factor taking into account the EFE of the Milky Way. The difference between them is not large but a test with galactic rotation curves marginally prefers AQUAL \citep{chae2022b}. Thus, I will test specifically the AQUAL model.

For the AQUAL model, numerical solutions for the elliptical motions of binaries under the external field of the Milky Way are not available at present. However, numerical solutions for circular orbits of a test particle have been studied extensively by \cite{chae2022a} with various mass models of the gravitational source from point-like objects to realistic disks. \cite{chae2022a} report that AQUAL numerical solutions predict the following fitting function for the ratio of Milgromian gravity to Newtonian gravity
\begin{equation}
	\frac{g}{g_{\rm{N}}} = \nu ({y_{\beta}})\left[ 1 + \left\{{\rm{tanh}}{\left( {\frac{{\beta{g_{\rm{N,ext}}/a_0}}}{{{g_{{\rm{N}}}}/{a_0}}}} \right)} \right\}^{\gamma} \frac{{\hat \nu ({y_{\beta}})}}{3} \right],
    \label{eq:aqformula}
\end{equation}
where the definition ${y_{\beta}} \equiv \sqrt {{{({g_{{\rm{N}}}}/{a_0})}^2} + {{(\beta{g_{\rm{N,ext}}/a_0})}^2}}$ is used with a Newtonian external field $g_{\rm{N,ext}}$, $\nu(y)$ is the MOND interpolating function, and  $\hat\nu (y) \equiv d{\rm{ln}}\nu {\rm{(}}y{\rm{)/}}d\ln y$. \cite{chae2022a} use the simple function \citep{famaey2005} $\nu (y) = 0.5 + \sqrt {0.25 + {y^{ - 1}}}$ (for the acceleration range $g_{\rm{N}}\la 10^{-7}$~m~s$^{-2}$) for which the relation between the empirical external field $g_{\rm{ext}}$ and the Newtonian external field is given by $x^2/(1+x)=g_{\rm{N,ext}}/a_0$ where $x\equiv g_{\rm{ext}}/a_0$.

The AQUAL model described by Equation~(\ref{eq:aqformula}) is implemented as follows. I take $\beta=1.1$ and $\gamma=1.2$ that were estimated using orbits with an average inclination $60^\circ$ of the external field with respect to the orbital axis.\footnote{When the inclination is not zero, the orbit is actually not circular. \cite{chae2022a} estimated $g/g_{\rm{N}}$ at fixed $g_{\rm{N}}$ by taking an average over the entire range of the azimuthal angle. However, for circular orbits when the inclination is zero, the modification to the estimated $g/g_{\rm{N}}$ is minor and within the considered uncertainty.} For the eternal field I take $g_{\rm{ext}}/a_0=1.90\pm 0.19$ \citep{chae2023a,chae2024} with $a_0=1.2\times 10^{-10}$~m~s$^{-2}$. The 10\% uncertainty reasonably covers its current uncertainty.

The AQUAL model can be related to the velocity boost factor $\gamma_{\tilde{v}}^{\rm{model}}$ defined in Equation~(\ref{eq:gamma}) as follows.  From $g=v^2/r$ and $g_{\rm{pred}}=g_{\rm{N}}=v_c^2/r$ for circular orbits, the ratio $g/g_{\rm{N}}$ at $g_{\rm{N}}/a_0$ is equal to $(v/v_c)^2$ at $r/r_{\rm{M}}$. Now averaging over random orientations and phases, I get $\tilde{v}\approx v/v_c$ at $r\approx 1.2 s$. Thus, I have the approximate relation
\begin{equation}
  \gamma_{\tilde{v}}^{\rm{model}} \approx \sqrt{\frac{g}{g_{\rm{N}}}}\text{ with }\frac{s}{r_{\rm{M}}} \approx \frac{1}{1.2}\sqrt{\frac{a_0}{g_{\rm{N}}}},
  \label{eq:aqgamma}
  \end{equation}
where $g/g_{\rm{N}}$ is given by Equation~(\ref{eq:aqformula}). Thus, Equation~(\ref{eq:aqgamma}) gives an AQUAL prediction as a function of $s/r_{\rm{M}}$ to be tested through Equation~(\ref{eq:chisq}).

\section{Results} \label{sec:result}

\subsection{Test with a virtual Newtonian sample}  \label{sec:result_Newton}

Before presenting the results for the Gaia samples I here test the statistical analysis with a virtual Newtonian sample. This sample is similar to the Chae~(2023a) sample except that the measured PMs were replaced with simulated PMs as described in Section~\ref{sec:calcvt}. This sample was produced with a fixed $f_{\rm{multi}}=0.43$ that is the value required for the Chae~(2023a) sample based on the most up-to-date acceleration-plane analysis (Appendix~\ref{sec:accel}). 

\begin{figure*}
  \centering
  \includegraphics[width=0.9\linewidth]{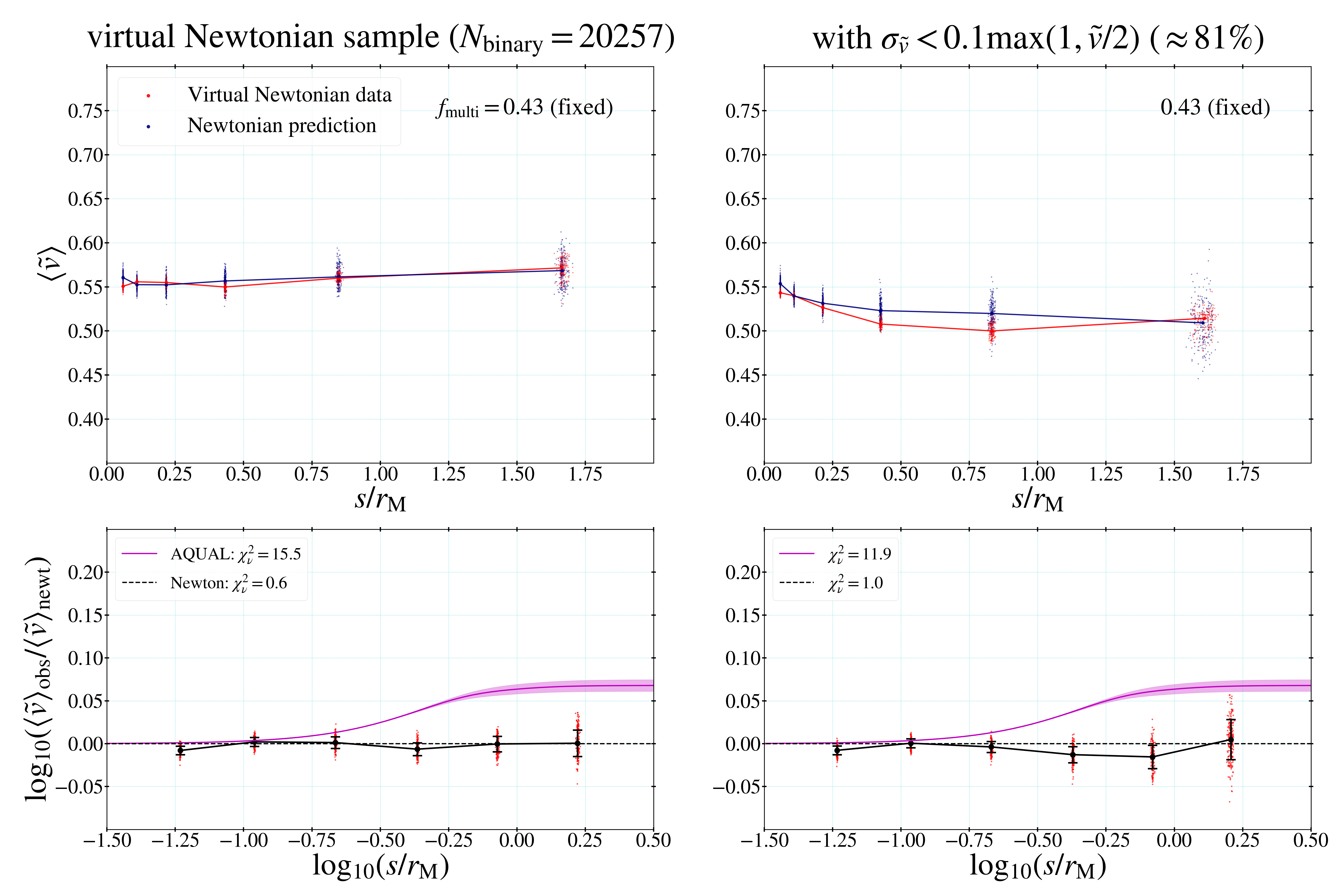}
    \vspace{-0.3truecm}
    \caption{\small 
    {The upper panels show the distributions of binned medians of $\tilde{v}$ from 200 MC results for the input sample and the corresponding virtual Newtonian sample. One set of binned medians along with all values of  $\tilde{v}$ from one MC are shown in Figure~\ref{vtilde_merged}. For this figure, the input sample itself is a virtual Newtonian sample. The bottom panels show the inferred values of $\Gamma (\equiv \log_{10}(\langle\tilde{v}\rangle_{\rm{obs}}/\langle\tilde{v}\rangle_{\rm{newt}})$, Equation~(\ref{eq:gamma})) where $\langle\tilde{v}\rangle_{\rm{obs}}$ and $\langle\tilde{v}\rangle_{\rm{newt}}$ are the values from the input and virtual Newtonian samples. Newtonian and AQUAL theories are tested through the reduced $\chi^2$ statistics (Equation~(\ref{eq:chisq})) with the measured values of $\Gamma$. The left column is for the full sample while the right column is for the subsample satisfying the kinematic cut. The Newtonian prediction is indicated by the horizontal black dashed line while the AQUAL prediction is indicated by the magenta curve with a band representing the uncertainty of the external field effect. As expected, the Newtonian model is well acceptable while the AQUAL model is ruled out by the virtual Newtonian sample. Note also that the kinematic cut does not change the $\chi^2$ testing significantly.}
    } 
   \label{vtest_sim}
\end{figure*}

Figure~\ref{vtest_sim} shows the results for the virtual Newtonian sample as the input data. The upper panels show the distributions of the binned medians of $\tilde{v}$ from $N_{\rm{draw}}=200$ MC sampling for the input data ($\langle\tilde{v}\rangle_{\rm{obs}}$) and the corresponding Newtonian simulated data ($\langle\tilde{v}\rangle_{\rm{newt}}$). The bottom panels show the distribution of $\Gamma$ ($\equiv\log_{10}(\langle\tilde{v}\rangle_{\rm{obs}}/\langle\tilde{v}\rangle_{\rm{newt}})$, Equation~(\ref{eq:gamma})). The left column shows the results for the full sample for which mass is not biased  (see Figure~\ref{scaling_sim}). The right column with the kinematic cut of Equation~(\ref{eq:vtildecut}) is considered to investigate its effects.

In the bottom panels of Figure~\ref{vtest_sim}, the Newtonian and AQUAL predictions are compared with the derived distributions of $\Gamma$. As expected, the distributions of $\Gamma$ agree well with the Newtonian prediction of $\Gamma=0$ whether the kinematic constraint is imposed or not. The derived values of $\chi^2_\nu=0.6$ and $1.0$ for the Newtonian model are well acceptable by the reduced $\chi^2_\nu$ static. In contrast, the AQUAL model has $\chi^2_\nu=15.5$ and $11.9$ for which the complementary (or survival) probability is $P_c\equiv 1-P(<\chi_\nu^2)\la 2\times 10^{-13}$.\footnote{In this paper, unless specified otherwise I use the capital letter $P$ to denote accumulated probability while the small letter $p$ is used to denote the probability density.} Thus, if the standard gravity holds for wide binaries, the AQUAL model is expected to be completely ruled out with a Gaussian equivalent confidence well above $5\sigma$.

\subsection{Results for Gaia samples with a broad dynamic range} \label{sec:result_gaia}
   
In this subsection I present the results for three Gaia samples (Table~\ref{tab:sample}) with the broad dynamic range $0.2<s<30$~kau. For the \cite{chae2024,chae2024b} pure binary sample, $f_{\rm{multi}}=0$ is fixed, so the analysis is the same as that for the virtual Newtonian sample presented in Section~\ref{sec:result_Newton}. For the Chae~(2023a) and the new samples, the best-fit value of $f_{\rm{multi}}$ can be determined with either low-$s/r_{\rm{M}}$ bins or high-acceleration data through the acceleration-plane analysis of \cite{chae2023a}. I find that both methods give consistent results except when mass-biased samples are used as in the cases with the \cite{banik2024} kinematic cut. Even in the latter cases, the differences are moderate and the results are not significantly affected whichever method is used. I will use the $f_{\rm{multi}}$ values determined through the acceleration-plane method as presented in Appendix~\ref{sec:accel} because it is more proper in principle. 

\begin{figure*}
  \centering
  \includegraphics[width=0.9\linewidth]{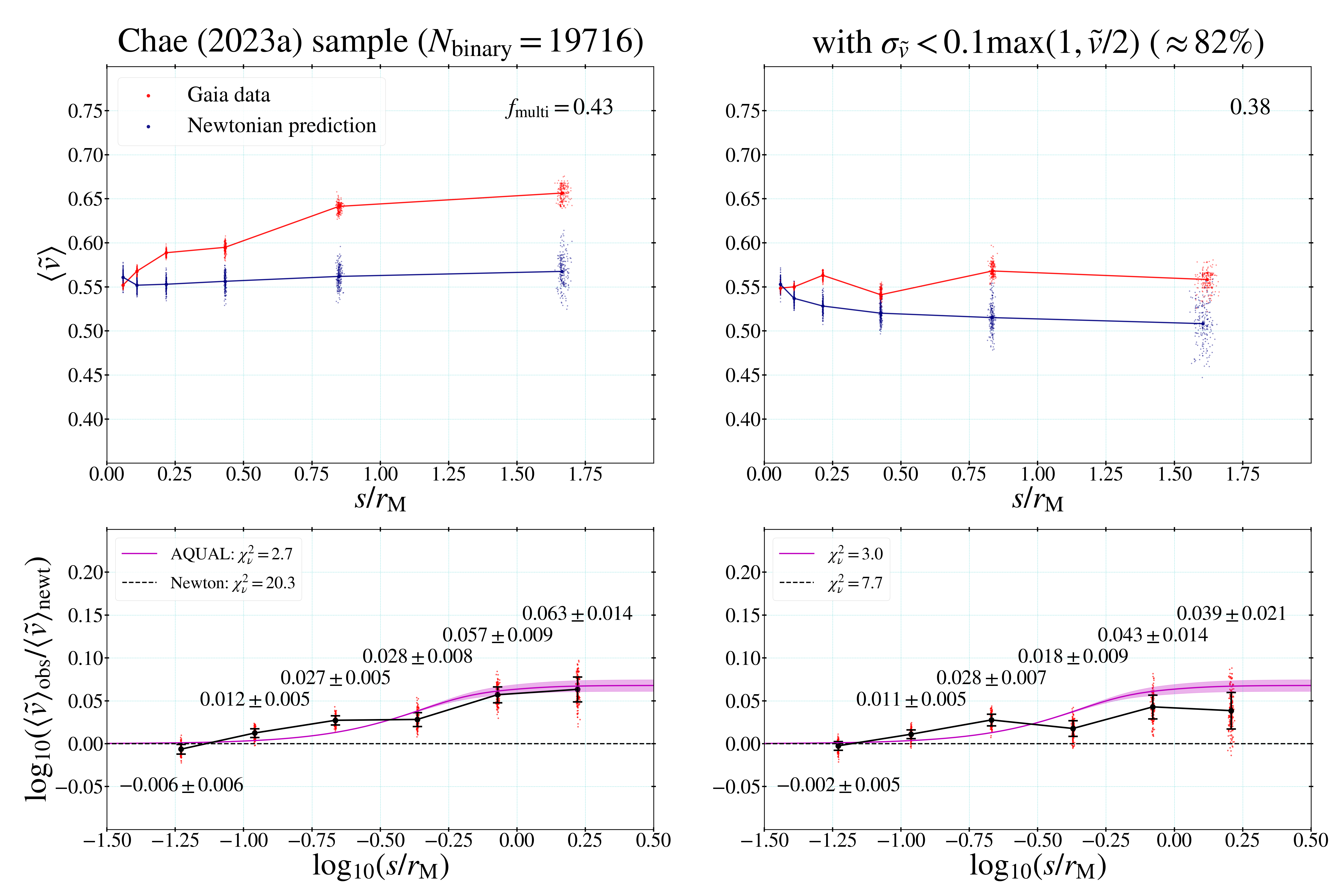}
    \vspace{-0.3truecm}
    \caption{\small 
    {The same as Figure~\ref{vtest_sim} but for a real sample of binaries, i.e.\ the Chae (2023a) sample. The $\chi^2$ testing outcomes are almost excatly opposite to those for the virtual Newtonian sample shown in Figure~\ref{vtest_sim}. In particular, even with the kinematic cut the $\chi^2$ testing conclusion is unaffected: the AQUAL model is adequate while the Newtonian model is ruled out. However, I emphasize that in the present MC approach the kinematic cut is not necessary or motivated as explained in detail in the text.  }
    } 
   \label{vtest_main}
\end{figure*} 

Figure~\ref{vtest_main} shows the distributions of $\langle\tilde{v}\rangle_{\rm{obs}}$, $\langle\tilde{v}\rangle_{\rm{newt}}$, and $\Gamma$ for the Chae~(2023a) sample. The best-fit value of $f_{\rm{multi}}$ for this sample is $0.43$. The distributions for the mass-unbiased sample (see Figure~\ref{scaling_main}) shown in the left column exhibit almost the exact opposite properties of those for the virtual Newtonian sample shown in the left column of Figure~\ref{vtest_sim}. The AQUAL model fits the binned data adequately with $\chi^2_\nu=2.7$ while the Newtonian model fails completely with $\chi^2_\nu=20.3$ for which the survival probability is $P_c \approx 2.6\times 10^{-20}$ with a Gaussian equivalent significance of $9.2\sigma$ that is similar to $10\sigma$ obtained by \cite{chae2023a} with a somewhat larger sample. 

If some binaries are removed by the kinematic cut of Equation~(\ref{eq:vtildecut}), two binned velocities, $\langle\tilde{v}\rangle_{\rm{obs}}$ and $\langle\tilde{v}\rangle_{\rm{newt}}$, are lowered progressively with $s/r_{\rm{M}}$ as already noted in Figure~\ref{scaling_main}. However, the AQUAL model is still not ruled out with $\chi^2_\nu=3.0$ ($P_c\approx 0.01$) by the conventional standards while the Newtonian model is ruled out with $\chi^2_\nu=7.7$ ($P_c\approx 3.0\times 10^{-7}$).

More importantly, I reemphasize that this cut is not necessary in the current MC approach (as in the acceleration-plane MC approach of \cite{chae2023a}) because all uncertainties and likely ranges are taken into account already. As the right column of Figure~\ref{vtest_main} shows, the main role of the cut is to systematically twist the normalized velocity profiles in an unnatural way. This is the consequence of the cut-introduced bias in the mass profile. For this reason, the result for the cut-introduced biased subsample should not be regarded as a representative or preferred result. It is presented here merely to clarify the effects of such a cut. For the rest of this work, I will not consider the cut-introduced subsample of the Chae (2023a) sample.

If a sample satisfying the cut such as Equation~(\ref{eq:vtildecut}) is desired, it is necessary to apply, in an overall way, stricter selection criteria that naturally select an unbiased sample automatically satisfying the cut while not biasing the mass profile. It turns out that the new sample largely meets the cut while the pure binary sample meets the cut almost completely as already noted in \cite{chae2024}. 

\begin{figure*}
  \centering
  \includegraphics[width=0.9\linewidth]{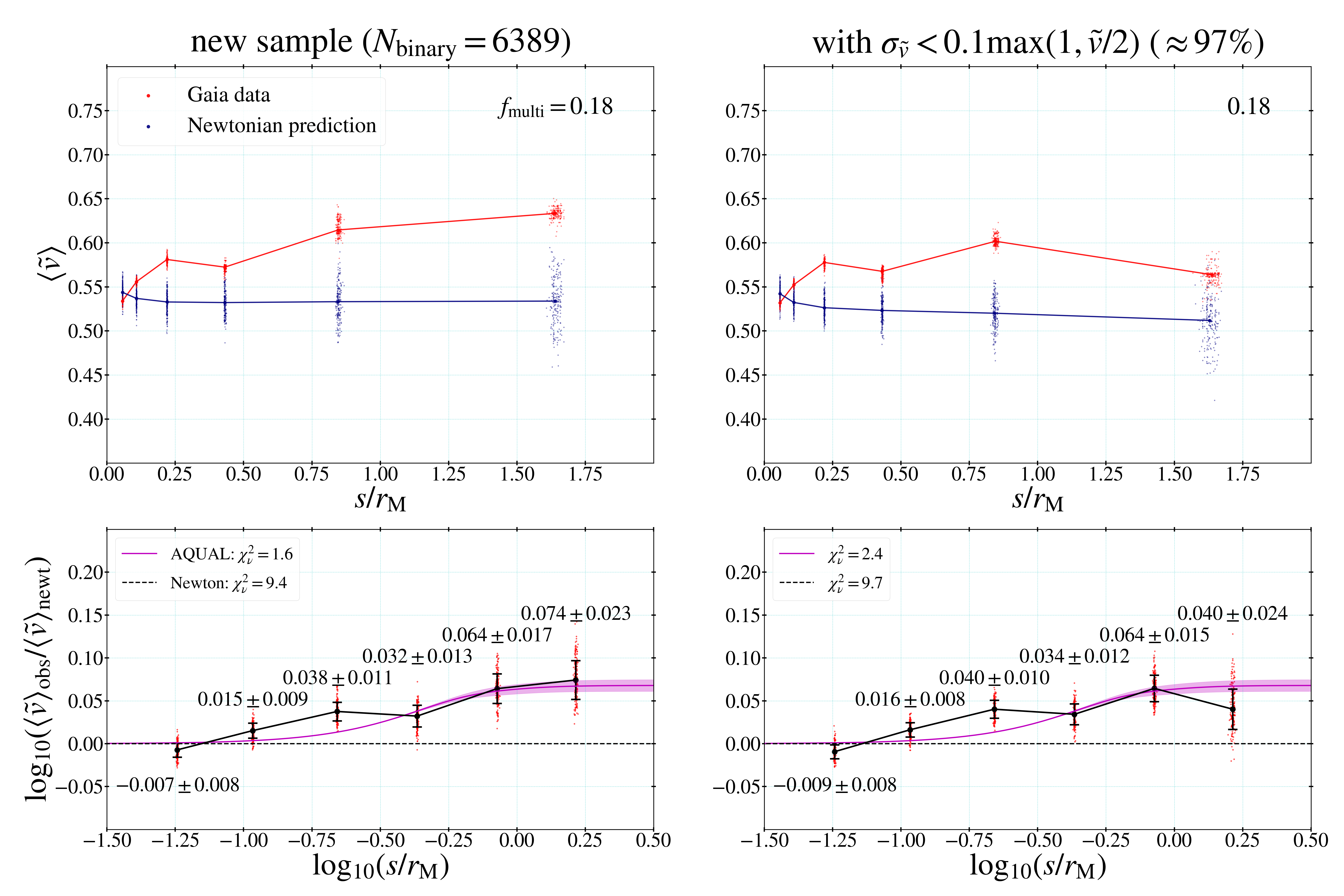}
    \vspace{-0.3truecm}
    \caption{\small 
    {The same as Figure~\ref{vtest_main} but for the new sample defined in this work. The $\chi^2$ testing outcomes are qualitatively the same as those for the Chae (2023a) sample. In particular, the AQUAL model is well acceptable while the Newtonian model is ruled out at $>5\sigma$ Gaussian equivalent significance. The kinematic cut has negligible effects because the sample naturally satisfies the cut thanks to the overall higher quality requirements without biasing the mass distribution. }
    } 
   \label{vtest_new}
\end{figure*}

Figure~\ref{vtest_new} shows the results for the new sample. This sample has been selected with different selection criteria and has a significantly lower fraction of hierarchical systems with $f_{\rm{multi}}=0.18$. About 97\% of this sample already satisfy the kinematic cut. Nevertheless, the observed normalized velocity profile from the full sample (the left column) is statistically equivalent to that from the Chae (2023a) full sample. The subsample satisfying the kinematic cut gives a very similar profile. The AQUAL model is well acceptable by both profiles with $\chi_\nu^2 = 1.6$ and $2.4$ while the Newtonian model is ruled out with $\chi_\nu^2 = 9.4$ and $9.7$ with a Gaussian equivalent significance of $\approx 5.8\sigma$.

The above results are based on the individual eccentricity ranges as described in Section~\ref{sec:input}. Now I present results based on less informative eccentricities from statistical distributions to investigate the effects of possible variations of the measured eccentricities. As described in Section~\ref{sec:input}, I consider ``statistical eccentricities'' from an observationally-inferred varying power-law distribution and ``thermal eccentricities'' from a fixed power-law distribution.

\begin{figure*}
  \centering
  \includegraphics[width=0.9\linewidth]{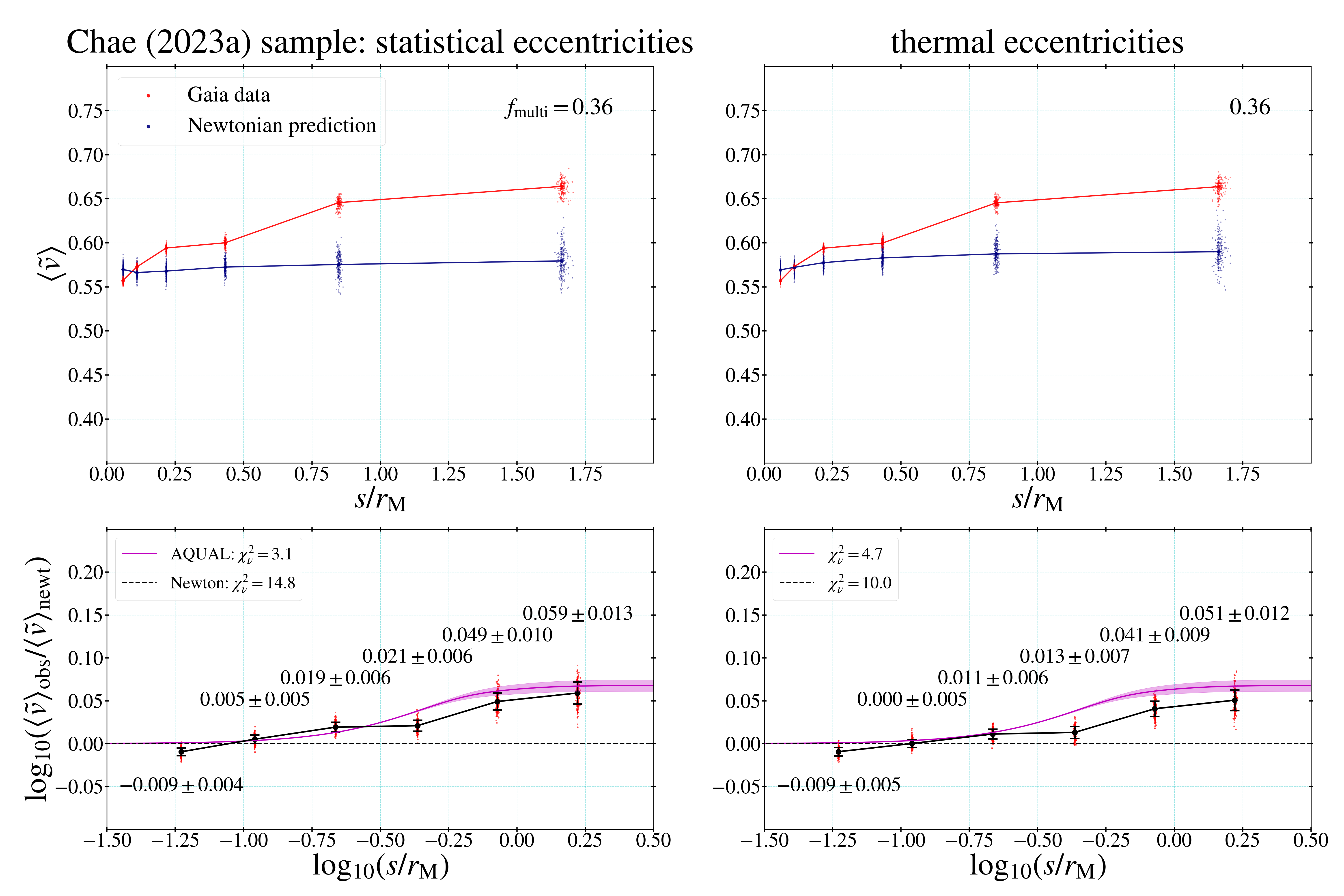}
    \vspace{-0.3truecm}
    \caption{\small 
    {Each column is the same as the left column of Figure~\ref{vtest_main} for the Chae (2023a) sample except that statistical or thermal eccentricity distributions are used instead of the individual eccentricities. The result with statistical eccentricities is well consistent with that with the individual eccentricities. Because thermal eccentricities are not based on pertinent observations for the binary sample, the result with them should not be used to test gravity. It is given here merely as a systematically different one. Not surprisingly, the $\Gamma$ values in the bottom panel of the right column result in an acceptably large value of $\chi^2_\nu=4.7$ for the AQUAL model although the Newtonian model has a much larger value of $10.0$. }
    } 
   \label{vtest_main_stateccen}
\end{figure*} 

\begin{figure*}
  \centering
  \includegraphics[width=0.9\linewidth]{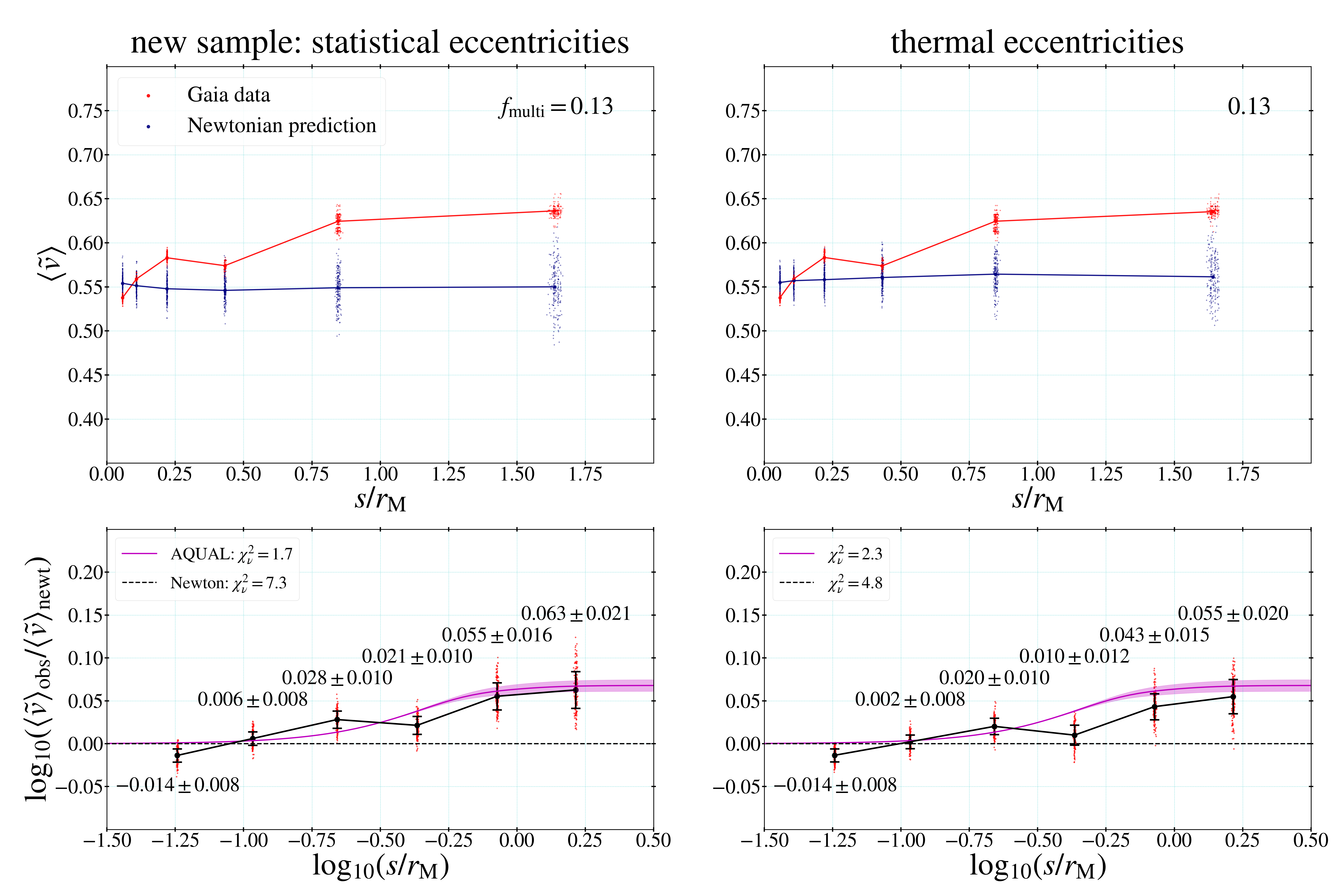}
    \vspace{-0.3truecm}
    \caption{\small 
    {The same as Figure~\ref{vtest_main_stateccen} but for the new sample. The results are qualitatively the same as those shown in Figure~\ref{vtest_main_stateccen}. }
    } 
   \label{vtest_new_stateccen}
\end{figure*} 

Figure~\ref{vtest_main_stateccen} and Figure~\ref{vtest_new_stateccen} respectively show the results for the Chae~(2023a) sample and the new sample based on statistical and thermal eccentricities. With the statistical/thermal eccentricities, the determined values of $f_{\rm{multi}}$ are somewhat lower. For both samples, the inferred $\Gamma$ values with statistical eccentricities agree well (or adequately) with the AQUAL model and rule out the Newtonian model. Even with thermal eccentricities, the AQUAL model is clearly preferred in a relative sense.

\begin{figure*}
  \centering
  \includegraphics[width=0.9\linewidth]{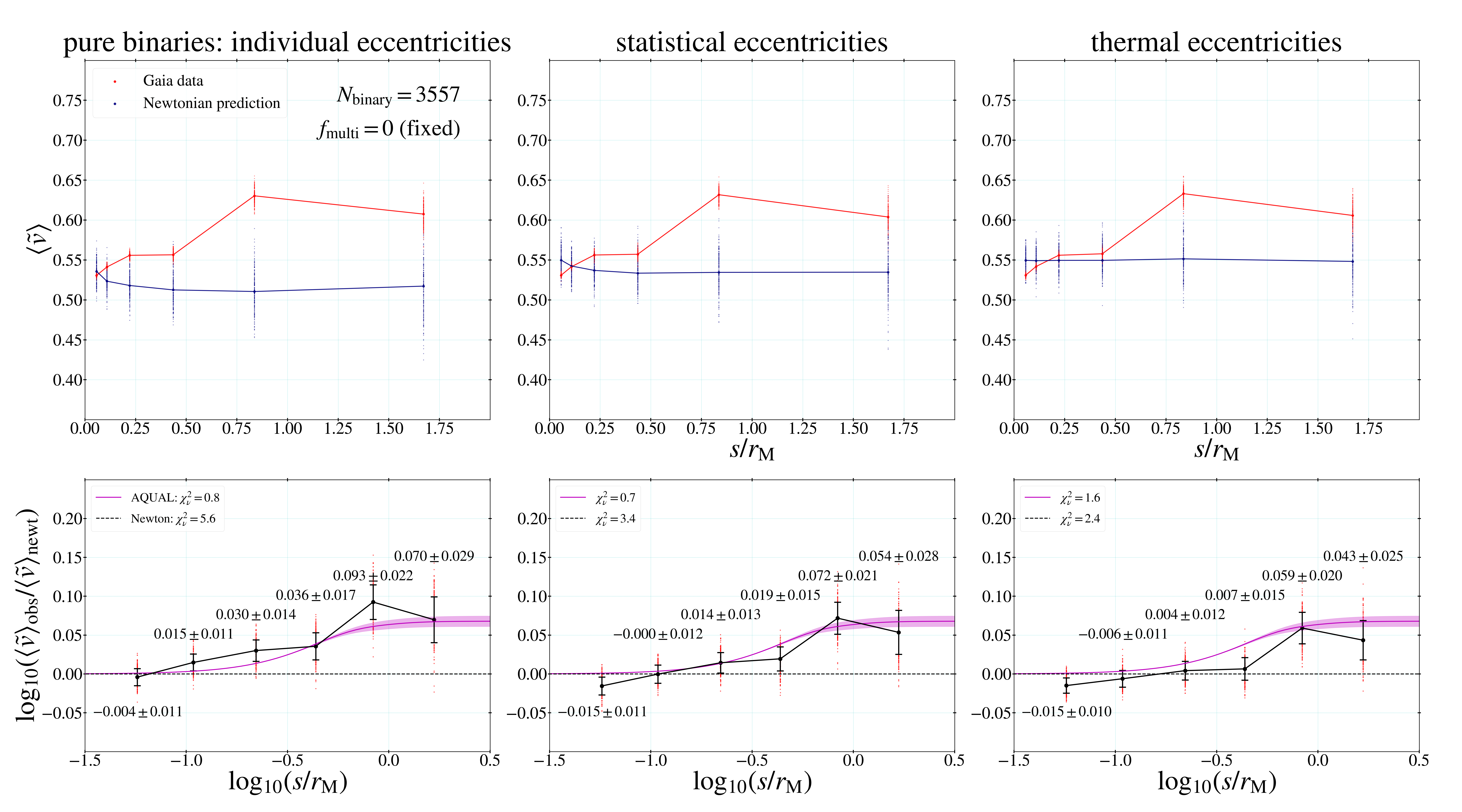}
    \vspace{-0.3truecm}
    \caption{\small 
    {This figure shows the results for the pure binary sample with three choices of eccentricities. This highest quality sample selected with very stringent selection criteria \citep{chae2024,chae2024b} automatically satisfies the kinematic cut almost completely. The results are well consistent with those with the ``impure'' samples shown in Figure~\ref{vtest_main}, \ref{vtest_main_stateccen}, \ref{vtest_new}, and \ref{vtest_new_stateccen} showing that varying degree of data quality requirements does not affect the result of gravity test as long as an MC taking fully into account possible parameter ranges and uncertainties is used. }
    } 
   \label{vtest_pure}
\end{figure*} 

Finally, in Figure~\ref{vtest_pure} I present results for the \citep{chae2024,chae2024b} pure binary sample. The results are well consistent with those for the Chae (2023) sample and the new sample. With the individual eccentricities the AQUAL model is well acceptable by the inferred $\Gamma$ values, but the Newtonian model fails with a high statistical significance ($P_c \approx 8\times 10^{-6}$ for $\chi^2_\nu=5.6$) despite the relatively small sample size. Also, with statistical or thermal eccentricities, the AQUAL model is preferred over the Newtonian model.

 From the above results the logarithmic velocity boost factor for $s/r_{\rm{M}}\ga 0.7$ is measured to be
 \begin{equation}
   \Gamma = \left\{ \begin{array}{lll} 0.059\pm 0.009 & \text{Chae (2023a)} & \\ 0.068\pm 0.015 & \text{new}& (\text{individual }e) \\ 0.092\pm 0.023 & \text{pure binary} &  \\ \end{array} \right.
  \label{eq:Gamma_values}
 \end{equation}
 with the individual eccentricities and
 \begin{equation}
   \Gamma = \left\{ \begin{array}{lll} 0.053\pm 0.008 & \text{Chae (2023a)} & \\ 0.058\pm 0.014 & \text{new}& (\text{statistical }e) \\ 0.077\pm 0.022 & \text{pure binary} &  \\ \end{array} \right.
  \label{eq:Gamma_values}
 \end{equation} 
 with statistical eccentricities. Here the quoted uncertainties include the uncertainties of $f_{\rm{multi}}$. These results alone rule out the Newtonian model with a significance ranging from $3.9-6.6\sigma$. I do not quote values from the results with thermal eccentricities because they are not based on observational inference but considered merely as a possible systematic variation.
 
To summarize, three samples with various sample sizes (from $\approx 3500$ to $20000$), different selection criteria, and different fractions of hierarchical systems (from $f_{\rm{multi}}\approx 0$ to $\approx 0.4$) are all well consistent with the AQUAL model and rule out the Newtonian model with a varying degree of confidence depending on the sample size. Table~\ref{tab:result} gives a summary of the reduced $\chi^2$ values, the survival probabilities, and the logarithmic values of the velocity boost $\Gamma$ (Equation~(\ref{eq:gamma})). From the inferred values of $\Gamma$, the gravitational acceleration boost factor or gravitational anomaly is found to be $10^{2\Gamma}\approx 1.4\pm 0.1$ well consistent with recent results \citep{chae2023a,chae2024,hernandez2023,hernandez2024}.
   
\begin{table*}
  \caption{The inferred velocity boost factors and the $\chi^2$ test  Results}\label{tab:result}
\begin{center}
  \begin{tabular}{cccccccc}
  \hline
 sample   & eccentricities & $f_{\rm{multi}}$  & $\Gamma$ & $10^{2\Gamma}$ & model  &  $\chi^2_\nu$ & $P_c[\equiv 1-P(<\chi^2_\nu)]$ \\
 \hline
 Chae (2023a) & individual  & $0.43\pm 0.05$ &  $0.059\pm 0.009$  & $1.31^{+0.06}_{-0.05}$  &  $\left\{\begin{array}{c} \text{AQUAL} \\ \text{Newton} \\ \end{array}\right.$  & $\begin{array}{c} 2.7 \\ 20.3  \\ \end{array}$ &  $\begin{array}{c} 0.02\\ 2.6\times 10^{-20}  \\ \end{array}$  \\
 Chae (2023a) & statistical & $0.36\pm 0.05$  &  $0.053\pm 0.008$   &  $1.28^{+0.05}_{-0.05}$   &  $\left\{\begin{array}{c} \text{AQUAL} \\ \text{Newton} \\ \end{array}\right.$  & $\begin{array}{c} 3.1 \\ 14.8  \\ \end{array}$ &  $\begin{array}{c} 0.0084 \\ 1.5\times 10^{-14}  \\ \end{array}$  \\
 new      & individual & $0.18\pm 0.09$  &   $0.068\pm 0.015$  &   $1.37^{+0.10}_{-0.09}$   &  $\left\{\begin{array}{c} \text{AQUAL} \\ \text{Newton} \\ \end{array}\right.$  & $\begin{array}{c} 1.6 \\ 9.4  \\ \end{array}$ &  $\begin{array}{c} 0.16 \\  5.7\times 10^{-9} \\ \end{array}$    \\
 new      & statistical & $0.13\pm 0.09$  &   $0.058\pm 0.014$  &  $1.31^{+0.09}_{-0.08}$    &  $\left\{\begin{array}{c} \text{AQUAL} \\ \text{Newton} \\ \end{array}\right.$  & $\begin{array}{c} 1.7 \\ 7.3  \\ \end{array}$ &  $\begin{array}{c} 0.13\\ 7.5\times 10^{-7}  \\ \end{array}$    \\
 pure binary & individual & 0  &  $0.085\pm 0.018$  &  $1.48^{+0.13}_{-0.12}$    &  $\left\{\begin{array}{c} \text{AQUAL} \\ \text{Newton} \\ \end{array}\right.$  & $\begin{array}{c} 0.8 \\ 5.6  \\ \end{array}$ &  $\begin{array}{c} 0.6\\ 8.0\times 10^{-6}  \\ \end{array}$    \\
 pure binary & statistical & 0  &  $0.066\pm 0.017$  &  $1.36^{+0.11}_{-0.10}$    &  $\left\{\begin{array}{c} \text{AQUAL} \\ \text{Newton} \\ \end{array}\right.$  & $\begin{array}{c} 0.7 \\ 3.4  \\ \end{array}$ &  $\begin{array}{c} 0.6\\ 2.3\times 10^{-3}  \\ \end{array}$    \\
 \hline
\end{tabular}
\end{center}
Note. Parameter $\Gamma$ is the logarithmic value of the velocity boot factor defined by Equation~(\ref{eq:gamma}). Quantity $10^{2\Gamma}$ corresponds to the gravitational acceleration boost factor or gravitational anomaly.
\end{table*}

\subsection{Test with a Gaia sample with a narrow dynamic range}  \label{sec:result_narrow}

Here I carry out a quantitative analysis of the subsample with the limited dynamic range $2<s<30$~kau from the Chae (2023a) sample. This is the range considered by \cite{banik2024} while \cite{pittordis2023} considered an even narrower range of $5<s<20$~kau. Comparison of Figure~\ref{scaling_sub} with Figure~\ref{scaling_main} already indicates that the subsample with the limited dynamic range has a nearly flat $\tilde{v}$ profile with $s/r_{\rm{M}}$ and cannot easily distinguish between AQUAL and Newton because AQUAL predicts a nearly flat profile for $s/r_{\rm{M}}\ga 0.7$. Then, the different medians (not the profiles) of the two models may be used to distinguish the models. For this test to work, $f_{\rm{multi}}$ must be known accurately for the test sample because of the inherent degeneracy between total mass and gravity in binaries. However, for this narrow range the acceleration-plane analysis cannot be used to determine $f_{\rm{multi}}$ because the sample lacks the Newtonian regime ($\ga 10^{-8}$~m~s$^{-2}$) data.

If $f_{\rm{}}=0.43\pm 0.05$ determined from the Chae (2023a) sample with $0.2<s<30$~kau is used, the results for the subsample will of course be consistent with those for the full sample. It is then interesting to ask whether the limited range sample can prefer Newton as claimed by \cite{banik2024}. Thus, I consider a fixed value of $f_{\rm{multi}}=0.65$ that is a typical value suggested by \cite{banik2024}. Note, however, that the \cite{banik2024} sample and the present sample cannot be considered statistically equivalent because the \cite{banik2024} sample includes kinematic contaminants such as chance alignments (and thus their analysis had to introduce extra free parameter(s).)

\begin{figure*}
  \centering
  \includegraphics[width=0.9\linewidth]{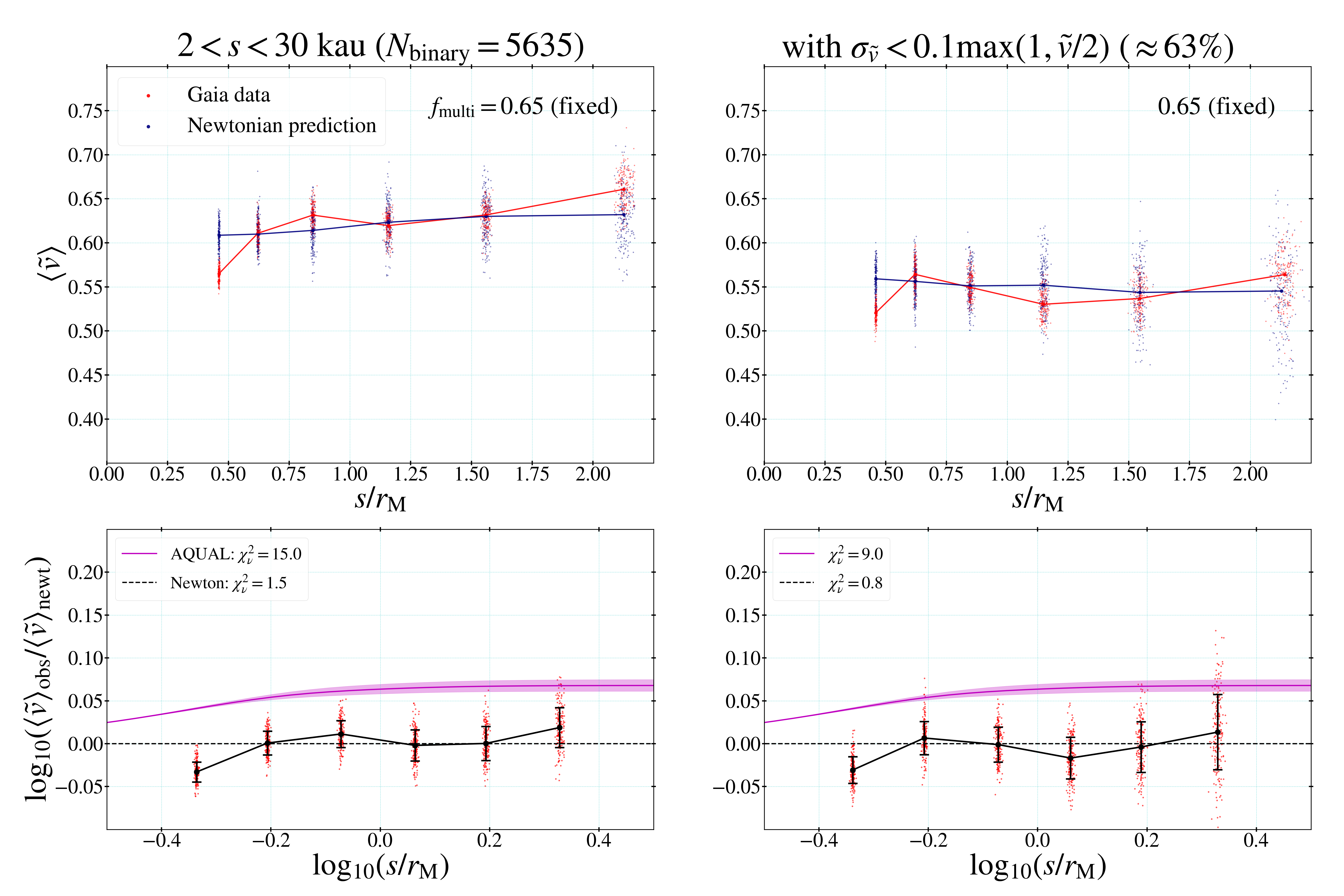}
    \vspace{-0.3truecm}
    \caption{\small 
    {This figure shows gravity test results for the `Chae (2023a) limited sample' the limited dynamic range $s>2$~kau. Because it lacks the Newtonian regime binaries and thus $f_{\rm{multi}}$ could not be determined,  a fixed value of $f_{\rm{multi}}=0.65$ (a value from \cite{banik2024}) is used. For this test, statistical eccentricities are used considering that \cite{banik2024} do not have individual eccentricities for their binaries. Indeed, for this high value of $f_{\rm{multi}}$ the Newtonian model is acceptable while the AQUAL model is ruled out to agree with the \cite{banik2024} conclusion. However, the input value of $f_{\rm{multi}}=0.65$ is ruled out by the determined value of $f_{\rm{multi}}=0.36\pm 0.05$ with statistical eccentricities for the sample of the full range $0.2<s<30$~kau. }
    } 
   \label{vtest_narrow}
\end{figure*} 

Figure~\ref{vtest_narrow} shows the $\tilde{v}$ profiles, the binned measurements of $\Gamma$, and the $\chi^2_\nu$ tests of the AQUAL and the Newton models. Remarkably, with $f_{\rm{multi}}=0.65$ this narrow range data are perfectly consistent with the Newtonian prediction with $\chi^2_\nu = 1.5$ or $0.8$. In contrast, the AQUAL model is ruled out with $\chi^2_\nu = 15.0$ or $9.0$ (up to a Gaussian equivalent significance of $\approx 8\sigma$). These results are fully consistent with those obtained by \cite{banik2024} though formal statistical significance is not as high as their $19\sigma$. Thus, though with a different sample, this test with the same narrow dynamic range confirms the \cite{banik2024} conclusion. Indeed, this conclusion is the consequence of the narrow dynamic range and the high occurrence rate of hierarchical systems.

\subsection{ {A Deeper Test with Distributions of $\tilde{v}$ in Bins of $s/r_{\rm{M}}$}}  \label{sec:result_hist}

So far all tests have been done with parameter $\Gamma$ (Equation~(\ref{eq:gamma})), i.e.\ the median of $\log_{10}(\tilde{v}_{\rm{obs}}/\tilde{v}_{\rm{newt}})$, in bins of $s/r_{\rm{M}}$. The median is relatively robust against any potential outliers that may arise from some unknown systematic in $\tilde{v}_{\rm{obs}}$ and/or $\tilde{v}_{\rm{newt}}$. The Newtonian model has already failed the test with $\Gamma$ while the MOND model represented by a generic numerical solution of the AQUAL field equation is acceptable. When a model fails a first-order/tier test, there is little point of considering a deeper or fuller test.

Thus, a deeper test beyond $\Gamma$ may be warranted only for MOND gravity based on the results of this work so far. However, I also consider testing Newton for the purpose of comparison. A popular test in the recent literature of wide binary gravity tests has been to compare the distribution/histogram of $\tilde{v}_{\rm{obs}}$ (the observed value) with that of $\tilde{v}_{\rm{mod}}$ (the prediction of the model under consideration) in some ranges of $s$ (e.g.\ \citealt{pittordis2019,pittordis2023,clarke2020,banik2024,hernandez2024}). However, because binary systems have different total masses, it is more accurate to consider bins of $s/r_{\rm{M}}$.

While it is straightforward to obtain $\tilde{v}_{\rm{mod}}$ from physically possible orbits for Newtonian gravity as described in Section~\ref{sec:calcvt}, it is not the case for Milgromian gravity, in particular for a fully non-linear field equation such as AQUAL. Here I take advantage of the characteristics of the obtained gravitational anomaly from the studies of this and previous works \citep{chae2023a,chae2024,hernandez2024}, which show consistently that the observed gravity anomaly occurs in a rather narrow range $1\la s\la 7$~kau (mostly in $2\la s\la 5$~kau) for binaries with typical total masses $M_{\rm{tot}}\approx 1.0-1.5 M_\odot$, and then becomes pseudo-Newtonian with a boosted effective gravitational constant $G_{\rm{eff}} \approx 1.4 G$. In the context of MOND gravity, this pseudo-Newtonian behavior is the consequence of the specific gravitational environment thanks to a super-critical external field of the Milky Way.

In this specific environment common\footnote{There are second-order differences due to the inclinations of the orbital axes with respect to the the Galactic external field and small gradients of the Galactic field, which are minor and can be assumed to be statistically averaged out in the present study.} to all nearby binaries, I approximate Milgromian gravity by a model with varying Newton's constant depending on $s/r_{\rm{M}}$. The effective Newton's constant depends on $s/r_{\rm{M}}$ through $G_{\rm{eff}}=\gamma_g(g_{\rm{N}})G$ with $g_{\rm{N}}\approx a_0(1.2 s/r_{\rm{M}})^{-2}$ (see Equation~(\ref{eq:aqgamma})), where the function $\gamma_g(g_{\rm{N}})$ represents the right-hand side of Equation~(\ref{eq:aqformula}). This model becomes Newtonian as $s/r_{\rm{M}}\rightarrow 0$ and pseudo-Newtonian with $G_{\rm{eff}}=1.37$ for $s/r_{\rm{M}}\ga 1$. In the transition range $0.1\la s/r_{\rm{M}} \la 0.7$, this model is less accurate and should be taken with a grain of salt.

I consider three  {representative} bins of $s/r_{\rm{M}}$ to test the distribution/histogram of $\tilde{v}$: $0.040<s/r_{\rm{M}}<0.055$ (deep Newtonian), $0.3<s/r_{\rm{M}}<0.5$ (transition), and $0.8<s/r_{\rm{M}}<1.6$ (MOND). The first bin can be used to test the validity of the MC algorithm itself because gravity is known a priori for the bin. In other words, all gravity models (whether Newtonian or Milgromian) must agree with the Gaia data. Note that for this purpose I consider a narrow range of $s/r_{\rm{M}}\ll 1$  {with the upper limit 0.055 chosen empirically} to minimize the possibility of including systems with $r\ga r_{\rm{M}}$ (where $r$ is the 3D separation deprojected from $s$) as a result of projection effect,  {and the lower limit 0.040 chosen to avoid the edge effect (see the range shown in Figure~\ref{scaling_main}).}  Note also that for the last bin the projection effect is not a concern because all systems with $s/r_{\rm{M}}\ga 1$ are guaranteed to be in the MOND regime since $r\ge s$. The upper limit of $1.6$ (corresponding to $s\approx 11$ kau for a one solar mass binary) in the last bin is chosen as a compromise between number statistics and the width of the range. Fortunately, this limit also helps preventing any possible contamination of the kinematic data for widest binaries.  {(However, below I will also consider a full MOND range of $0.75<s/r_{\rm{M}}<2.50$.)}  {The middle bin of $0.3<s/r_{\rm{M}}<0.5$ is taken from the Newton-MOND transition regime $0.1 \la s/r_{\rm{M}} \la 0.7$.}

\begin{figure*}
  \centering
  \includegraphics[width=1.\linewidth]{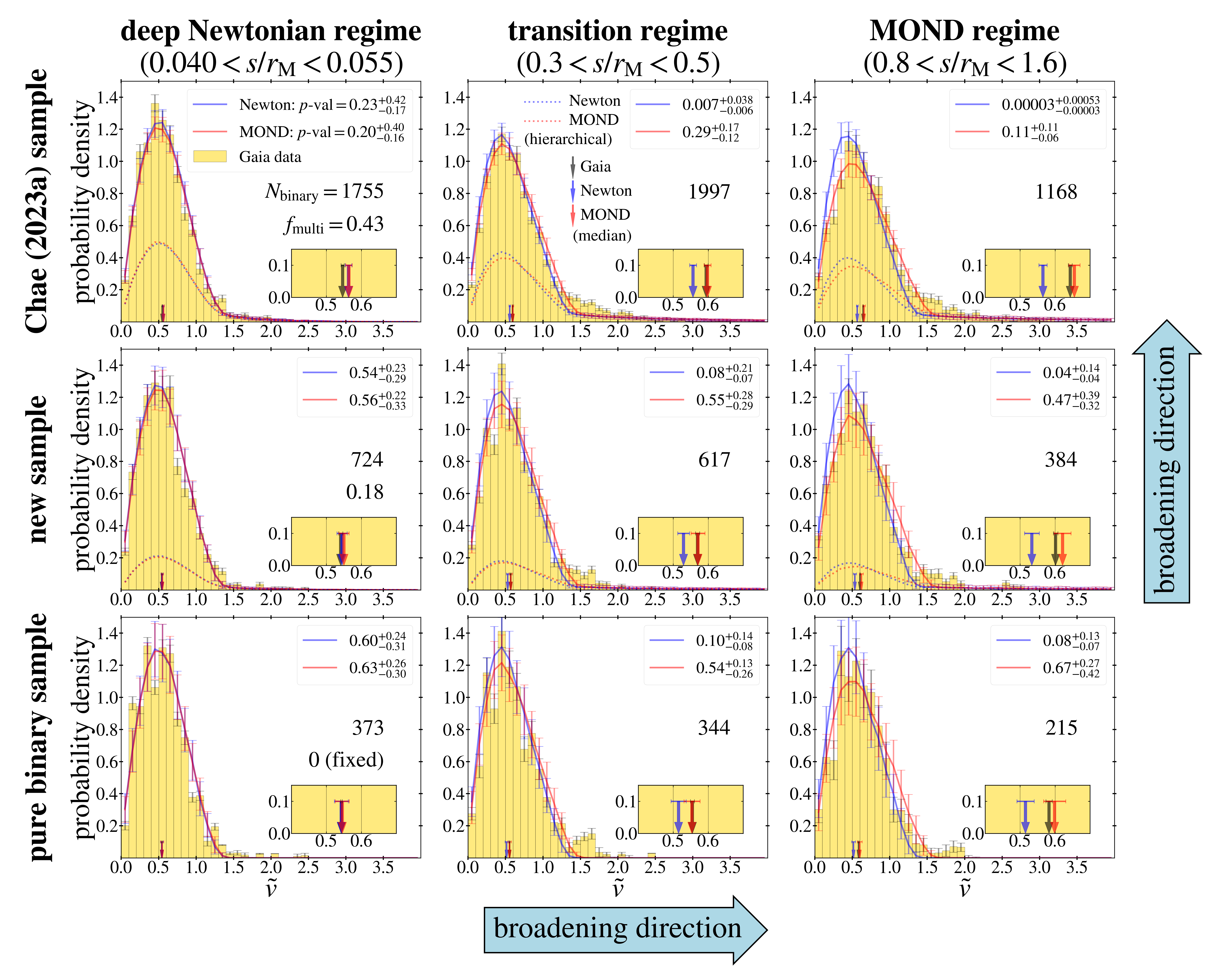}
    \vspace{-0.3truecm}
    \caption{\small 
    {This figure compares the Gaia observed distribution of $\tilde{v}$ with the Newtonian and the MOND/pseudo-Newtonian predictions in three  {representative} bins of $s/r_{\rm{M}}$ for the three samples used in this work. The left column shows the results for the deep Newtonian regime. The shape of the distribution is gradually broadened from the statistically pure binary sample with $f_{\rm{multi}}=0$ to the largest sample with $f_{\rm{multi}}=0.43$ which is well reproduced by the MC algorithms of both models that are indistinguishable in this regime. As the gravity regime changes from the deep Newton to MOND at fixed $f_{\rm{multi}}$, the distribution of $\tilde{v}$ is also broadened. In the MOND regime shown in the right column, the broadened distribution is adequately reproduced by the pseudo-Newtonian model with a boosted gravitational constant $G_{\rm{eff}}\approx 1.37$ for each of the three samples.  {The inset shows the arrows indicating the medians. In the MOND regime the medians of the MOND distributions (red arrows) agree well with those of the Gaia data while the Newtonian medians do not.  Also, in the transition regime the MOND model performs better than the Newtonian model in reproducing the overall shape of the observed distribution. The legend indicates the K-S $p$-value representing goodness of fit of each model.} (see the texts for further).  {The dotted curves indicate the contributions of the hierarchical systems only. They dominate the probability distributions for $\tilde{v}> 1.5$.} }
    } 
   \label{vt_hist}
\end{figure*}

Figure~\ref{vt_hist} shows the results of comparing the Newtonian and MOND predictions with the observed distribution of $\tilde{v}$ in three samples that include different fractions of hierarchical systems in the range $0\le f_{\rm{multi}}\le 0.43$. For the deep Newtonian regime, the Newton/MOND predicted shape agrees nearly perfectly with the Gaia distribution for all samples regardless of $f_{\rm{multi}}$. It can be seen clearly that the distribution gets gradually broadened with a longer tail as $f_{\rm{multi}}$ increases.  {As the dotted curves indicate, the broadening and the tail arise from the hierarchical systems.} This gradual broadening appears to be accurately predicted by the MC modeling.  {I consider the K-S test (see, e.g., \citealt{wall2012}) to check the overall visual agreement quantitatively. The K-S test is performed with the Python module {\tt scipy.stats.ks{\underline{ }}2samp} between the model (Newton or MOND) MC set and the observed MC set, which includes the random scatter due to hierarchical systems (see below for the results even including the small PM measurement scatters). The K-S $p$-value provides a survival probability for the null hypothesis that the two samples are drawn from the same underlying distribution. The high values ($0.2\la$ $p$-value $\la 0.6$) confirm the agreement between the observed data and the MC-produced model data. Also, the Newton/MOND medians indicated by downward-pointing arrows match well the observed medians.} Thus, these results for the deep Newtonian regime robustly verify the validity of the MC algorithm including the kinematic effects of hidden close companions.

 {In the transition regime ($0.3<s/r_{\rm{M}}<0.5$) shown in the middle column of Figure~\ref{vt_hist}, the MOND prediction starts to deviate from the Newtonian prediction as the former is more broadened than the latter. Because the observed distribution is broadened with respect to the deep Newtonian case, the MOND model is preferred over the Newtonian model for all the samples. The MOND predicted medians agree well with the Gaia medians while the Newton predicted medians start to show deviations in agreement with the analyses of $\Gamma$ in the previous subsections. The overall shapes of the observed distributions are adequately described by the MOND model as the high $p$-values indicate (see below for the discussion on the bumps in the high $\tilde{v}$ tail). On the other hand, the low $p$-values represent problems for Newton at least for the largest Chae (2023a) sample.}

I consider next the results for the MOND regime ($0.8<s/r_{\rm{M}}<1.6$) shown in the right column of Figure~\ref{vt_hist}. It can be clearly seen that the observed distribution is  {now further} broadened compared with the corresponding Newtonian prediction for all the samples regardless of $f_{\rm{multi}}$. This inconsistency with the Newtonian prediction was expected  {from} the median test results with $\Gamma$ in the previous subsections. The more interesting question is how well the MOND/pseudo-Newtonian distribution would match the Gaia distribution. Compared with the Newtonian distribution, the MOND distribution is  {significantly} broadened with a more slowly declining tail in all the samples, and consequently the median is shifted to a higher value in good agreement with the observed median. 

Moreover, the shape of the MOND distribution for $0.8<s/r_{\rm{M}}<1.6$ captures reasonably well the overall shape of the broadened Gaia distribution for each of the samples,  {as the high $p$-values suggest. On the other hand, the Newtonian model has some tension with the new and pure binary samples and is clearly discrepant with the largest Chae (2023a) sample with $p$-value of $< 0.0005$.} Considering that MOND gravity is represented by a simplified pseudo-Newtonian model rather than extensive sets of orbits derived by numerically solving the AQUAL field equation, the near perfect reproduction of the median and the overall adequate reproduction of the shape are striking.

 {While the medians and p-values in the transition and MOND regions indicate that the MOND predicted distributions agree overall with the Gaia observed distributions, there is some concern in the details of the distributions because the mild bumps in the range $1.5 < \tilde{v} < 2.0$ of the observed distributions are systematically above the MOND predictions. Considering that the MOND model represented by a varying $G$ in the transition regime is not satisfactory even in principle, any further investigation of the $\tilde{v}$ distributions within the present study may be well motivated only for the MOND regime where the boosted gravitational constant is expected to be universal. }

 {To understand the mild bumps in $1.5 < \tilde{v} < 2.0$ and fully test the Newtonian and MOND models with the $\tilde{v}$ distributions, I further investigate the full range of the MOND regime given by $0.75 < s/r_{\rm{M}} < 2.50$ where the lower limit is chosen to insure a uniform gravitational boost factor (see, e.g., Figure~\ref{vtest_main}) and the upper limit is chosen to avoid the mass selection effect (see the range shown in Figure~\ref{scaling_main}).}

\begin{figure*}
  \centering
  \includegraphics[width=1.\linewidth]{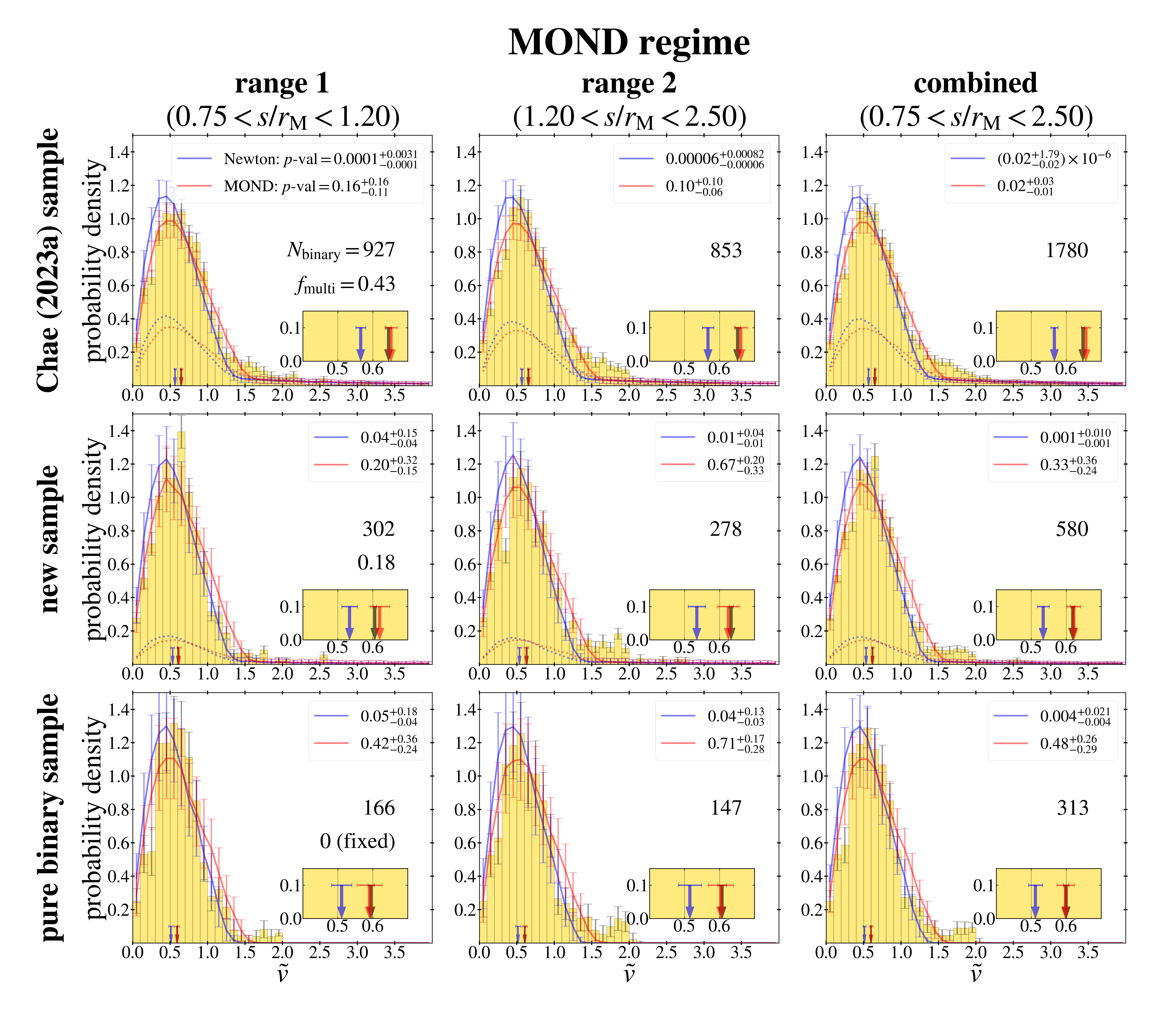}
    \vspace{-0.3truecm}
    \caption{\small 
     {In this figure, a full MOND range of $0.75< s/r_{\rm{M}} <2.50$ is further investigated. The results for the split ranges and the full range are presented in the same format as Figure~\ref{vt_hist}.  }
    } 
   \label{vt_hist_mond}
\end{figure*}

 {Figure~\ref{vt_hist_mond} shows the results for the MOND regime. I consider the split ranges of $0.75< s/r_{\rm{M}} <1.20$ (range 1: median = 0.93) and $1.20< s/r_{\rm{M}} <2.50$ (range 2: median = 1.6) as well as the combined range $0.75< s/r_{\rm{M}} <2.50$. The two split ranges provide independent data to test MOND and can be used to investigate any variation of the shape within the MOND range. It can be seen that the medians and the shapes are similar between the two ranges. Most significantly, the difference between the Newtonian median and the observed/MOND median is almost the same regardless of the range or the sample. This is exactly what MOND predicts. All the $p$-values from either range indicate that the MOND model is acceptable while the Newtonian model is strongly disfavored, in particular by the largest Chae (2023a) sample. The combined range shown in the right column reveals that the Newtonian model is ruled out with $p$-value from $<1.8\times 10^{-6}$ (Chae (2023a) sample) to $<0.02$(pure binary sample) while the MOND model is well acceptable by the new and pure binary samples and acceptable within 95\% by the Chae (2023a) sample. }

 {However, the concern of the mild bumps in the range $1.5 < \tilde{v} < 2.0$ remains in the results for the combined range. Comparison of the results for the split ranges indicates that the bumps are largely contributed by the data from range~2. At larger $s/r_{\rm{M}}$, the sky-projected velocities $v_p$ are lower and thus have larger relative uncertainties for the same required precision of PMs. For those low values of $v_p$ even the small fractional uncertainty ($<0.005$) of PMs could matter. In the main median tests with $\Gamma$ carried out in this work, the small uncertainties of PMs are largely irrelevant (see Appendix~\ref{sec:pmscat} for an explicit demonstration) compared with the dominant MC scatters contributed by hierarchical systems and thus were ignored. Here I investigate whether the ignored PM scatters can have any relevance with the bumps. }

\begin{figure*}
  \centering
  \includegraphics[width=1.\linewidth]{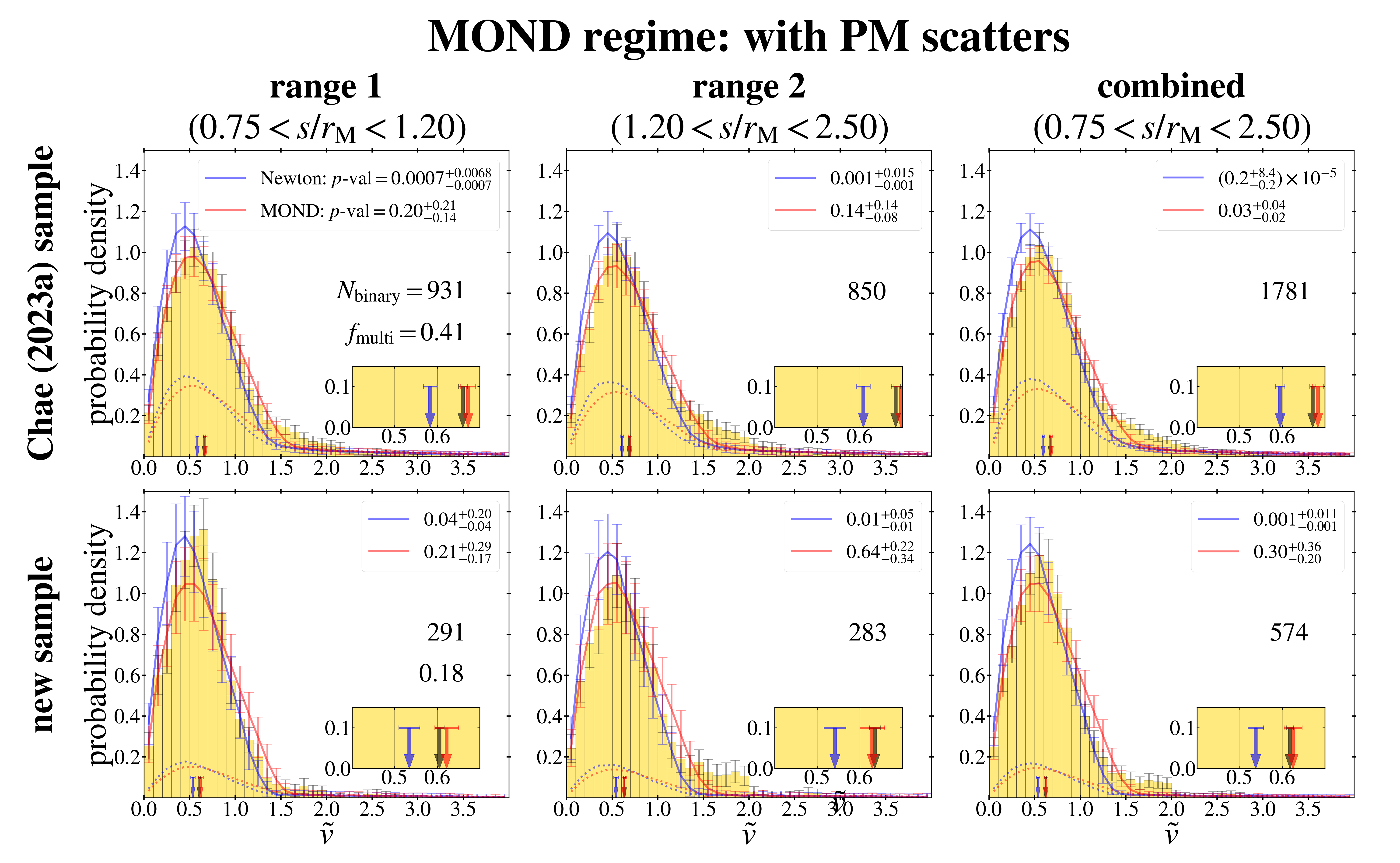}
    \vspace{-0.3truecm}
    \caption{\small 
     {Same as the first two rows of Figure~\ref{vt_hist_mond} but with the small measurement errors ($<0.005$) of PMs included in the MC procedure. The high-$\tilde{v}$ tails are somewhat affected by the  measurement errors. However, the Newtonian model remains excluded while the MOND model is now slightly more acceptable by the Chae (2023a) sample.  }
    } 
   \label{vt_hist_pmscat_mond}
\end{figure*}

 {Figure~\ref{vt_hist_pmscat_mond} shows the results with PM scatters included in the MC procedure for the Chae (2023a) and new samples. (Note that the PM scatters were already included in producing the results for the pure binary sample shown in Figure~\ref{vt_hist_mond} because the sample has no hierarchical systems by the statistical definition.) Compared with Figure~\ref{vt_hist_mond} all medians are somewhat shifted to the right, but the difference between the Newtonian median and the observed/MOND median remains unchanged in all cases. The $p$-values for the Newtonian model are still very low essentially ruling it out with these tests alone. Now the tension of the bumps for the MOND model is significantly reduced for the new sample. For the Chae (2023a) sample some tension still remains in the combined range. The remaining minor tension may be due to various factors such as some mismatch between the Chae (2023a) sample and the hierarchical modeling, some unremoved kinematic contaminants at large separations, etc. However, it is clear that the tension is not serious enough in a statistical sense. }

\begin{table*}
  {\scriptsize
  \caption{ {The binned medians of $\tilde{v}$: Gaia observed values versus Newton and MOND predictions}}\label{tab:vtval}
\begin{center}
  \begin{tabular}{ccccccccc}
  \hline
 sample   & variable  & MC set   & median  &  median  &  median  &  median  &  median &  median \\
 \hline
 C23a  &  $\begin{array}{c} s/r_{\rm{M}} \\  \\  \\  \\  \\ \tilde{v} \\ \\ \\ \\  \\  \\ \end{array}$ &  $ \begin{array}{l} -----  \\ \text{Gaia/No} \\ \text{Gaia/Yes} \\ \text{N-bin/No} \\ \text{N-bin/Yes} \\ \text{N-multi/No} \\ \text{N-multi/Yes} \\ \text{M-bin/No} \\ \text{M-bin/Yes} \\ \text{M-multi/No}\\ \text{M-multi/Yes} \\ \end{array} $  &
 $\begin{array}{c} 0.059 \\ 0.552\pm 0.003 \\ 0.554\pm 0.003 \\ 0.536\pm 0.006 \\ 0.538\pm 0.006 \\ 0.592\pm 0.006 \\ 0.592\pm 0.007 \\ 0.538\pm 0.006 \\  0.540\pm 0.005 \\ 0.592\pm 0.007 \\ 0.595\pm 0.006 \\ \end{array}$  &
 $\begin{array}{c} 0.110 \\ 0.568\pm 0.002 \\ 0.571\pm 0.003 \\ 0.524\pm 0.005 \\ 0.528\pm 0.006 \\ 0.599\pm 0.007 \\ 0.604\pm 0.007 \\ 0.527\pm 0.005 \\  0.532\pm 0.006 \\ 0.605\pm 0.008 \\ 0.608\pm 0.007 \\ \end{array}$  &
 $\begin{array}{c} 0.217 \\ 0.589\pm 0.003 \\ 0.594\pm 0.004 \\ 0.516\pm 0.006 \\ 0.524\pm 0.006 \\ 0.617\pm 0.009 \\ 0.623\pm 0.009 \\ 0.535\pm 0.006 \\  0.540\pm 0.007 \\ 0.638\pm 0.009 \\ 0.645\pm 0.009 \\ \end{array}$  &
 $\begin{array}{c} 0.432 \\ 0.595\pm 0.004 \\ 0.606\pm 0.005 \\ 0.511\pm 0.008 \\ 0.525\pm 0.008 \\ 0.640\pm 0.012 \\ 0.655\pm 0.012 \\ 0.558\pm 0.008 \\  0.573\pm 0.009 \\ 0.700\pm 0.013 \\ 0.715\pm 0.013 \\ \end{array}$  &
 $\begin{array}{c} 0.848 \\ 0.642\pm 0.006 \\ 0.658\pm 0.008 \\ 0.511\pm 0.011 \\ 0.534\pm 0.010 \\ 0.669\pm 0.016 \\ 0.693\pm 0.018 \\ 0.588\pm 0.011 \\  0.608\pm 0.014 \\ 0.771\pm 0.019 \\ 0.791\pm 0.020 \\ \end{array}$  &
 $\begin{array}{c} 1.665 \\ 0.657\pm 0.008 \\ 0.687\pm 0.012 \\ 0.510\pm 0.015 \\ 0.552\pm 0.014 \\ 0.700\pm 0.025 \\ 0.750\pm 0.024 \\ 0.596\pm 0.016 \\  0.636\pm 0.016 \\ 0.812\pm 0.031 \\ 0.857\pm 0.029 \\ \end{array}$ \\
 \hline
 new  &  $\begin{array}{c} s/r_{\rm{M}} \\  \\  \\  \\  \\ \tilde{v} \\ \\ \\ \\  \\  \\ \end{array}$ &  $ \begin{array}{l} -----  \\ \text{Gaia/No} \\ \text{Gaia/Yes} \\ \text{N-bin/No} \\ \text{N-bin/Yes} \\ \text{N-multi/No} \\ \text{N-multi/Yes} \\ \text{M-bin/No} \\ \text{M-bin/Yes} \\ \text{M-multi/No}\\ \text{M-multi/Yes} \\ \end{array} $  &
 $\begin{array}{c} 0.057 \\ 0.534\pm 0.004 \\ 0.534\pm 0.003 \\ 0.533\pm 0.010 \\ 0.535\pm 0.010 \\ 0.588\pm 0.012 \\ 0.587\pm 0.010 \\ 0.535\pm 0.010 \\  0.535\pm 0.009 \\ 0.590\pm 0.011 \\ 0.589\pm 0.011 \\ \end{array}$   &
 $\begin{array}{c} 0.108 \\ 0.556\pm 0.004 \\ 0.556\pm 0.004 \\ 0.526\pm 0.010 \\ 0.527\pm 0.010 \\ 0.602\pm 0.012 \\ 0.600\pm 0.012 \\ 0.529\pm 0.009 \\  0.530\pm 0.010 \\ 0.607\pm 0.012 \\ 0.605\pm 0.013 \\ \end{array}$  &
 $\begin{array}{c} 0.219 \\ 0.581\pm 0.004 \\ 0.580\pm 0.005 \\ 0.517\pm 0.012 \\ 0.516\pm 0.011 \\ 0.620\pm 0.015 \\ 0.618\pm 0.014 \\ 0.536\pm 0.013 \\  0.534\pm 0.012 \\ 0.638\pm 0.017 \\ 0.638\pm 0.017 \\ \end{array}$  &
 $\begin{array}{c} 0.431 \\ 0.572\pm 0.005 \\ 0.572\pm 0.006 \\ 0.513\pm 0.014 \\ 0.515\pm 0.014 \\ 0.644\pm 0.019 \\ 0.646\pm 0.021 \\ 0.560\pm 0.016 \\  0.562\pm 0.016 \\ 0.705\pm 0.023 \\ 0.703\pm 0.021 \\ \end{array}$  &
 $\begin{array}{c} 0.848 \\ 0.614\pm 0.009 \\ 0.619\pm 0.009 \\ 0.511\pm 0.017 \\ 0.518\pm 0.017 \\ 0.673\pm 0.029 \\ 0.671\pm 0.029 \\ 0.589\pm 0.021 \\  0.595\pm 0.020 \\ 0.767\pm 0.036 \\ 0.774\pm 0.035 \\ \end{array}$ &
 $\begin{array}{c} 1.642 \\ 0.633\pm 0.007 \\ 0.635\pm 0.014 \\ 0.511\pm 0.024 \\ 0.519\pm 0.024 \\ 0.696\pm 0.040 \\ 0.706\pm 0.042 \\ 0.596\pm 0.027 \\  0.602\pm 0.027 \\ 0.826\pm 0.056 \\ 0.823\pm 0.056 \\ \end{array}$ \\
 \hline
\end{tabular}
\end{center}
 {\footnotesize Note. `C23a' and `new' refer to the Chae (2023a) sample of 19716 systems and the new sample of 6389 systems defined in Table~\ref{tab:sample}. The medians of $\tilde{v}$ are given for the six bins of $s/r_{\rm{M}}$ uniformly spaced in the log space as defined in the upper left panel of Figure~\ref{vtilde_merged}. The Gaia observed medians are compared with Newton(`N') and MOND(`M') predictions. The medians are given separately for the binary (`-bin') and hierarchical (`-multi') possibilities for each sample. Because it is unknown which ones are binary or hierarchical, all systems in each sample are used to derive `-bin' and `-multi' medians. In deriving each MC set, the medians are calculated either including the individual Gaia PM scatters (`/Yes') or not (`/No'). In comparing the model predictions with the Gaia observed medians, $f_{\rm{multi}} =0.43/0.41$ and $0.18$ should be used for the `C23a'(No/Yes) and `new' samples, respectively.}
}
\end{table*}

 {Table~\ref{tab:vtval} presents the median $\tilde{v}$ values in the six bins of $s/r_{\rm{M}}$ used in the previous subsections. The Newton and MOND predicted medians are separately given for the binary and hierarchical possibilities. When these medians are compared with the Gaia observed medians, the given fractions of hierarchical systems should be used to take an average of the binary and hierarchical predictions.}

\section{Discussion} \label{sec:disc}

\subsection{Comparison with Previous Results} \label{sec:comparison}

In this work I have considered the normalized velocity profile as a function of the normalized separation, i.e.\ $\tilde{v}=\tilde{v}(s/r_{\rm{M}})$. This is a new independent statistical analysis in the plane spanned by two dimensionless quantities $\tilde{v}$ and $s/r_{\rm{M}}$. Unlike \cite{chae2023a,chae2024}, the present method is based on the popular parameter $\tilde{v}$ that has been widely used in the recent literature of wide binary gravity tests (e.g., \citealt{banik2018,banik2024,pittordis2018,pittordis2019,pittordis2023,hernandez2024}). However, this work analyzes $\tilde{v}$ in the bins of $s/r_{\rm{M}}$ rather than $s$. This is because the scaling of $\tilde{v}$ with $s/r_{\rm{M}}$ is equivalent to the scaling of $g/g_{\rm{N}}$ with $g_{\rm{N}}/a_0$ as described in Section~\ref{sec:relations}.

The present results unambiguously reveal the boost of $\tilde{v}$ for $s/r_{\rm{M}}\ga 0.7$ with respect to the Newtonian prediction. The logarithmic value of the velocity boost factor ranges from $\Gamma=0.05-0.09$ depending on the sample and the observational input (in particular eccentricity). The implied boost factor of the observed gravity with respect to Newtonian gravity ranges from $\gamma_g = 10^{2\Gamma}=1.3-1.5$. This is in excellent agreement with several recent results \citep{chae2023a,chae2024,hernandez2023,hernandez2024}.

The aforementioned studies reporting gravitational anomaly all considered a broad dynamic range from the fully Newtonian regime to the fully MOND regime and identified the anomaly in the MOND regime after checking/requiring no anomaly in the Newtonian regime. In contrast, \cite{banik2024} considered a narrow dynamic range excluding the fully Newtonian regime and claimed no anomaly using a MOND model as the benchmark of modified gravity. To understand the origin of their result I have investigated a subsample with the same dynamic range of $2<s<30$~kau taken from the Chae (2023a) sample. With the normalized velocity profile method, I find it unfeasible to simultaneously constrain $f_{\rm{multi}}$ and the boost/anomaly factor ($\Gamma$) with such a narrow dynamic range because MOND gravity models predict a nearly flat $\Gamma$ for such a narrow dynamic range mostly in the MOND regime and thus there is no clear distinction with Newton.

Nevertheless, \cite{banik2024} claimed that their methodology allowed a simultaneous constraint on $f_{\rm{multi}}$ and $\Gamma$\footnote{\cite{banik2024} introduced their own parameter denoted by $\alpha_{\rm{grav}}$ so that $\alpha_{\rm{grav}}=0$ and $1$ for Newtonian and MOND models.} based on counts-in-cells in the plane spanned by $\tilde{v}$ and $s$, not $s/r_{\rm{M}}$. However, because their binary masses are in the broad range $0.5\la M_{\rm{tot}}/{\rm{M}}_\odot \la 3$ and thus the MOND radii are in the corresponding broad range, their distribution of binaries in the $s-\tilde{v}$ plane mixed gravity data in a complex way. Further methodological problems of \cite{banik2024} including too small (compared with data errors) cell sizes are discussed in detail in \cite{hernandez2024a}.

Although $f_{\rm{multi}}$ cannot be determined for the sample with the range $2<s<30$~kau, I have considered a fixed $f_{\rm{multi}}=0.65$ that is a value ``fitted'' by \cite{banik2024}. For this high value, the flat $\tilde{v}$ profile indeed ``agrees well'' with the Newtonian prediction and ``rules out'' the AQUAL model because AQUAL predicts a higher median value of $\tilde{v}$ (Figure~\ref{vtest_narrow}). Thus, though  two (Chae~(2023a) and \cite{banik2024}) samples were selected with different criteria, wide binaries selected within a narrow dynamic range appear to confirm standard gravity if $f_{\rm{multi}}\approx 0.65$ is really the case for typical Gaia DR3 samples. This is, of course, not the case because $f_{\rm{multi}}\approx 0.65$ is ruled out by the Newtonian regime subsample. See the next subsections for further discussions.

\subsection{Possible Sources of Systematic Errors?} \label{sec:systematic}

In this work (and \cite{chae2023a,chae2024}) I have considered various samples and observational inputs and showed that the gravitational anomaly is robust against sample/input variations. Here I carry out some auxiliary analyses to address any remaining possible concerns.

For the analyses of $\tilde{v}$ I have considered 6 bins uniformly spaced in $\log_{10}(s/r_{\rm{M}})$ that are sufficiently wide to have good number statistics and can still capture any possible variation from the flat Newtonian line in the Newton-MOND transition range of $0.8\la s\la 6$~kau. To investigate the effects of binning on the statistical significance of the inferred gravitational anomaly, here I consider variations in binning. For this test, I consider 4 and 9 bins uniformly spaced in $\log_{10}(s/r_{\rm{M}})$ using the new sample that is intermediate in number statistics.  

\begin{figure*}
  \centering
  \includegraphics[width=0.9\linewidth]{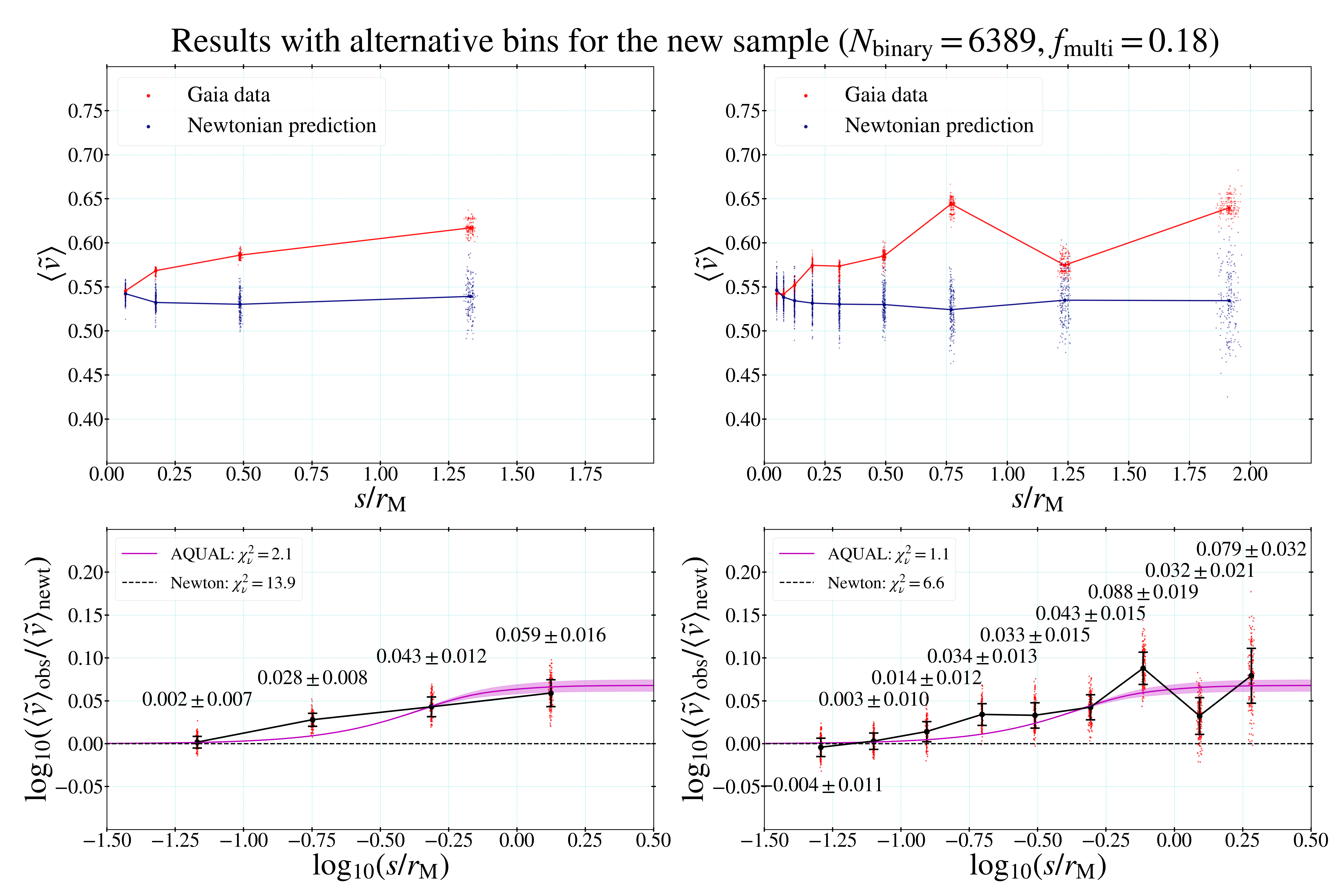}
    \vspace{-0.3truecm}
    \caption{\small 
    {This figure shows the results with different numbers of bins than the standard choice $N_{\rm{bins}}=6$ for the new sample shown in the left column of Figure~\ref{vtest_new}. The $\chi^2_\nu$ values of the models change but the survival probabilities remain consistent showing that the gravity test results are not affected by the number of bins in the MC approach.}
    } 
   \label{vtest_altbins}
\end{figure*} 

Figure~\ref{vtest_altbins} shows the results with the alternative bins. These results are qualitatively very similar to the result with 6 bins. As for $\chi^2$ statistics, the AQUAL model has $1.1\le \chi^2_\nu\le 2.1$ and survival probabilities are in the range $0.10\le P_c \le 0.36$. The Newtonian model has $\chi^2_\nu=13.9$, $9.4$, and $6.6$ for $\nu= 3$, 5, and 8, respectively. The survival probabilities are similar within the range $(0.5\la P_c \la 1)\times 10^{-8}$. These results show that the reduced $\chi^2$ test is largely independent of the specific choice of $N_{\rm{bins}}$ unless unreasonably small or large numbers are used.

One possible issue that \cite{chae2023a,chae2024} have not dealt with explicitly is the possible effects of the binary system (i.e.\ center of mass) velocity $\mathbf{v}_{\rm{sys}}$ relative to the observer. If this velocity is sufficiently large, it can create non-negligible spurious relative velocities between the binary components that are comparable to the intrinsic relative velocities due to the orbital motions. They are called the perspective effects in the literature \citep{shaya2011,pittordis2018}. For the perspective effects to be important, $v_{\rm{sys}}s/d$ must be comparable to the relative orbital velocity between the binary stars that has the same order-of-magnitude as the Newtonian circular velocity $v_c(s)$.

\begin{figure*}
  \centering
  \includegraphics[width=0.9\linewidth]{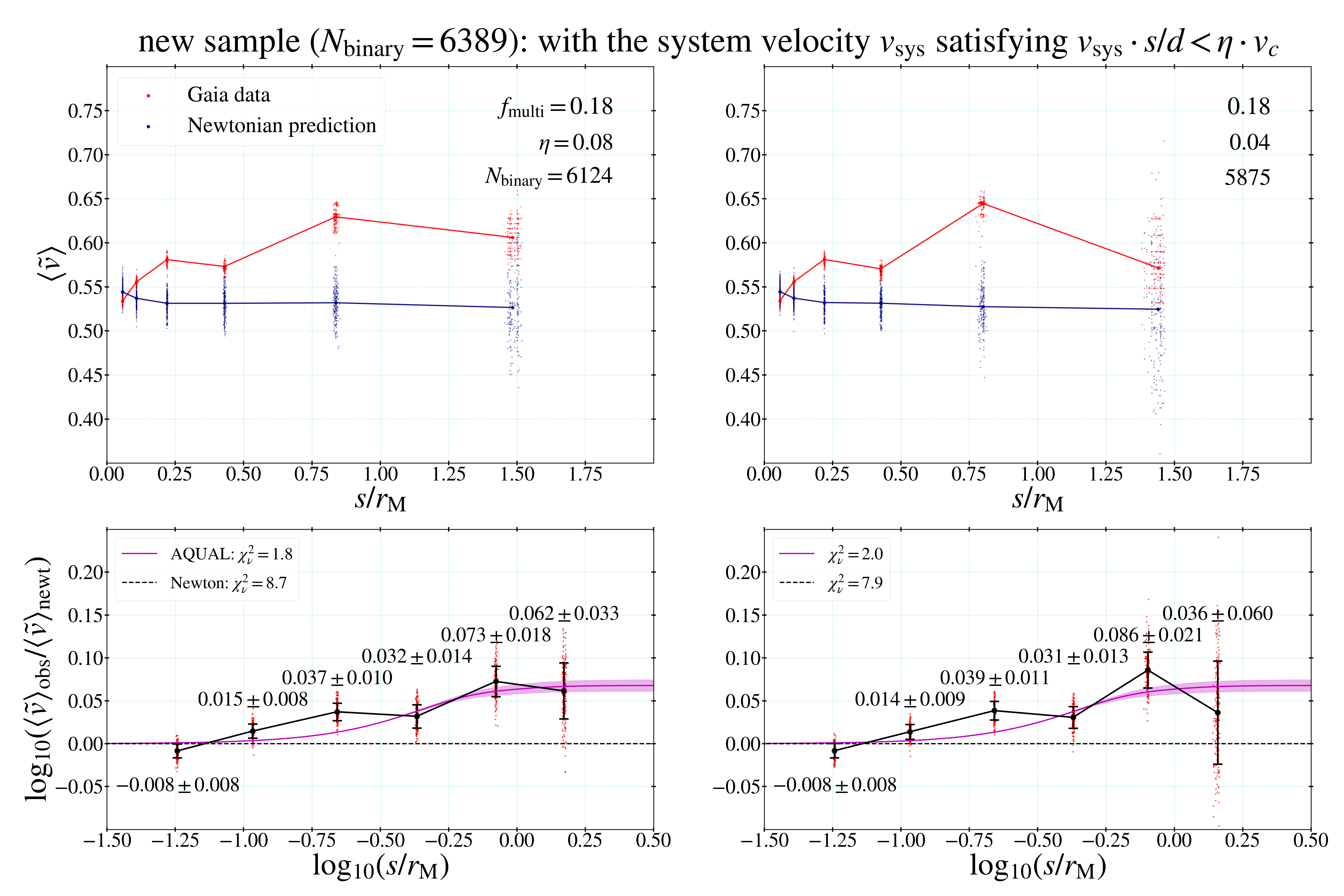}
    \vspace{-0.3truecm}
    \caption{\small 
    {In this figure I consider subsamples of the new sample (shown in the left column of Figure~\ref{vtest_new}) that explicitly exclude binaries with relatively larger system (i.e. center pf mass) velocities $v_{\rm{sys}}$ with respect to the Sun. Because systems with $v_{\rm{sys}}s/d$ comparable to the internal circular velocity $v_c$ can have spurious relative velocities between the components (the so-called ``perspective effects''), I consider $\eta=0.08$ and $0.04$ with the requirement $v_{\rm{sys}}s/d< \eta v_c$. With these requirements only the last bin is somewhat affected, but gravity test results are unaffected in any significant way because the Newton-MOND transition range $0.15\la s/r_{\rm{M}} \la 0.8$ are free of the perspective effects. Note also that my samples already large meet a reasonable requirement (e.g.\ 96/92\% of the new sample satisfy the tested requirement with $\eta=0.08/0.04$). }
    } 
   \label{vtest_sysvlimited}
\end{figure*}

For most binaries in my samples $v_{\rm{sys}}s/d\ll v_c(s)$ is already met, so the perspective effects are not a significant concern. Nevertheless, to explicitly test any possible perspective effects, I consider subsamples with the explicit condition $v_{\rm{sys}}s/d<\eta v_c(s)$ for $\eta=0.08$ and $0.04$. For this purpose, I use the new sample because RVs as well as PMs are needed. Figure~\ref{vtest_sysvlimited} shows the results. It is clearly seen that the magnitude of the gravitational anomaly and $\chi^2_\nu$ statistics are little affected by the constraints. This is not surprising because 92\% of the binaries satisfy even the strong requirement with $\eta=0.04$.

Finally, I revisit the issue of whether variations of $f_{\rm{multi}}$ and/or eccentricity with $s$ can be responsible for the measured gravitational anomaly. This issue was discussed in \cite{chae2023a}. In light of the pure binary sample of \cite{chae2024} and the new sample selected in this work, I discuss further here.

Observational inference of eccentricities using the relative displacement and the relative velocity vector on the sky is now routinely done (e.g.\ \citealt{tokovinin2016,hwang2022}), and there is no reason to suspect the \cite{hwang2022} results on the \cite{elbadry2021} binaries. Nevertheless, I have considered not only individually inferred values but also values from statistical distributions. As a further precaution, I have also considered the thermal distribution. Can there still be a wild variation from these? If any, it may depend on projected separation $s$ and/or the hierarchy. The latter concern may be lessened by considering a sample with a low value of $f_{\rm{multi}}$ as in the new sample or the pure binary sample. The former concern may be largely removed by considering a sufficiently narrow range of $s$.

The occurrence rate of hierarchical systems $f_{\rm{multi}}$ has been assumed to be constant across binary subsamples in this work and other wide binary tests of gravity in the literature. When the range of $s$ is broad as in my samples, this could raise some concern. In particular, if $f_{\rm{multi}}$ dramatically increases with $s$ across the Newton-MOND transition separation $\approx 2$~kau, that could remove the gravitational anomaly in principle (see the next subsection further).  {However, such a possibility is unlikely for two reasons. First, $f_{\rm{multi}}$ concerns only unresolved hidden components because all resolved triples and higher-order multiples with detectable brightness have already been excluded from the binary samples. But, all binary stars satisfy the same photometric, astrometric, and kinematic criteria by the selection requirement, and thus the probability of harboring a hidden companion in a star should not depend on the separation $s$ because measured stellar masses are not correlated with $s$. Second, observed statistics even including resolved multiples do not indicate a strong variation of $f_{\rm{multi}}$ with $s$ (see, e.g., Figure~11 of \cite{tokovinin2014}, Figure~14 of \cite{hartman2022}).}

To test robustly the gravitational anomaly largely free of the above possible remaining concerns, I consider a narrow range of $s$ that precisely covers the Newton-MOND transition range. From a Milgromian gravity model such as AQUAL, the Newton-MOND transition is expected to occur within the acceleration range $10^{-10}\la g_{\rm{N}}\la 10^{-8}$~m~s$^{-2}$. For a binary system with $M_{\rm{tot}}=1{\rm{M}}_\odot$, the transition range is approximately $0.8<s<6$~kau or $0.1\la s/r_{\rm{M}}\la 0.8$ for $r_{\rm{M}}\approx 7$~kau. This range is sufficiently narrow that the eccentricity distribution  {is nearly uniform. Moreover,} any possible variation of $f_{\rm{multi}}$ must be tiny and so I take the uniform value already determined for each source/full sample.

As a further precaution, I consider the limited distance range $100 < d< 200$~pc for the current test. There are two reasons for this choice. First, this range out of the full range $0<d<200$~pc allows the maximum number of binaries within a factor of 2 for the distance difference between binaries. Second, for the combination of $s\la 6$~kau and $d>100$~pc, $s/d\la 3\times 10^{-4}$ and thus any possibility of the perspective effect can be safely avoided.

For the above specifications, I consider a robust and simple slope test without reference to any modified gravity model. The motivation is clear. If the standard gravity holds for this narrow range of $0.8\la s \la 5$~kau, the slope of the anomaly parameter $\Gamma$ (Equation~(\ref{eq:gamma})) must be consistent with zero regardless of the overall amplitude. A nonzero slope is a sure signature of gravity breaking down. As mentioned above, for this test the new and the pure binary samples are preferable because $f_{\rm{multi}}$ is small or negligible. However, I also include the Chae (2023a) sample.

\begin{figure*}
  \centering
  \includegraphics[width=1.0\linewidth]{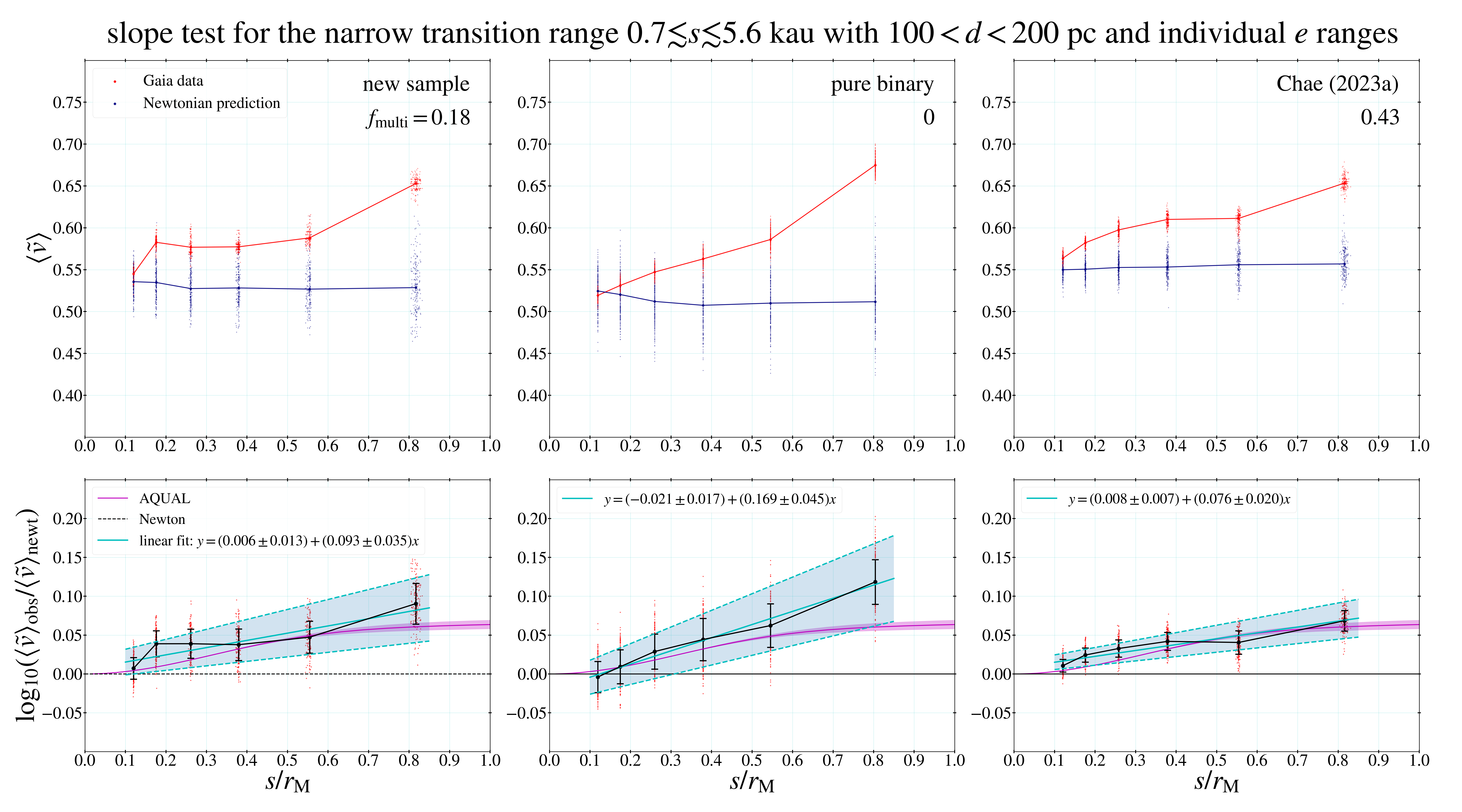}
    \vspace{-0.5truecm}
    \caption{\small 
    {This figure shows the results from a robust slope test for binaries in the narrow range of the Newton-MOND transition regime. For this narrow range of $0.1 \la  s/r_{\rm{M}} \la 0.8$ ($0.7 \la s \la 5.6$~kau for the median mass of $1{\rm{M}}_\odot$), any possible concerns in variation of $f_{\rm{multi}}$ and/or eccentricity with $s$ are minimal. As a further precaution, I also consider the narrow distance range of $100<d<200$~pc to minimize any possible dependence with distance. These results are pure measurements without reference to any modified gravity theory. The results are particularly robust for the new and pure binary samples because $f_{\rm{multi}}$ is very low. The measured slopes rule out the Newtonian expectation of zero with statistical significance ranging from $2.7-3.8\sigma$ estimated with an MCMC analysis as shown in Figure~\ref{linfit}. The light blue bands in the bottom panels represent the combined uncertainties of the slope and the $y$-intercept. The AQUAL model is well within this band while the Newtonian model is not.}
    } 
   \label{vtest_trans}
\end{figure*} 

\begin{figure}
  \centering
  \includegraphics[width=0.9\linewidth]{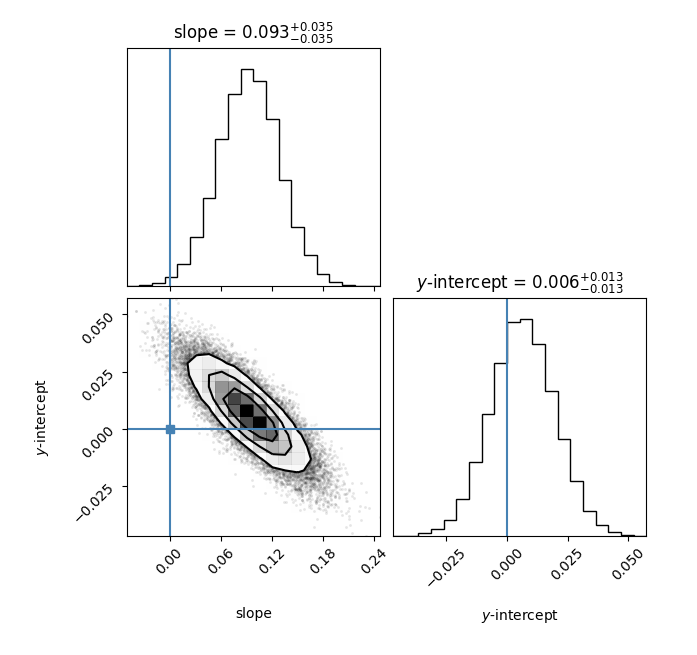}
    \vspace{-0.5truecm}
    \caption{\small 
    {An example of the MCMC fit of the measured $\Gamma$ values in the narrow transition range shown in Figure~\ref{vtest_trans} is exhibited for the new sample. The public Python MCMC package {\tt emcee} \citep{emcee} is used. The Newtonian model (center of the blue cross) is well outside the probable parameter space. }
    } 
   \label{linfit}
\end{figure}

Figure~\ref{vtest_trans} shows the slope test results for the range of $0.1 <  s/r_{\rm{M}} < 1$. I consider 6 uniform bins in the logarithmic space $-1<\log_{10}(s/r_{\rm{M}})<0$. In all three cases, there is a clear systematic rise from $\Gamma=0$ at $s/r_{\rm{M}}\approx 0.1$ to $\Gamma>0$ at $s/r_{\rm{M}}\approx 0.8$. The bands representing the combined uncertainties of the slope and the y-intercept do not encompass the Newtonian flat line. The estimated slopes are $b=0.093\pm 0.035$ (new sample), $0.169\pm 0.045$ (pure binary), and $0.076\pm 0.020$ (Chae 2023a).  These slopes are discrepant with zero at a significance ranging from $2.7-3.8\sigma$. Here the Bayesian uncertainties are estimated with the Markov chain MC method. An example for the new sample is shown in Figure~\ref{linfit}.

The slope test results robustly confirm the breakdown of standard gravity. For the narrow range of $0.8\la s \la 6$~kau
 (within an order of magnitude) there is no observational reason that the systematic rise of $\Gamma$ should occur. Although I do not show it here (because it is obvious from various results shown in the results section of this work and the results from \cite{chae2023a,chae2024}), if a similar test with another narrow range such as $-1.4<\log_{10}(s/r_{\rm{M}})<-0.4$ (the same width as above), the measured slopes are well consistent with zero.

\subsection{Newtonian ``Prediction'' on the Statistical Properties of Wide Binaries} \label{sec:newton}

All the above results including the narrow range slope test results shown in Section~\ref{sec:systematic} have already shown that the breakdown of standard gravity is inevitable. As an auxiliary argument and a matter of completeness, here I quantify what standard gravity requires for Gaia wide binaries. These requirements can be considered as the Newtonian ``predictions'' on observations.

Standard gravity requires a dramatic change in $f_{\rm{multi}}$ and/or eccentricity distribution in the transition range of $1\la s \la 5$~kau. The required changes from the regime $s<1$~kau to the regime $s\ga 5$~kau are summarized in Table~\ref{tab:newton}. Unless the current estimate of eccentricities by \cite{hwang2022} is grossly in error, $f_{\rm{multi}}$ must increase from $s\la 1$~kau to $s\ga 5$~kau by $\Delta f_{\rm{multi}}\approx +0.4$ in all three samples.  {This seems impossible mainly because $f_{\rm{multi}}$ concerns only unresolved hidden companions and all stars satisfy the same photometric, astrometric, and kinematic criteria regardless of $s$, so that stars should have nearly equal probabilities of hiding unresolved companions as long as masses are independent of $s$ as is the case for the samples. Moreover, the observed statistics even including resolved companions (e.g. \citealt{tokovinin2014,hartman2022}) do not indicate at all such a dramatic increase.}

If $f_{\rm{multi}}$ is assumed to be not increasing with $s$ for the whole probed range $0.2<s<30$~kau, eccentricity distribution (and consequently the mean value) must become subthermal ($\alpha<1$) for large separation systems, in the opposite direction to the current estimate by \cite{hwang2022}. The mean eccentricity must change from $s\la 1$~kau to $s\ga 5$~kau by $\Delta\langle e\rangle \approx -0.10$. However, as Figure~24 of \cite{chae2023a} summarizes various survey results, there is a clear observational trend that $\langle e\rangle$ increases with $s$.

But, could triples and higher-order multiples affect the eccentricity distribution in a complex way? For example, \cite{tokovinin2022} reports a mean eccentricity of $0.54\pm 0.02$ for wide outer orbits of nearby resolves triples. This value is similar to the required mean values written in the fourth column of Table~\ref{tab:newton}. However, such an argument cannot work because even if one half of the Chae (2023a) sample are triples the average over pure binaries and triples must be $>0.60$. Also, it is unreasonable to imagine that resolved triples affect the eccentricity distribution preferentially for $s\ga 5$~kau. Moreover, the pure binary sample requires a similar or even lower mean eccentricity although the possible effects of triples are negligible or minimal.

\begin{table*}
  \caption{Newtonian ``prediction'' on the wide binary statistics for $5\la s\la 30$~kau}\label{tab:newton}
\begin{center}
  \begin{tabular}{ccccc}
  \hline
 sample   & change in $f_{\rm{multi}}$ &  change in $\alpha$  & change in mean $e$ & current literature estimate \\
 \hline
 Chae (2023a) &  $0.36\pm 0.05\rightarrow 0.78\pm 0.10$  & $1.0\rightarrow 0.3$  & $0.67\rightarrow 0.57$ & $\Delta\alpha\ga+0.3$, $\Delta\langle e\rangle\ga+0.03$  \\
 new    &  $0.13\pm 0.09\rightarrow 0.60\pm 0.15$  & $1.0\rightarrow 0.3$    &  $0.67\rightarrow 0.57$ &  same as above  \\
 pure binary &  $0 \rightarrow 0.47\pm 0.25$  & $1.0\rightarrow 0.2$    &  $0.67\rightarrow 0.55$ & same as above  \\
\hline
\end{tabular}
\end{center}
Note. (1) The required change is with respect to the current estimate for $0.2<s\la 1$~kau. (2) $f_{\rm{multi}}$ is estimated for the ``statistical eccentricity distribution'' with the acceleration-plane method. (3) $\alpha$ refers to the index in the power-law eccentricity distribution $p(e)=(1+\alpha)e^\alpha$. (4) The last column quote is based on \cite{hwang2022}.

\end{table*}

\section{Summary and Conclusions} \label{sec:conc}

In this work I have investigated the normalized velocity profile $\tilde{v}$ as a function of the normalized radius $s/r_{\rm{M}}$ as a new study complementary to and independent of two recent studies \citep{chae2023a,chae2024}. I have considered three samples of binaries including a new independent sample selected with requirements on distances and RVs without using the \cite{elbadry2021} $\mathcal{R}$ values. The samples cover a broad range in number from 3500 to 20000 and in the implied fraction of hierarchical systems from $f_{\rm{multi}}\approx 0$ to $\approx 0.4$. Through the MC method binaries were distributed probabilistically in the plane spanned by two dimensionless quantities $s/r_{\rm{M}}$ and $\tilde{v}$ and at the same time corresponding mock binaries from the virtual Newtonian world were also distributed in the same plane. Then, I derived values of the logarithmic velocity boost parameter $\Gamma$ (Equation~(\ref{eq:gamma})) in bins of $s/r_{\rm{M}}$ using the binned median $\tilde{v}$ values from the two MC distributions. The uncertainties of $\Gamma$ were naturally estimated from a number of MC runs.

The derived values of $\Gamma$ are pure measurements without reference to any modified gravity theory because only Newtonian reference calculations were used to derive them. As shown with a sample of realistic virtual Newtonian binaries, measured values of $\Gamma$ and their slope as a function of $s/r_{\rm{M}}$ are expected to be consistent with zero in a universe obeying standard gravity. Because the MC distributions of $\Gamma$ are consistent with the Gaussian probability density function and each bin has a good number of data points, I have used the widely used $\chi^2$ statistics to test gravity models. I specifically tested the Newtonian and AQUAL models. I have also done a simple slope test with a narrow range of $s/r_{\rm{M}}$.

From the measured values of $\Gamma$ and statistical tests based on them the following have been found.
\begin{enumerate}
\item The measured values of $\Gamma$ show a systematic variation with $s/r_{\rm{M}}$ with respect to the Newtonian flat line. The trend can be easily recognizable even before detailed statistical analyses. $\Gamma$ varies from $\approx 0$ at $s/r_{\rm{M}}\la 0.15$ to $0.068\pm 0.015$ (stat) $_{-0.015}^{+0.024}$ (syst) for $s/r_{\rm{M}} \ga 0.7$ or for $s\ga 5$~kau. Here the representative value is derived from the new sample of 6389 binaries and systematic uncertainties are due to variations in samples and eccentricities.

\item The gravitational anomaly (i.e.\ acceleration boost) factor given by $\gamma_g = 10^{2\Gamma}$ is measured to be $\gamma_g = 1.37_{-0.09}^{+0.10}$ (stat) $_{-0.09}^{+0.16}$ (syst).

\item The Newtonian model has $\chi^2_\nu=9.4$ for the new sample with the survival probability of $P_c=5.7\times 10^{-9}$ that corresponds to a Gaussian equivalent significance of $5.8\sigma$. For the largest sample of $\approx 2\times 10^4$ binaries selected with somewhat relaxed criteria, the Newtonian model has $\chi^2_\nu=20.3$ with a Gaussian equivalent significance of $9.2\sigma$.

\item The AQUAL model has $\chi^2_\nu=1.6$ for the new sample with the survival probability of $P_c= 0.16$ that is fully acceptable. For sample and eccentricity variations, it has $0.7\le \chi^2_\nu \le 3.1$ that are acceptable by conventional standards.

\item The variation of $\Gamma$ occurs in the narrow range of $0.8\la s\la 6$~kau. For best-controlled samples of binaries from this range and the limited distance range $100<d<200$~pc, a simple slope test robustly rules out the Newtonian null value in line with the $\chi^2$ test results.

  \item  {The shape of the probability density distribution of the normalized velocity $\tilde{v}$ in a deep Newtonian regime of strong acceleration is well reproduced by the Newtonian model. However, in a MOND regime of low acceleration, the observed distribution mismatches severely the Newtonian prediction but matches the MOND (represented by a pseudo-Newtonian model with a boosted effective gravitational constant $G_{\rm{eff}}=1.37 G$) prediction (Figure~\ref{vt_hist}).}
\end{enumerate}

From this work and \cite{chae2023a,chae2024} I have now considered three independent methods with numerous samples. All results consistently point to the immovable gravitational anomaly as long as a sufficiently wide dynamic range is used. The null result by \cite{banik2024} with $f_{\rm{multi}}>0.6$ can be largely attributed to their choice of the narrow dynamic range  {not encompassing the Newtonian regime and a significant portion of the transition regime}\footnote{There was an agreement on the need for a large dynamic range after an intense discussion at the June 2023 St Andrews conference.} (in particular considering the edge effect), but there are other methodological issues as well \citep{hernandez2024a}. Indeed, an independent group uses a sufficiently wide dynamic range and reports essentially the same anomaly through independent analyses based on independent samples \citep{hernandez2023,hernandez2024}. The current evidence for the gravitational anomaly is well above $5\sigma$ and the evidence comes from multiple analyses.

The reported gravitational anomaly is a pure measurement without reference to any modified gravity model. It is striking that the gravitational anomaly agrees remarkably well with a Milgormian (MOND) gravity model. There are no observational reasons that gravity suddenly appears to change from $s\la 1$~kau to $s\ga 5$~kau by 40\%. It appears that Milgromian gravity is verified as a phenomenological model from current observations. In this respect, it is interesting to note two nearly concurrent independent studies that reached the same conclusion through observations of asymmetrical tidal tails of open clusters \citep{kroupa2024} and indefinitely flat rotation curves of rotationally supported galaxies \citep{mistele2024}. Perhaps, what is even more mysterious is that the world does not know what the underlying fundamental theory (presumably encompassing general relativity; \citealt{einstein1916}) is for the low-acceleration MOND phenomenology.

\section*{Acknowledgments}
Preliminary results of this work were originally presented by an invited talk in an online meeting `Challenges of modern cosmology' held on January 18th, 2024.\footnote{https://sites.google.com/view/cmc2024-challengesofmoderncosm/}  {The author thanks the anonymous referee for suggesting an additional analysis presented in Section~\ref{sec:result_hist} and valuable comments to improve the presentations. The author also thanks Xavier Hernandez for suggesting an improved presentation of figures, and Kareem El-Badry for pointing \cite{tokovinin2022} to the author for the statistics of resolved triples.} This work was supported by the National Research Foundation of Korea (grant No. NRF-2022R1A2C1092306). 

\bibliographystyle{aasjournal}

\appendix

\section{Results from the acceleration plane analysis} \label{sec:accel}

\begin{figure*}
  \centering
  \includegraphics[width=0.9\linewidth]{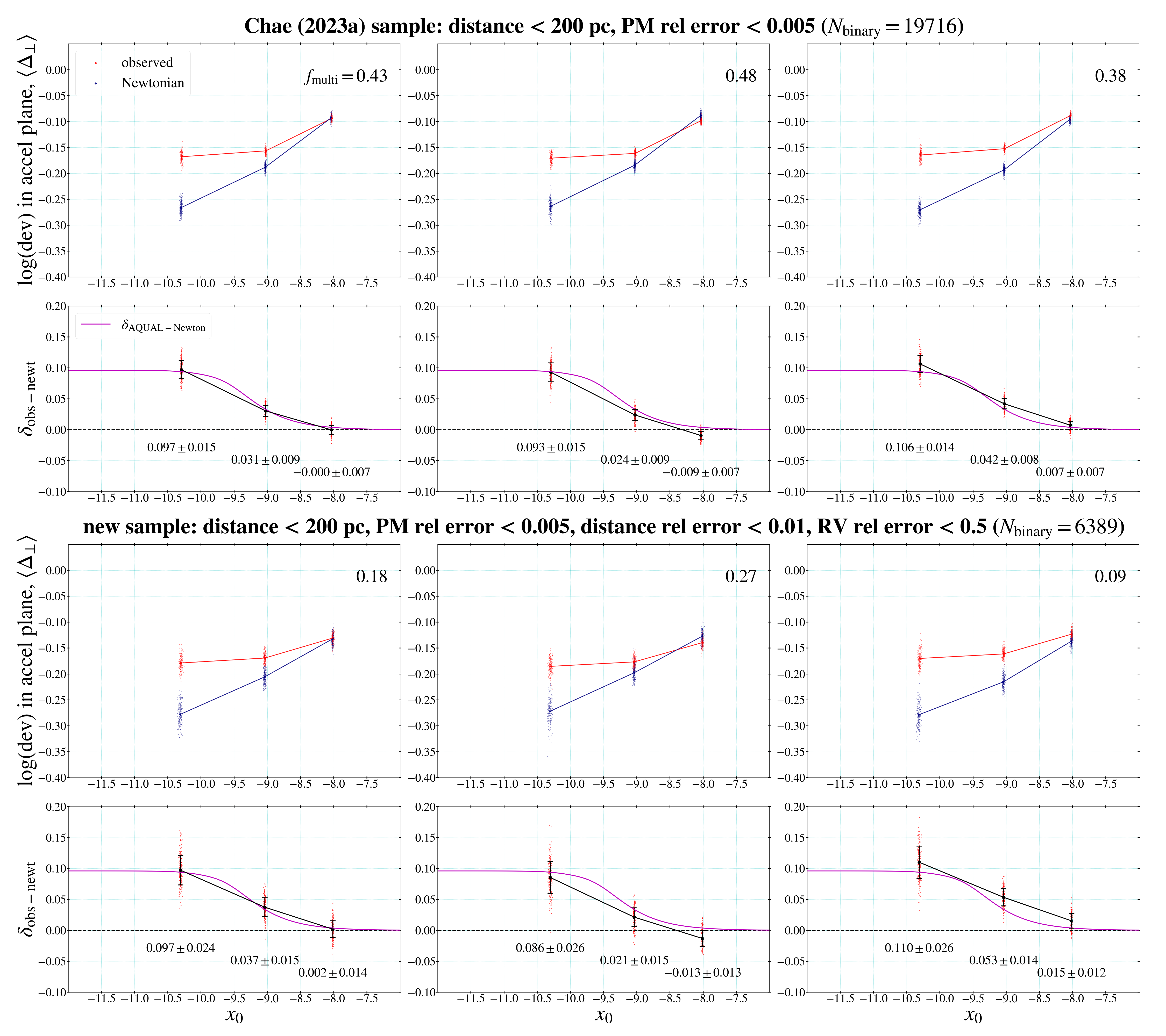}
    \vspace{-0.3truecm}
    \caption{\small 
    {Red and blue points represent the median orthogonal deviations $\langle\Delta_\bot\rangle$ in three orthogonal bins in the acceleration plane from MC results for the Gaia observed binaries and the corresponding virtual Newtonian binaries. The reader is referred to \cite{chae2023a} for the definition of the acceleration plane and $\langle\Delta_\bot\rangle$. The parameter $\delta_{\rm{obs-newt}}$ represents the difference in $\langle\Delta_\bot\rangle$ between the observed and virtual binaries. The left columns are the results for the best-fit values of $f_{\rm{multi}}$ while the other columns represent the uncertainties.}
    } 
   \label{residual}
\end{figure*}

Here I present the results from the acceleration-plane analysis  {in light of a revision of the code from \cite{chae2023a} and samples}. The code from \cite{chae2023a} is slightly updated \citep{chae2023b} in drawing individual eccentricities based on the \cite{hwang2022} measurements as described in Section~\ref{sec:input}. Figure~\ref{residual} shows the orthogonal deviations from the $g=g_{\rm{N}}$ line in the plane spanned by logarithmic accelerations for the Gaia observed binaries and the corresponding virtual Newtonian binaries and the gravitational anomaly parameter $\delta_{\rm{obs-newt}}$ defined by the difference of the two. While these results represent measurements of gravitational anomaly, in the present study they mainly serve to provide measurements of $f_{\rm{multi}}$ to be used in the normalized velocity profile analysis. I find $f_{\rm{multi}}=0.43\pm 0.05$ and $f_{\rm{multi}}=0.18\pm 0.09$ for the two samples \text{(Chae 2023a and new)} with individual eccentricities. For statistical eccentricities, they are somewhat lower. 

\begin{figure*}
  \centering
  \includegraphics[width=0.65\linewidth]{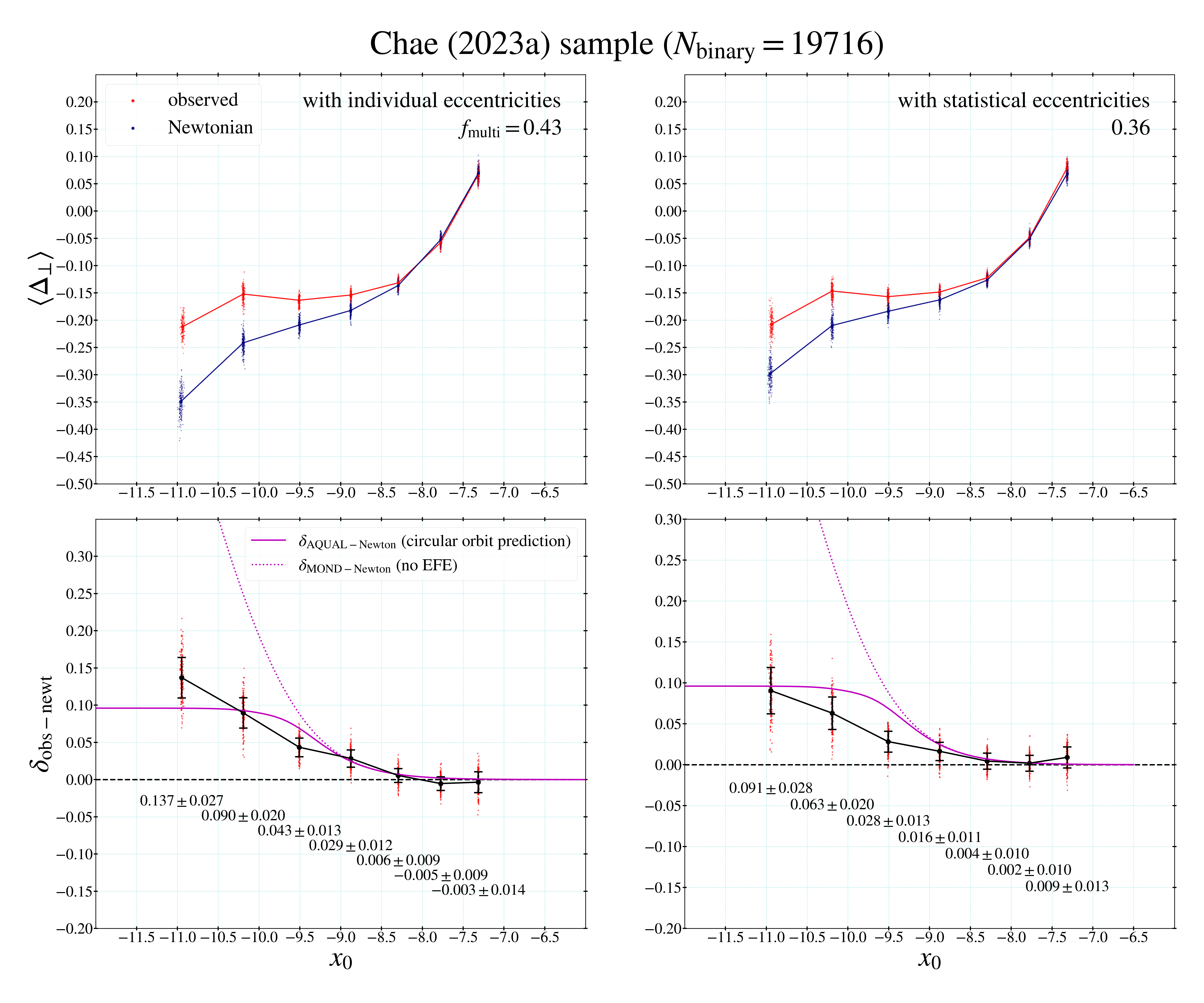}
    \vspace{-0.5truecm}
    \caption{\small 
    {This figure shows the acceleration-plane results in 7 bins for the `Chae (2023a)' sample. The best-fit cases are shown with individual or statistical eccentricities. Note that the algebraic MOND model without an EFE represented by the dotted curve is extremely highly excluded. Any modified gravity model phenomenologically mimicking the algebraic MOND is thus ruled out already. }
    } 
   \label{residual_main_7bins}
\end{figure*}

\begin{figure*}
  \centering
  \includegraphics[width=0.65\linewidth]{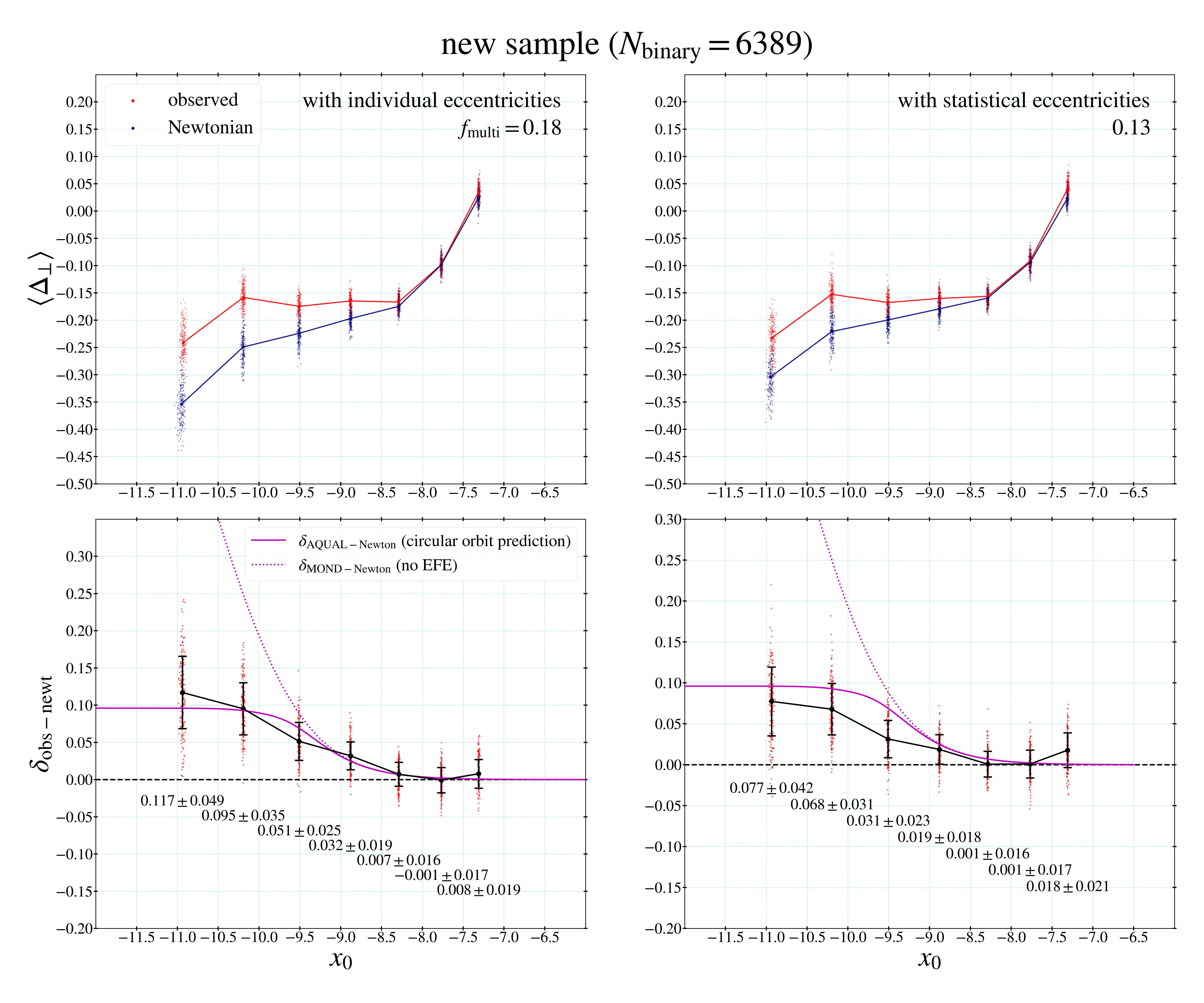}
    \vspace{-0.5truecm}
    \caption{\small 
    {Same as Figure~\ref{residual_main_7bins} but for the `new sample'. }
    } 
   \label{residual_new_7bins}
\end{figure*}

I also present the results for 7 bins in Figure~\ref{residual_main_7bins} and \ref{residual_new_7bins} for the two samples. In these figures, only best-fit $f_{\rm{multi}}$ cases are shown but for both individual and statistical eccentricities.  {Finally, Figure~\ref{residual_main_7bins} shows the result for the pure binary sample updating Figure~11 of \cite{chae2024}.}

\begin{figure*}
  \centering
  \includegraphics[width=0.65\linewidth]{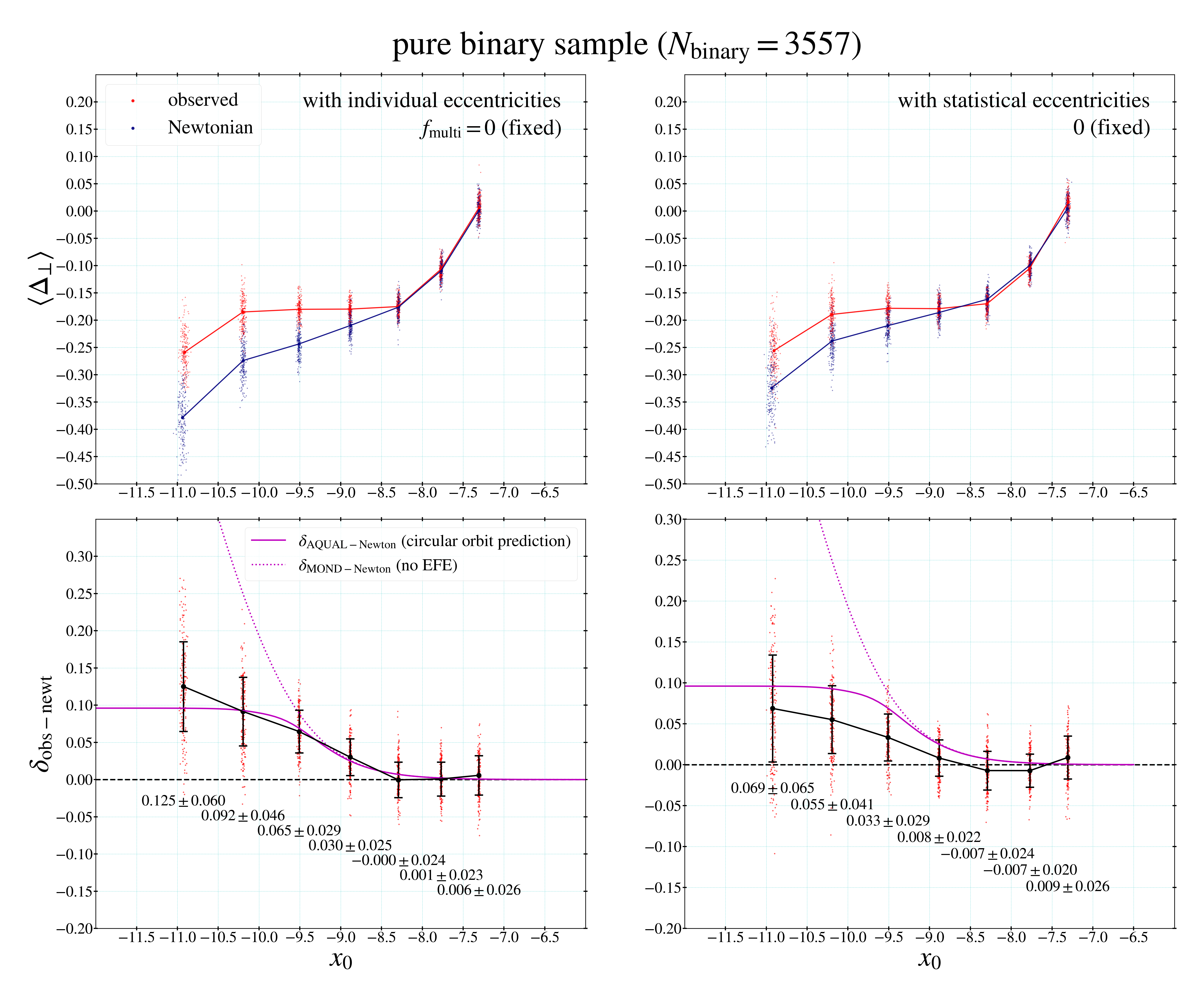}
    \vspace{-0.5truecm}
    \caption{\small 
    {Same as Figure~\ref{residual_main_7bins} but for the `pure binary sample'. }
    } 
   \label{residual_pure_7bins}
\end{figure*}

\section{ {Effects of PM scatters}} \label{sec:pmscat}

 {In this work, all samples have been selected with the requirement that the measurement uncertainties of all PM components are smaller than 0.5\%. In the main parts of testing gravity with median quantities, the tiny measurement  uncertainties of PMs have been ignored since scatters of hierarchical systems dominate the uncertainties of $\tilde{v}$. Here I present essential results of testing gravity with the acceleration-plane analysis and the $\Gamma$ (Equation~(\ref{eq:gamma})) analysis including PM scatters on top of the scatters of hierarchical systems as a matter of completeness and because PM scatters are considered in investigating the tail of the $\tilde{v}$ distribution in a study beyond the analyses with $\Gamma$ as discussed in Section~\ref{sec:result_hist}. When PM scatters are included, it is important to endow the same uncertainties to the mock PMs as to the measured PMs. I find that whenever this is done, the MC procedure works well in reproducing the input gravity law when tested with mock samples.}

\begin{figure*}
  \centering
  \includegraphics[width=0.65\linewidth]{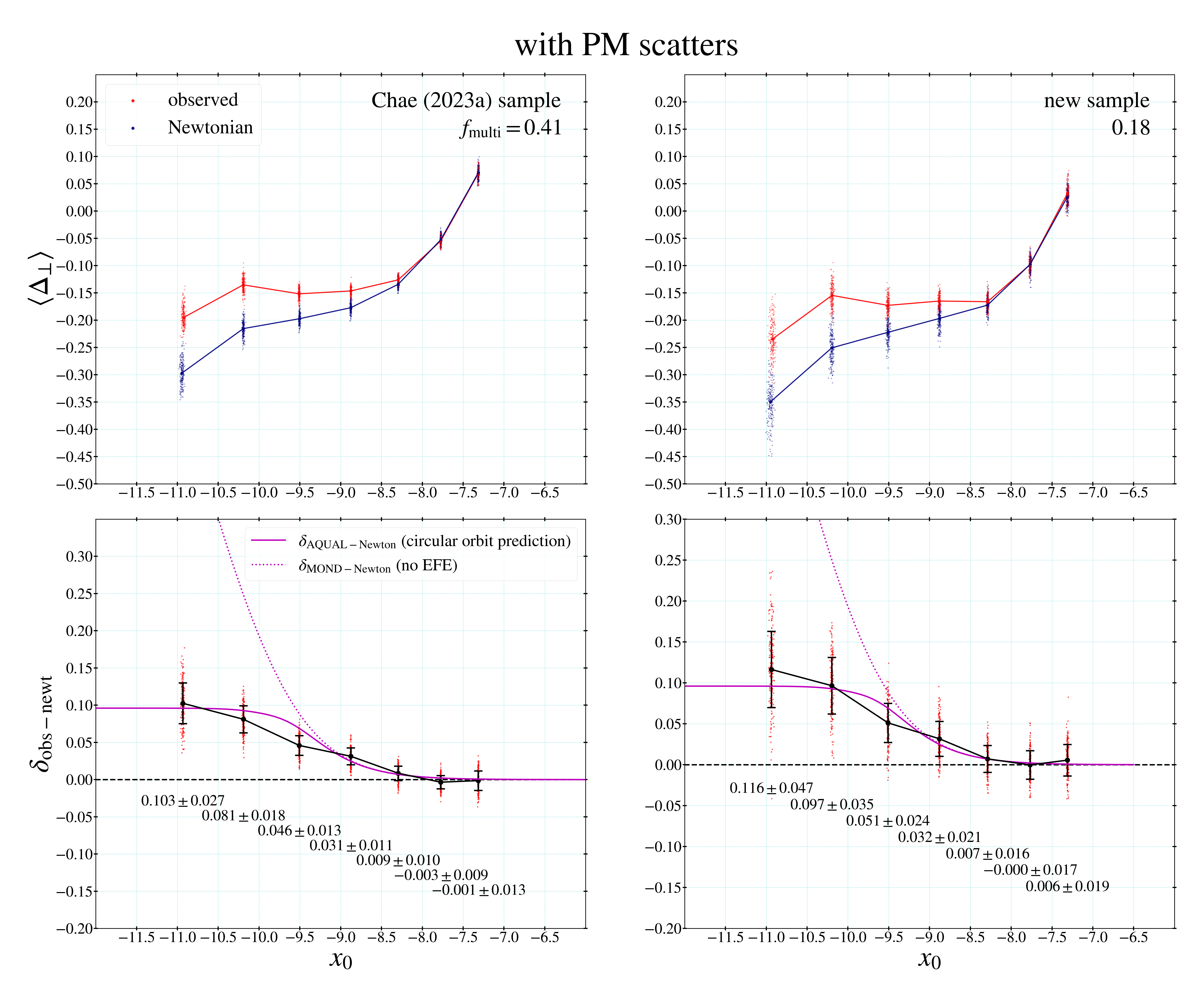}
    \vspace{-0.5truecm}
    \caption{\small 
     {Same as the left columns of Figure~\ref{residual_main_7bins} and Figure~\ref{residual_new_7bins} but with PM scatters included in the MC procedure. }
    } 
   \label{residual_pmscat_7bins}
\end{figure*}

 {Figure~\ref{residual_pmscat_7bins} shows the acceleration-plane results for the Chae (2023a) and new samples with PM scatters included in the MC procedure. It can be seen that both the observed and the Newton-predicted accelerations are somewhat shifted upward at lower accelerations due to the added scatters, but the gravitational anomaly (i.e.\ the difference between the two) remains essentially unaffected compared with the results of Figure~\ref{residual_main_7bins} and Figure~\ref{residual_new_7bins}. In detail, I note that the fitted value of $f_{\rm{multi}}$ for the Chae (2023a) sample is slightly lower.}

\begin{figure*}
  \centering
  \includegraphics[width=0.65\linewidth]{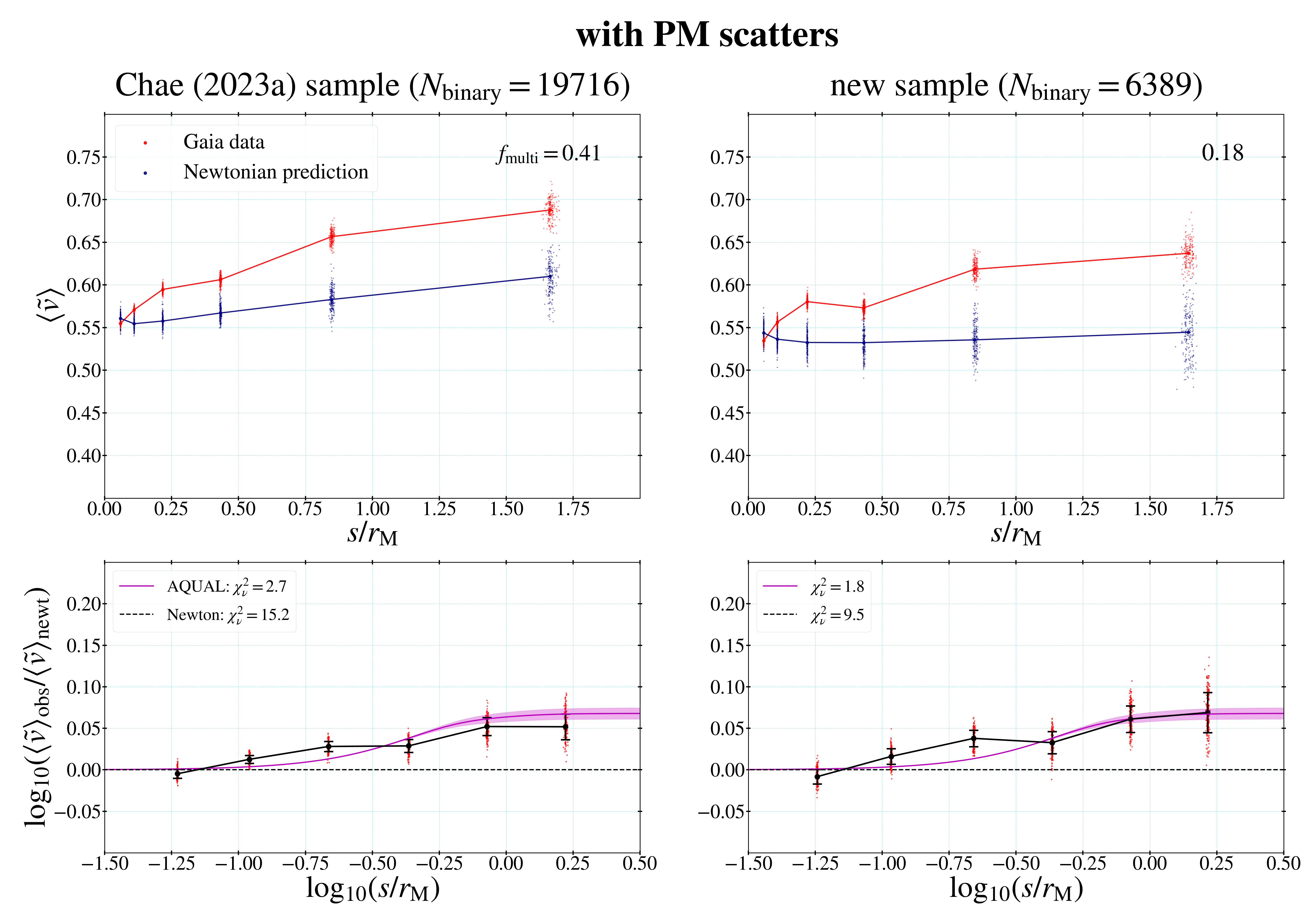}
    \vspace{0.truecm}
    \caption{\small 
     {Same as the left columns of Figure~\ref{vtest_main} and Figure~\ref{vtest_new} but with PM scatters included in the MC procedure. }
    } 
   \label{vtest_pmscat}
\end{figure*}

 {Figure~\ref{vtest_pmscat} further shows the results of testing gravity with $\Gamma$. For the Chae (2023a) sample the Newton-predicted $\langle\tilde{v}\rangle$ now rises more clearly with $s/r_{\rm{M}}$, but the gravitational anomaly is little affected because the observed  $\langle\tilde{v}\rangle$ also rises with $s/r_{\rm{M}}$ by an extra amount thanks to the included PM scatters. Also, the statistical significance measured by $\chi^2_\nu$ is only weakly affected.}

\end{document}